\documentclass[journal]{IEEEtran}

\usepackage{indentfirst}
\usepackage{url}
\usepackage{cite}

\ifCLASSINFOpdf
    \usepackage[pdftex]{graphicx}
\else

   \usepackage[dvips]{graphicx}
\fi

\usepackage{newtxtext,newtxmath}

\usepackage{float}

\usepackage{enumitem}

\usepackage{adjustbox}
\usepackage{booktabs}
\usepackage{multirow}
\usepackage{array}
\usepackage{stfloats}
\usepackage{threeparttable}
\usepackage[utf8]{inputenc}

\usepackage{algorithmicx}
\usepackage{algorithm} 
\usepackage{algpseudocode} 

\algnewcommand\algorithmicinput{\textbf{INPUT: }}
\algnewcommand\Input{\item[\algorithmicinput]}
\algnewcommand\algorithmicoutput{\textbf{OUTPUT: }}
\algnewcommand\Output{\item[\algorithmicoutput]}

\usepackage{color}

\usepackage[table]{xcolor}

\usepackage{hyperref}

\usepackage{chemscheme}
\usepackage{setspace}

\usepackage{titlesec}

\titleformat{\subparagraph}[runin]{}{}{0pt}{}

\renewcommand{\baselinestretch}{1.0}
\usepackage[utf8]{inputenc}
\usepackage{nomencl}
\titleformat{\subparagraph}[runin]{}{}{0pt}{}

\makenomenclature

\begin{document}
\bstctlcite{MyBSTcontrol}
\title{Non-Intrusive Load Monitoring in Smart Grids: A Comprehensive Review}

\author{Yinyan~Liu, ~\IEEEmembership{Member,~IEEE},  Yi~Wang, ~\IEEEmembership{Member,~IEEE},  ~Jin~Ma,~\IEEEmembership{Member,~IEEE} 
        }


\markboth{Proceedings of the IEEE,~Vol.~, No.~,}%
{Yinyan \MakeLowercase{\textit{et al.}}: Non-Intrusive Load Monitoring in Smart Grids: A Comprehensive Review}

\maketitle

\begin{abstract}
Non-Intrusive Load Monitoring (NILM) is pivotal in today's energy landscape, offering vital solutions for energy conservation and efficient management. Its growing importance in enhancing energy savings and understanding consumer behavior makes it a pivotal technology for addressing global energy challenges. This paper delivers an in-depth review of NILM, highlighting its critical role in smart homes and smart grids. The significant contributions of this study are threefold: Firstly, it compiles a comprehensive global dataset table, providing a valuable tool for researchers and engineers to select appropriate datasets for their NILM studies. Secondly, it categorizes NILM approaches, simplifying the understanding of various algorithms by focusing on technologies, label data requirements, feature usage, and monitoring states. Lastly, by identifying gaps in current NILM research, this work sets a clear direction for future studies, discussing potential areas of innovation. 
\end{abstract}


\section{Introduction}
\IEEEPARstart{A}{s} the global community grapples with the challenges of reducing carbon emissions and transitioning toward renewable energy sources, smart grids have emerged as a linchpin for achieving these ambitious goals. Smart grids leverage advanced communication, sensing, and control technologies to create a dynamic and responsive energy ecosystem \cite{cimen_online_2022}. By integrating renewable energy, energy storage, and intelligent demand-side management, smart grids enhance the efficiency, reliability, and resilience of the entire energy infrastructure.

\begin{figure}[ht]
\vspace*{-6pt}
\setlength{\abovecaptionskip}{-.1cm} 
\setlength{\belowcaptionskip}{-2cm} 
    \centering
    \includegraphics[width=\linewidth,scale=1.0]{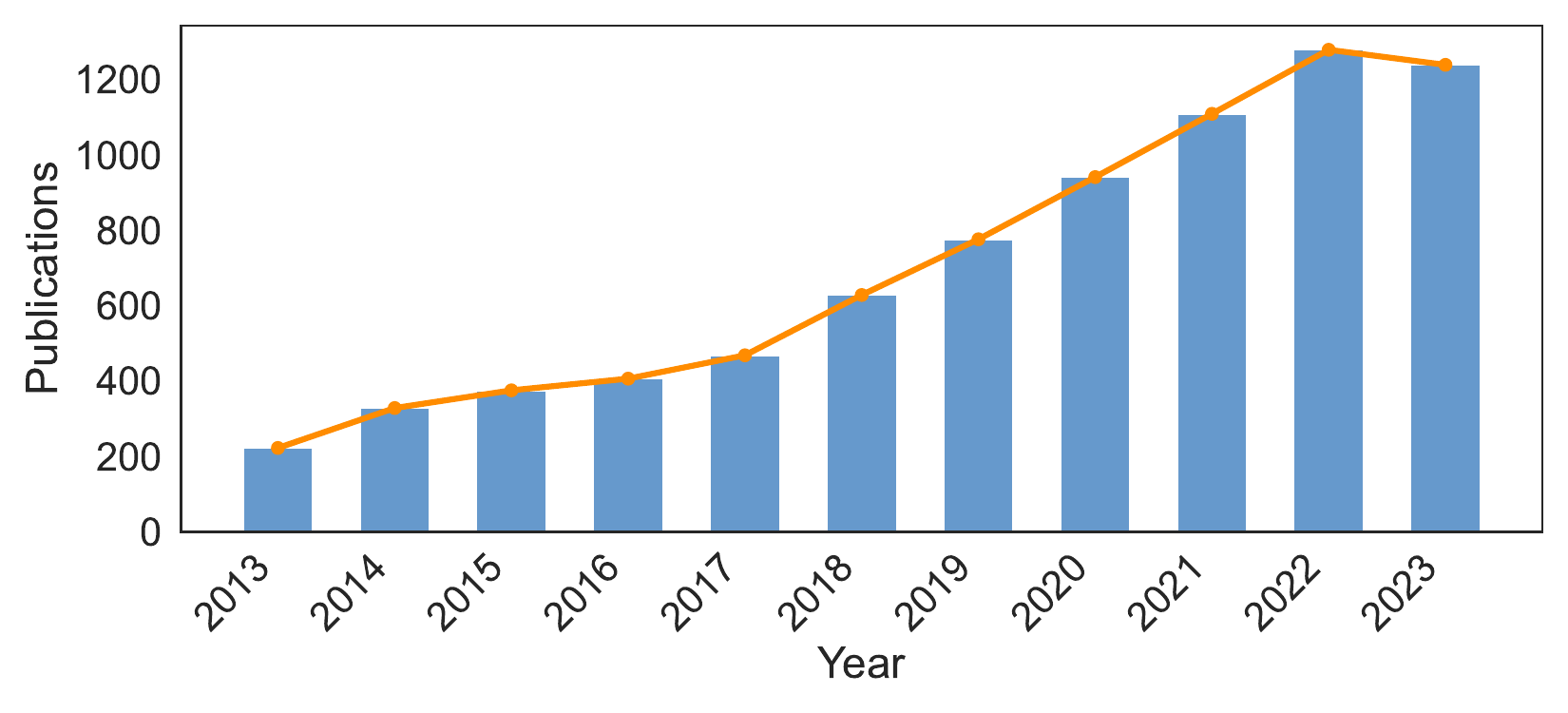}
    \caption{Annual publications on NILM from 2013 to 2023, with data sourced from Google Scholar.}
    \label{fig:annual_publications}
\vspace*{-6pt}
\end{figure}

Non-Intrusive Load Monitoring (NILM) has become a pivotal technology in the context of smart grids, offering innovative solutions to monitor and understand household energy consumption without the need for intrusive hardware installations \cite{tanoni_multilabel_2023}. This technology enables the disaggregation of total energy consumption into individual appliances' energy profiles, providing valuable insights for both consumers and utility providers. As the demand for sustainable and efficient energy use continues to grow, the development of NILM has gained prominence, as evidenced by the increasing number of annual publications (see Fig. \ref{fig:annual_publications}). The evident substantial growth in NILM research output since 2018 can be attributed to advancements in machine learning technologies.

By employing advanced algorithms and data analytics, NILM enables real-time disaggregation of overall energy usage, providing consumers, utilities, and grid operators with detailed information on the energy consumption patterns of specific appliances \cite{imran_power_2018}. This granular understanding facilitates informed decision-making, empowers users to adopt energy-efficient behaviors, and supports utilities in optimizing grid operations \cite{ruano2019nilm, you2022non, luo_atyd_2022}. Moreover, NILM aligns with the principles of demand response, allowing for the strategic management of energy loads during peak periods \cite{zhao2019quantifying}. As the integration of renewable energy sources becomes increasingly prevalent, NILM's ability to discern consumption patterns contributes to the seamless integration of sustainable practices within the smart grid framework. 

The existing review works on NILM have provided valuable insights into its methodologies, applications, and challenges. These reviews cover a broad range of topics, including machine learning-based algorithms \cite{huber_reviewdeep_2021, georgios_nilm_2022, schirmer2022non}, challenges \cite{georgios_nilm_2022, yan2022challenges, himeur2022recent}, and NILM applications across different domains \cite{ruano2019nilm, sayed_nonintrusive_2017, gopinath_energy_2020, rafati2022fault, georgios_nilm_2022, schirmer2022non}. However, a comprehensive review of NILM, encompassing traditional event-based and state-based detection to deep learning-based classification and energy disaggregation, is yet to be undertaken. Unlike existing review papers that focus on specific aspects of NILM, a comprehensive review should discuss the interdisciplinary nature of datasets, feature selection, feature extraction, technologies for monitoring, and metrics for model evaluation. Furthermore, there is a lack of up-to-date analyses of applications and challenges, considering the rapid advancements in NILM technologies. To address these limitations, a need exists for a comprehensive and contemporary review that integrates various dimensions of NILM, providing a more holistic understanding and identifying future research directions in this dynamic field.

\subsection{Scope of This Paper}
 The comprehensive scope can become a valuable resource for researchers, practitioners, and policymakers aiming to harness the full potential of NILM in creating smarter and more efficient energy systems. The scope of this work includes the following aspects:

 \begin{enumerate}[label=\arabic*).]
  \item \textbf{Users}: This work explores the extensive landscape of NILM with a focus on its diverse applications across residential, commercial, and industrial sectors. 
  \item \textbf{Datasets}: The investigation spans datasets sourced from government, public, universities, and scientific repositories
  \item \textbf{Methods}: A methodological spectrum encompassing traditional statistical algorithms, classical machine learning approaches (e.g., K-means), and cutting-edge deep learning models tailored for NILM. 
\end{enumerate}

\subsection{Contributions and Outline} 

\begin{figure}[ht]
\vspace*{-6pt}
\setlength{\abovecaptionskip}{-.1cm} 
\setlength{\belowcaptionskip}{-2cm} 
    \centering
    \includegraphics[width=\linewidth,scale=1.0]{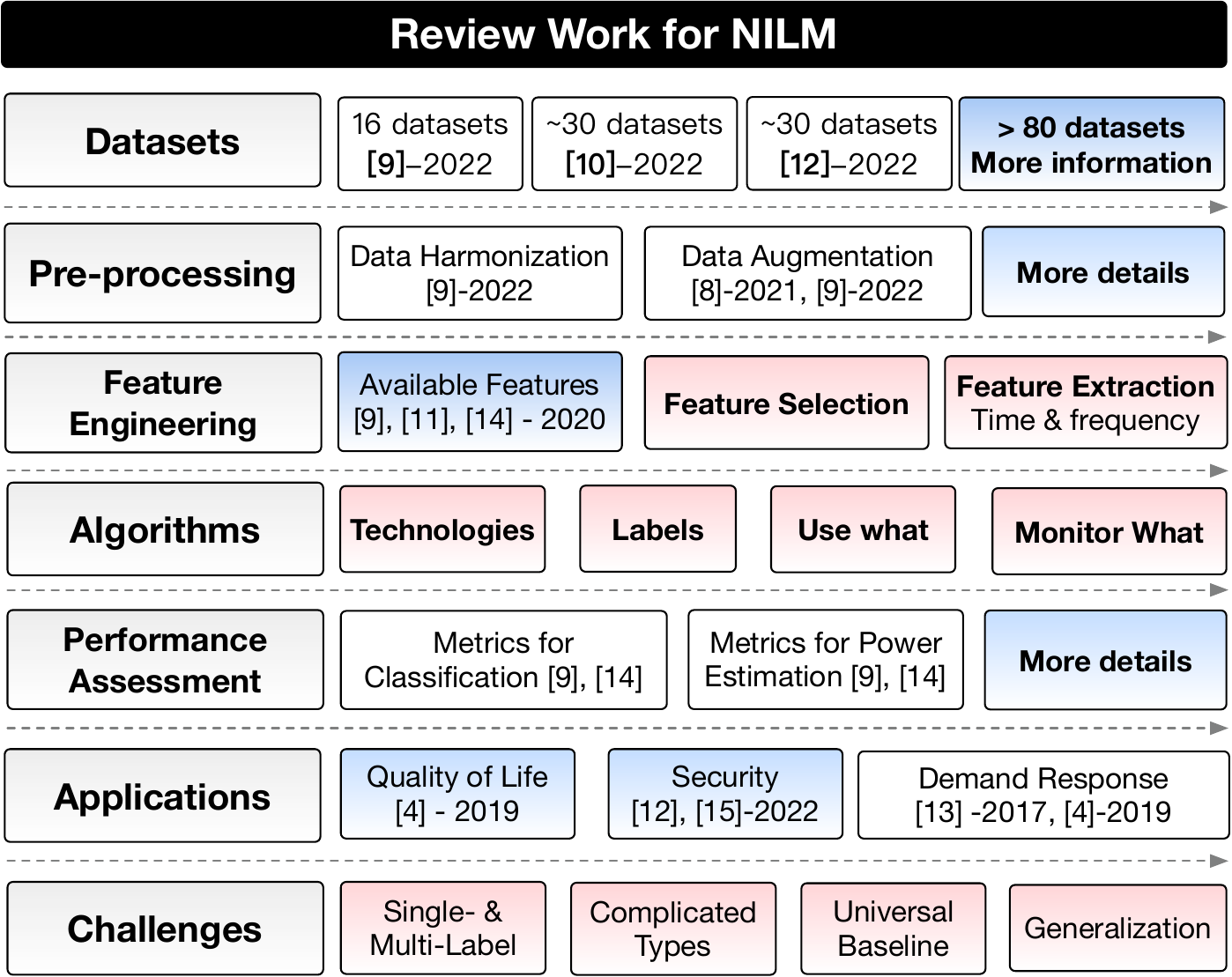}
    \caption{The contributions and novelties of this work are compared to the existing review work. The red color highlights the novelties of this work, while the blue color emphasizes the additional details and insights provided by this work.}
    \label{fig:reveiw}
\vspace*{-6pt}
\end{figure}

This comprehensive overview not only enhances the understanding of NILM methodologies but also guides future research directions in this rapidly evolving field. The contributions and outline of this work are summarized as follows (also see Fig. \ref{fig:reveiw}),
\begin{enumerate}[leftmargin = 1em]
\item{This work presents a thorough and comprehensive review of existing studies on NILM. It includes an extensive discussion of various critical components such as datasets, pre-processing algorithms, feature engineering, and methodologies from diverse perspectives. Additionally, the paper delves into the applications, challenges, and future prospects in the field of NILM, distinguishing it as a significant contribution to the current body of research. } 

\item{We have compiled a comprehensive table for NILM datasets from published papers and websites, providing a crucial resource for researchers and engineers in NILM to select the most suitable dataset for their work.}

\item{This work uniquely categorizes NILM approaches from various perspectives, including label data requirements, feature usage, and monitoring states or energy consumption, enabling a thorough discussion of NILM algorithms. We also delve into the advantages and challenges of each method, providing clear insights into their practical applications and limitations.}

\item{The review of NILM studies brings forth creative insights and identifies existing gaps in the literature, leading to a discussion of prospective research directions for future advancements in the field.}
\end{enumerate}

\section{Overview of NILM Framework}
This section provides a comprehensive overview of the NILM methodology. First, we establish the NILM problem formulation. Then, data preprocessing steps, including dataset availability, data harmonization, and data augmentation for NILM, are introduced. 

\subsection{NILM Problem Formulation}
\begin{figure}[ht]
\vspace*{-10pt}
\setlength{\abovecaptionskip}{-.1cm} 
\setlength{\belowcaptionskip}{-2cm} 
    \centering
    \includegraphics[width=\linewidth,scale=1.0]{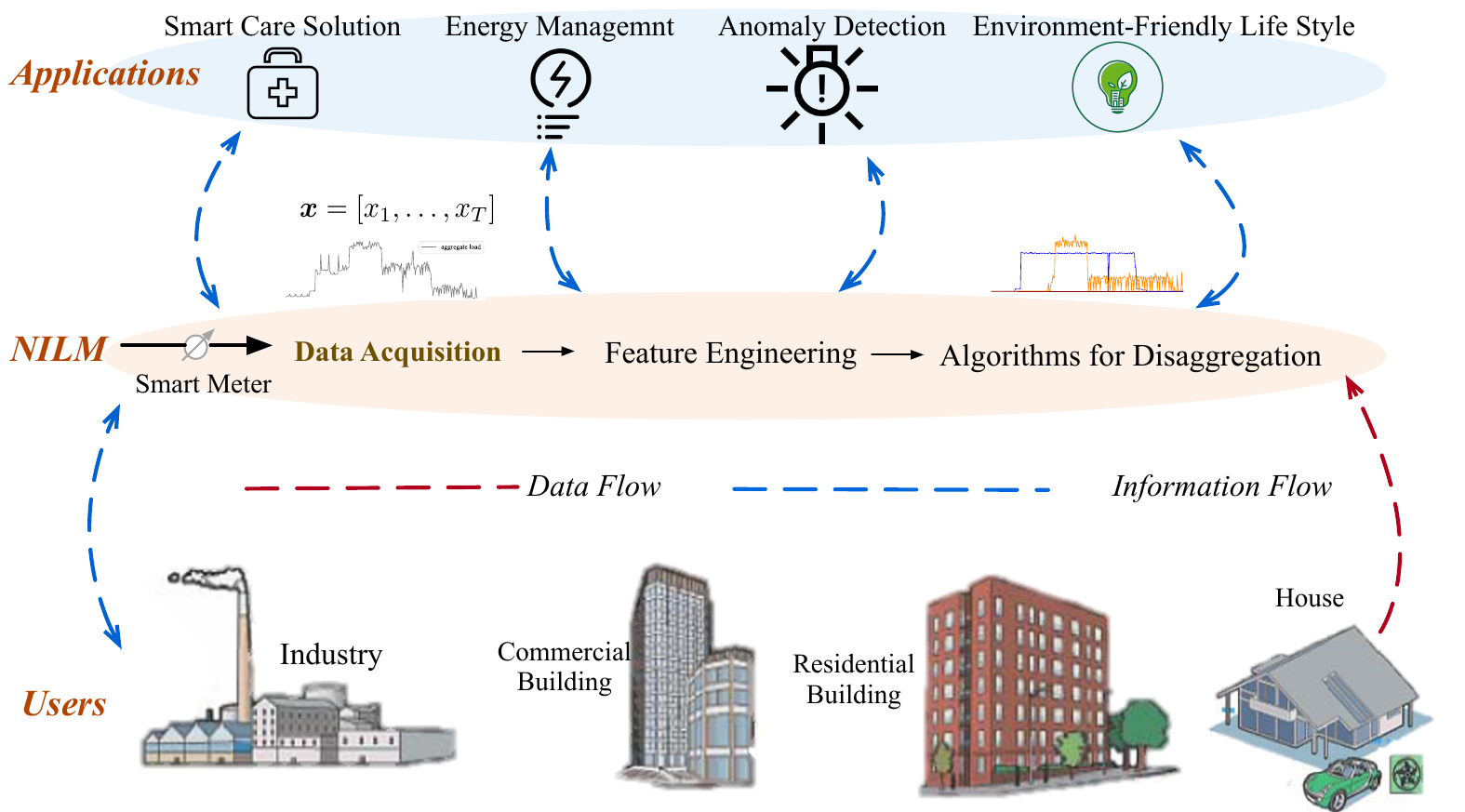}
    \caption{Overview of NILM framework and its applications.}
    \label{fig:overall}
\vspace*{-10pt}
\end{figure}

NILM is a technology that identifies the individual power consumption and other characteristics of electrical appliances from a single measurement point at the power entry. Suppose the load consumption of the $i$th electrical appliance in a home or building is $\boldsymbol{y}_{i}=(y_1, \dots, y_T)\in \mathcal{Y}$. In that case, the noisy aggregate load consumption $\boldsymbol{x}=(x_1, \dots, x_T)\in \mathcal{X}$ measured by the smart meter, can be modeled as follows,
\begin{equation}
    x_t = \sum_{i=1}^{N}(s_{i,t} \cdot y_{i,t}) + \epsilon_t \label{eq:nilm}
\end{equation}
where $t$ is the time slot, $T$ is the time slot number in a sample $\boldsymbol{x}\in \mathbb{R}^{T\times K}$, $K$ is the measured feature number of the aggregate load, $\mathcal{Y}$ represents the labeled data in the dataset, $N$ is the number of electrical appliances in a house/building, $\mathcal{F}(\cdot)$ is the aggregation function, $s_{i,t}$ is a binary variable representing the ON/OFF operating state of $i$th appliance, and $\epsilon_t$ generally indicates measurement noise or error item with uncounted contribution of appliances \cite{hart_non_1992}. The primary goal of NILM is to provide the disaggregation method $\mathcal{A}(\cdot)$ to estimate the power consumption $y_{i,t}$ or $\mathcal{S}(\cdot)$ to monitor the states (e.g., ON/OFF state) of the electric appliances,
\begin{align}
    \{\hat{y}_{1,t},\dots, \hat{y}_{N,t}\} = \mathcal{A}(\boldsymbol{x})\label{eq:prodisaggregation}\\
    \{\hat{s}_{1,t},\dots, \hat{s}_{N,t}\} = \mathcal{S}(\boldsymbol{x})\label{eq:proevent_detection}
\end{align}
s.t.
\begin{align}
\mathcal{A}(\boldsymbol{x}) &= \min\left(\sum_{i=1}^{N}(\hat{y}_{i,t} -  {y}_{i,t}) \right)\label{eq:pro_constraint1}\\
\mathcal{S}(\boldsymbol{x}) &= \min\left(\sum_{i=1}^{N}(\hat{s}_{i,t} -  {s}_{i,t})\right)\label{eq:pro_constraint2}
\end{align}
where $\hat{y}_{i,t}, \hat{s}_{i,t}$ are the estimations of ${y}_{i,t}, {s}_{i,t}$, respectively. Generally, the ON/OFF state ${s}_{i,t}$ of an appliance can be obtained with a threshold value $\xi$ from the load consumption ${y}_{i,t}$,
\begin{equation}
\label{eq: state}
    {s}_{i,t}=\left\{
\begin{aligned}
\textit{OFF}, & ~~ 0\le {y}_{i,t} \le \xi \\
\textit{ON}, & ~~ {y}_{i,t}> \xi
\end{aligned}
\right.
\end{equation}
Note that the threshold value $\xi$ varies for different appliances. The overall structure of the NILM methodology is illustrated in Figure \ref{fig:overall}. Generally, NILM encompasses data acquisition, data preprocessing, feature engineering, and learning approaches to monitor the states or energy consumption of appliances. 

\subsection{Data Acquisition}
Data acquisition involves acquiring the aggregate load $\textbf{x}_t$, achievable using various power meters \cite{hart_non_1984,wang_residential_2012,zoha_non_2012,sreev_design_2019}. Before proposing a specific approach to NILM, researchers generally need to determine the purpose of their study and the required features, and then select and download a dataset that meets these requirements. In recent years, there has been a notable increase in the number of datasets released for NILM studies. The authors in \cite{schirmer2022non} recently discussed 28 available public datasets on customer electricity use, highlighting their limitations. 

To provide a comprehensive overview or a checklist tool of the existing datasets for NILM, Table \ref{tab:datasets} summarizes more extensive datasets with diverse features and parameters. This summary briefly outlines the characteristics of each dataset, including features such as active power (P), reactive power (Q), apparent power (S), voltage (V), current (I), gas energy (G), frequency (f), phase angle (\(\theta\)), power factor (pf), energy cost (EC), weather (Wt) including indoor or outdoor temperature, customer's occupancy (O), Photovoltaic (PV), electric vehicle (EV), and ON-OFF event (Et). The 'access' column in Table \ref{tab:datasets} indicates the dataset's available access type. \textbf{'Reg.' and 'Acade.'} mean that accessing the dataset requires registration or an academic email, respectively. 'No-acc' signifies that detailed information about the dataset is not readily available, but descriptions and analyses of the dataset can be accessed.

\begin{table*}[htbp]
\vspace*{-10pt}
\setlength{\abovecaptionskip}{-.1cm} 
\setlength{\belowcaptionskip}{-2cm} 
\footnotesize{
    \centering
    \caption{Electricity consumption datasets from published articles or projects, including residential (R), commercial (C), and industrial (I) types.}
    \label{tab:datasets}
    \begin{adjustbox}{width=\linewidth, center}
    \begin{tabular}{l| l| c| l |l |l |c| c |c |l| c |c |c |c}
    \hline
    \toprule
    \hline
      \multirow{2}{*}{\#}  &\multirow{2}{*}{Name} & \multirow{2}{*}{Release}&\multirow{2}{*}{Type}&\multirow{2}{*}{Access} & \multirow{2}{*}{Country}& \multirow{2}{*}{\# Homes}& \multirow{2}{*}{\begin{tabular}[c]{@{}c@{}} \#Sub\\ Meters\end{tabular}} &\multirow{2}{*}{Duration} & \multirow{2}{*}{Features} & \multirow{2}{*}{\begin{tabular}[c]{@{}c@{}} Optional\\ Features\end{tabular}} & \multirow{2}{*}{\begin{tabular}[c]{@{}c@{}} Measured\\ Parameters\end{tabular}} & \multicolumn{2}{c}{Sampling Rate} \\
    \cline{13-14}
      & & & & & & & &  &  &  &   & Agg & App\\
     \hline
   1&	CTAMT \cite{ctamt_2006}&	2006&	C&	Reg.&	UK&	200+&	-&	3 years&	P&	-&	Wt&	30 mins&	-\\
    2&	TUVA \cite{einfalt_TUVA_2008}&	2008&	R&	Req.&	Upper Austria&	30&	-&	1 week&	P, Q, V&	-&	PV&	1 sec&	-\\
    3&	OMRDEUD \cite{thomson_ukda_2009}&	2009&	R, C&	Reg.&	UK&	20&	24&	2 years&	P&	-&	O&	1 min&	1 min\\
    4&	ADRES \cite{fur_ADRES_2011}&	2011&	R&	Acade.&	Austria&	30&	-&	2 weeks&	V&	-&	-&	1 sec&	-\\
    5&	DECC \cite{decc_2011}&	2011&	R, C&	Public&	Japan&	299&	-&	4 years&	P&	-&	O&	1 hour&	-\\
    6&	REDD \cite{kolter_redd_2011}&	2011&	R&	Public&	USA&	6&	9-24&	2-4 weeks&	P, V, I&	-&	-&	15k/1 Hz&	3 secs\\
    7&	BLUED \cite{kyle_blued_2012}&	2012&	R&	Public&	USA&	1&	1&	1 week&	P, Q, V, I&	-&	-&	12 kHz&	1 Hz\\
    8&	CLNR \cite{clan_2014}&	2012&	R, C&	Reg.&	England&	9000&	-&	1+ years&	P&	-&	-&	30 mins&	-\\
    9&	IHEPCDS \cite{georeges_uci_2012}&	2012&	R&	Public&	France&	1&	3&	4 years&	P, Q, V, I&	-&	-&	1 min&	1 min\\
    10&	MEULPv1 \cite{neil_MEULPv1_2012}&	2012&	R&	Public&	Canada &	11&	8&	1 year&	P&	-&	-&	1 min&	1 min\\
    11&	Smart* \cite{Mishra2012Smart}&	2012&	R&	Public&	USA&	400+&	25&	3+ years&	P, S&	-&	Wt&	1 min&	1 min\\
    12&	Tracebace \cite{andreas_tracebase_2012}&	2012&	R&	Public&	Germany&	14&	31&	6-34 days&	P&	-&	-&	1-3 secs&	1-3 secs\\
    13&	ACS-F1 \cite{gisler_ACS-F1_2013}&	2013&	R&	Public&	Switzerland&	-&	100&	1 hour&	P, Q, V, I&	 f, $\theta$&	-&	-&	10 secs\\
    14&	AMPds \cite{DVNFIE0S4_2016}&	2013&	R&	Public&	Canada&	1&	19&	2 years&	P, Q, V, I, S&	pf, f&	-&	1 min&	1 min\\
    15&	AusgridSH \cite{Ausgrid_solardata_2013}&	2013&	R&	Public&	Australia&	300&	-&	3 years&	P&	-&	-&	30 mins&	-\\
    16&	BERDS \cite{mehdi_BERDS_2013}&	2013&	C&	Public&	USA&	1&	4&	1 year&	P, Q, S&	-&	-&	20 secs&	20 secs\\
    17&	CRHLP \cite{ong_tmy3_2012}&	2013&	C&	Public&	USA&	16&	-&	1 year&	P&	-&	-&	1 hour&	-\\
    18&	Dataport \cite{pecan_dataport_2013}&	2013&	R&	Acade.&	USA&	1000+&	70&	4 years&	P&	-&	Wt, PV, EV&	1 min&	1 min\\
    19&	iAWE \cite{batra_iame_2013}&	2013&	R&	Public&	Delhi, India&	1&	33&	73 days&	P, V, I, S&	f, $\theta$&	G&	1 Hz&	1 Hz\\
    20&	GREEND \cite{monacchi_2014_GREEND}&	2013&	R&	Public&	Austria \& Italy&	8&	8&	1 year&	P&	-&	-&	1 sec&	1 sec\\
    21&	COMBED \cite{batra_COMBED_2014}&	2014&	C&	Public&	India&	6&	200&	1 month&	P, V&	-&	-&	30 secs&	30 secs\\
    22&	ECB \cite{AU_ECB_2014}&	2014&	R&	Public&	Australia&	25&	-&	2 years&	P&	-&	-&	1 Hz&	-\\
    23&	ECO \cite{kleiminger_ECO_2015}&	2014&	R&	Public&	Swiss&	6&	7-12&	8 months&	P, Q, V, I&	$\theta$&	O&	1 Hz&	1 Hz\\
    24&	ECOD \cite{beckel_ECOD_2014}&	2014&	R&	Public&	Switzerland&	6&	-&	8 months&	V, I&	$\theta$&	-&	1 Hz&	-\\
    25&	LCLP \cite{james_LCLP_2014}&	2014&	R&	Public&	London, UK&	5567&	&	26 months&	P&	-&	O&	30 mins&	-\\
    26&	RBSAM \cite{ben_RBSAM_2014}&	2014&	R&	No-acc&	Pacific Northwest&	101&	-&	4 years&	P&	-&	Wt&	15 mins&	1 hour\\
    27&	SustData \cite{lucas_SustData_2014}&	2014&	R&	Public&	Portugal&	50&	24&	1144 days&	P, Q, V, I, S&	-&	Wt&	2 secs&	10 secs\\
    28&	DRED \cite{uttama_DRED_2015}&	2015&	R&	Public&	Netherlands&	1&	12&	6 months&	P&	-&	Wt&	1 sec&	1 sec\\
    29&	EUFINESCE \cite{rwth_FINESCE_2015}&	2015&	R, C&	CC Lic.&	Italy&	20&	-&	several days&	P&	-&	Wt, EV, PV&	1 hour&	1 hour\\
    30&	OCTES \cite{octes_2011}&	2015&	R&	No-acc&	Iceland&	33&	-&	13 months&	P&	-&	PV&	7 secs&	-\\
    31&	OPLD \cite{kalluri_opld_2015}&	2015&	C&	No-acc&	Singapore&	-&	-&	-&	P, Q, S, I&	-&	&	1 sec&	1 sec\\
    32&	I-BLEND \cite{rashid_2019}&	2015&	R, C&	No-acc&	Delhi, India&	7&	-&	52 months&	P, V, I&	-&	O&	-&	30 mins\\
    33&	REFIT \cite{kleimi_REFIT_2015}&	2015&	R&	Public&	UK&	20&	8&	18 months&	P&	pf&	Wt, O, EC&	8 secs&	8 secs\\
    34&	SGSC \cite{au_sgsc_2015}&	2015&	R, C&	Public&	Australia&	17000&	-&	21 months&	P&	-&	Wt, O, EC&	30 mins&	-\\
    35&	UK-DALE \cite{UK-DALE_2015}&	2015&	R&	Public&	UK&	5&	5-40&	655 days&	P, Q, V, I, S&	-&	Et&	16 kHz&	6 secs\\
    36&	ENLITEN \cite{ENLITEN_2016}&	2016&	R&	Public&	UK&	1-14&	-&	30 days&	P&	-&	Wt, O, Et&	1 min &	-\\
    37&	HES \cite{uk_hes_2011}&	2016&	R, C&	Reg.&	England&	250&	23&	1+ years&	P&	-&	O, PV&	2/10 mins&	2/10 mins\\
    38&	SustDataED \cite{miguel_SustDataED_2016}&	2016&	R&	Public&	Portugal&	-&	17&	10 days&	P, Q, V, I&	-&	-&	50-12.8k Hz&	2 secs\\
    39&	AMBEL \cite{buneeva_AMBAL_2017}&	2017&	R&	Public&	Simulation&	-&	14&	1 day&	P&	-&	-&	1 sec&	1 sec\\
    40&	BDG1 \cite{miller_BDGP_2017}&	2017&	C, I&	Public&	USA&	507&	-&	2+ years&	P&	-&	-&	1 hour&	-\\
    41&	BDG2 \cite{Miller2020-bdg2}&	2017&	R, C, I&	Public&	USA, Europe&	1636&	-&	2 years&	P&	-&	Wt&	1 hour&	-\\
    42&	EEUD \cite{Johnson_EEUD_2017}&	2017&	R&	No-acc&	Canada&	23&	-&	1 year&	P&	-&	-&	1 min&	1 min\\
    43&	ESHL \cite{xu_ESHL_2018}&	2017&	R&	No-acc&	Germany&	1&	-&	4-5 years&	P, V,I&	-&	-&	1-2 secs&	1-2 secs\\
    44&	MEULPv2 \cite{geoffrey_MEULPv2_2017}&	2017&	R&	Public&	Canada&	23&	5&	1 year&	P&	-&	-&	1 min&	1 min\\
    45&	DISEC \cite{victor_DISEC_2018}&	2018&	R&	Public&	India&	19&	-&	284 days&	P&	-&	-&	15 min&	-\\
    46&	EMBED \cite{jazi_EMBED_2018}&	2018&	R&	Public&	Los Angeles&	3&	17&	14-27 days&	P, Q, V, I&	-&	-&	60-12k Hz&	1-2 Hz\\
    47&	LEEDR \cite{LEEDR_2018}&	2018&	R&	Public&	UK&	20&	-&	3+ years&	P&	-&	Wt, G, O, Et&	1 min&	-\\
    48&	RAE \cite{makonin_rae_2017}&	2018&	R&	Public&	Canada&	2&	21-24&	72 days&	V, P, Q, S&	f&	O, G&	1 Hz&	1 Hz\\
    49&	PRECON \cite{Nadeem_precon_2019}&	2018&	R&	Public&	Pakistan&	42&	6-26&	1+ years&	P&	-&	-&	1 min&	1 min\\
    50&	SHED \cite{simon_SHED_2018}&	2018&	C&	Public&	Simulation&	8 &	66&	14 days&	V, I&	-&	Wt, O&	0.033 Hz&	0.033 Hz\\
    51&	BFDDD \cite{granderson_BFDDD_2020}&	2019&	C&	Public&	Simulation&	1&	-&	1 + years&	P&	-&	Falut, Wt&	-&	-\\
    52&	DEFACTO \cite{haines_DEFACTO_2019} &	2019&	R&	No-acc&	UK&	186&	-&	10 months&	P& -&	G, Wt, O&	10 mins&	-\\
    53&	ENERTALK \cite{shin_ENERTALK_2019}&	2019&	R&	Public&	South Korea&	22&	7&	29-122 days&	P, Q&	-&	-&	15Hz&	15Hz\\
    54&	HUE \cite{stephen_HUE_2019}&	2019&	R&	Public&	Canada&	22&	-&	3 years&	P&	-&	Wt, PV&	1 hour&	-\\
    55&	NextGen (NG) \cite{shaw_NextGen_2019}&	2019&	R, C&	Public&	Australia&	5000&	-&	6 + years&	P, Q, V&	-&	-&	5 mins&	-\\
    56&	RSBS \cite{baylon_RSBS_2019}&	2019&	R&	Reg.&	USA&	100&	30+&	27 months&	P&	-&	O&	15 mins&	15mins\\
    57&	SERL \cite{webborn_SERL_2019}&	2019&	R&	UK-Uni&	UK&	8000+&	-&	1+ years&	P&	-&	E&	30 mins&	-\\
    58&	CoSSMic \cite{open_CoSSMic_2020}&	2020&	R, C,&	Public&	Germany&	11&	5&	1-3 years&	P&	-&	PV, EV, Bty&	1 min&	1 min\\
    59&	FreshEnergy \cite{zenodo_freshenergy_2020}&	2020&	R&	CC Lic.&	Germany&	200&	-&	1 year&	P&	-&	O&	15 mins&	15 mins\\
    60&	IEEEPESOD \cite{IEEE_Opendatasets_2020}&	2020&	R, C&	Public&	USA&	15&	2&	2 weeks&	P&	-&	-&	1-5 mins&	1-5 mins\\
    61&	CU-BEMS \cite{manisa_cubems_2020}&	2020&	C&	Public&	Thailand&	7&	-&	18 months&	P&	-&	Wt&	-&	1 sec\\
    62&	QUD \cite{Himeur_QUD_2020}&	2020&	R&	Public&	Doha, Qatar &	3&	4&	3+ months&	P&	-&	Wt, O&	3+ secs +&	3 s-30 mins\\
    63&	SEIL-R \cite{Joshi_SEIL-R_2020}&	2020&	R&	Public&	Mumbai, India&	14&	-&	1 year&	P, Q, V, I&	-&	-&	1 sec&	-\\
    64&	SynD \cite{klemenjak_SysD_2020}&	2020&	R&	Public&	Austria&	-&	21&	180 days&	P&	-&	-&	5 Hz&	5 Hz\\
    65&	ABSD \cite{li_ABSD_2021}&	2021&	C&	Public&	Simulation/USA&	1395&	3&	1 year&	P&	-&	Wt, O&	10 mins&	10 mins\\
    66&	DEDDIAG \cite{marc_DEDDIAG_2021}&	2021&	R&	Public&	Germany&	15&	50&	3.5 years&	P& 	-&	Et&	1 Hz&	1 Hz\\
    67&	IDEAL \cite{pullinger_IDEAL_2021}&	2021&	R&	Public&	UK&	255&	46&	23 months&	P&	-&	Wt, O, G&	1 Hz&	1 Hz\\
    68&	OEDI \cite{oedi_2021}&	2021&	R, C&	Public&	USA&	-&	-&	3 years&	P&	-&	Wt&	15 mins&	15 mins\\
    69&	MFRED \cite{meinrenken_MFRED_2020}&	2022&	R&	Public&	USA&	390&	-&	1 years&	P, Q&	-&	O&	10 secs&	-\\
    70&	ATYD \cite{luo_atyd_2022}&	2022&	C&	Public&	Berkeley, USA&	1&	300&	3 years&	P&	-&	Wt&	15 mins&	15 mins\\
    \hline													
     \multirow{3}{*}{71} &	 \multirow{3}{*}{ECD-UY \cite{chavat_ecd-uy_2022}} &	\multirow{3}{*}{2022} &	\multirow{3}{*}{R} &	\multirow{3}{*}{Public} &	\multirow{3}{*}{Uruguay} &	110953&	-&	1 year&	P&	-&	O&	15 mins&	-\\
    &	&	&	&	&	&	268&	-&	4 months&	P&	-&	O&	1 mins&	-\\
    &	&	&	&	&	&	9&	3+&	21 days&	P&	-&	O&	1 min&	1 min\\
    \hline													
    72&	IEDL \cite{deepika_IEDL_2022}&	2022&	R&	Public&	India&	1&	-&	1 year&	P,Q,S, V, I&	f, $\Phi$&	-&	1 min&	1 min\\
    \hline
    \hline
    73&	PLAID \cite{gao_plaid_2014}&	2014&	R&	Public&	USA&	55&	11&	-&	P&	-&	-&	-&	30 kHz\\
    74&	HFED \cite{gulati14EMI}&	2015&	R&	Public&	India&	2&	15&	-&	-&	EMI&	-&	-&	5 MHz\\
    75&	COOLL \cite{thomas_cooll_2016}&	2016&	R&	Public&	France&	-&	12&	-&	V, I&	-&	Wt&	-&	100 kHz\\
    76&	WHITED \cite{kahl_whited_2016} &	2016&	R&	Public&	Germany&	-&	110&	5 secs&	V, I&	-&	-&	-&	44kHz\\
    77&	BLOND \cite{kriech_blond_2017}&	2018&	C&	Public&	Germany&	1&	53&	50-213 days&	V, I&	-&	-&	50-250 kHz&	6.4-50 kHz\\
    78&	IMD \cite{bandeira_IMD_2018}&	2018&	I&	Public&	Brazil&	1&	10&	111 days&	P, Q, V, I, S&	-&	-&	8 kHz&	8 kHz\\
    79&	LIT \cite{renaux_LIT_2018}&	2018&	R&	Reg.&	Brazil&	-&	-&	-&	V, I&	-&	Et&	15 kHz&	-\\
    80&	LILACD \cite{matthias_LILACD_2019} &	2019&	R&	Public&	Germany&	-&	15&	-&	V, I&	-&	-&	-&	50 kHz\\
    81&	CREAM \cite{krie_CREAM_2020} &	2020&	I&	Public&	Germany&	2&	8&	-&	V, I&	-&	-&	-&	6.4 kHz\\
    82&	FIRED \cite{volker_FIRED_2020} &	2020&	R&	Public&	Germany&	-&	21&	52 days&	V, I&	-&	Wt&	8 kHz&	2 kHz\\
    83&	SustDataED2 \cite{pereira_sustdataed2_2022}&	2022&	R&	Public&	USA&	-&	18&	96 days&	P, Q, V, I&	-&	Et&	12.8 kHz&	0.5 Hz\\
    \toprule
    \hline
    \multicolumn{14}{l}{* Reg.: registration, ~ Req.: requirement, ~ Acade: academic usage,~ No-acc: published dataset or project without available data, ~ UK-Uni: only access by universities in UK.}\\
    \multicolumn{14}{l}{* CC Lic.: Creative Commons Attribution Share-Alike.}\\
    \multicolumn{14}{l}{* P: active power, ~ Q: reactive power, ~S: apparent power, ~V:voltage, ~ I: current, ~ G: gas energy, ~ f: frequency, ~$\theta$: phase angle, ~pf: power factor, ~ EC: energy cost,}\\
     \multicolumn{14}{l}{* Wt: weather features, including indoor or outdoor temperature, humidity, etc., ~O: customer's occupancy, ~PV: photovoltaic, ~ EV: electric vehicle, and Et: ON-OFF event.}\\
     \multicolumn{14}{l}{* Bty: Battery information, including the charging and discharging states. }
    \end{tabular}
    \end{adjustbox}}
\vspace*{-6pt}
\end{table*}

 Differing from existing review papers on NILM \cite{sayed_nonintrusive_2017, reza_big_chapter_2018, gopinath_energy_2020, yassine_building_2020, georgios_nilm_2022}, Table \ref{tab:datasets} offers several times more datasets than those previously reviewed. Additionally, this table categorizes datasets by type, such as residential, commercial, or industrial, aiding researchers in studying NILM under various scenarios. Moreover, Table \ref{tab:datasets} provides not only general features like active and reactive power but also includes non-electrical features such as occupancy and indoor/outdoor temperature information for further NILM analysis. With the growing popularity of PV systems and EV among users, and their increasing relevance in advancing a low-carbon lifestyle, researchers can find datasets including PV, EV, and pricing information in Table \ref{tab:datasets} for related research.

\begin{figure}[ht]
\vspace*{-6pt}
\setlength{\abovecaptionskip}{-.1cm} 
\setlength{\belowcaptionskip}{-2cm} 
    \centering
    \includegraphics[width=0.86\linewidth,scale=1.0]{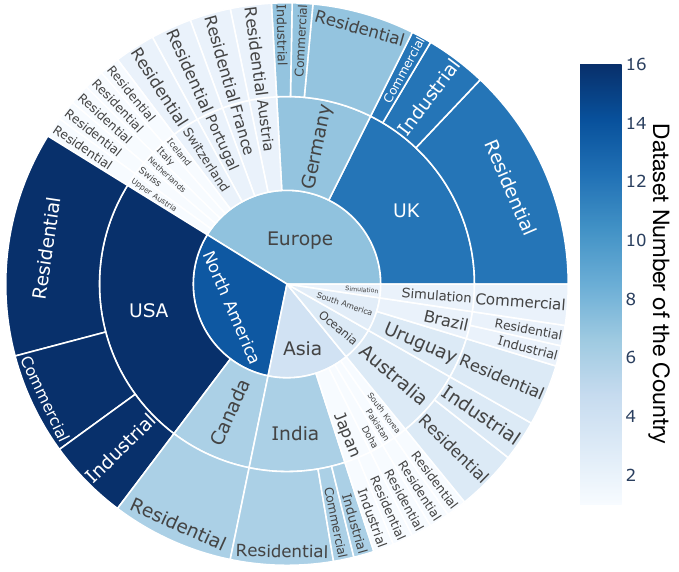}
    \caption{Statistics on the dataset number and type in different countries.}
    \label{fig:pub_dataset_sun}
\vspace*{-6pt}
\end{figure}

Compared to large number of residential datasets, the availability of industrial datasets is extremely limited, with only a few such as BDG1 \cite{miller_BDGP_2017}, BDG2 \cite{Miller2020-bdg2}, and CREAM \cite{krie_CREAM_2020}. In addition to features like active and reactive power, datasets such as OMRDEUD \cite{thomson_ukda_2009}, DECC \cite{decc_2011}, ECO \cite{kleiminger_ECO_2015}, LCLP \cite{james_LCLP_2014}, REFIT \cite{kleimi_REFIT_2015}, SGSC \cite{au_sgsc_2015}, HES \cite{uk_hes_2011}, RAE \cite{makonin_rae_2017}, SHED \cite{simon_SHED_2018}, and QUD \cite{Himeur_QUD_2020} also provide customer occupancy information, such as income and the number of adults in a household. Additionally, datasets like TUVA \cite{einfalt_TUVA_2008}, Dataport \cite{dataport_}, OCTES \cite{octes_2011}, and HUE \cite{stephen_HUE_2019} collect information on PV or EV for renewable energy analysis. Fig. \ref{fig:pub_dataset_sun} displays the residential, commercial, and industrial datasets available for NILM from different countries, with the USA contributing the largest number of datasets, totaling 13. The datasets with high sampling rates (more than 1 kHz) are primarily collected in Germany.

\begin{figure}[ht]
\vspace*{-6pt}
\setlength{\abovecaptionskip}{-.1cm} 
\setlength{\belowcaptionskip}{-2cm} 
    \centering
    \includegraphics[width=0.86\linewidth,scale=1.0]{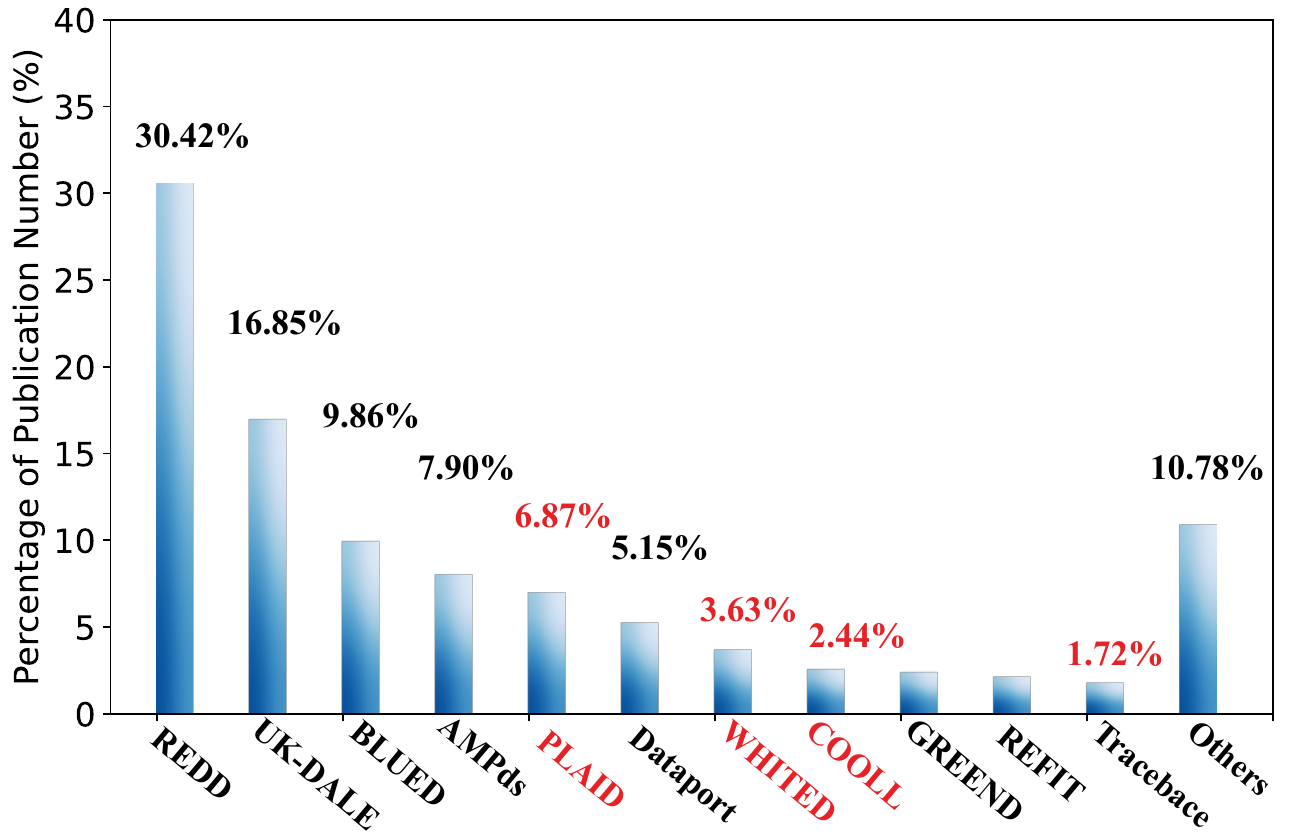}
    \caption{Percentage of publication numbers for NILM using various datasets on Scopus. The results were obtained using the keywords 'non-intrusive load monitoring' or 'energy disaggregation' (searched in title or abstract) combined with the dataset name and the term 'dataset' (searched throughout the document) as of Jan 2024.}
    \label{fig:pub_dataset}
\vspace*{-6pt}
\end{figure}

A comparison of publication numbers for NILM using different datasets, based on searches in the Scopus database with keywords 'load monitoring', 'energy disaggregation', and dataset names, is shown in Fig. \ref{fig:pub_dataset}. It reveals that the REDD \cite{kolter_redd_2011}, UK-DALE \cite{UK-DALE_2015}, and BLUED \cite{kyle_blued_2012} datasets are most frequently used for low-frequency NILM. \textbf{However}, the number of homes and the duration of monitoring in datasets like UK-DALE, REDD, and BLUED are very limited. This indicates that current NILM research is confined to a limited number of users and does not fully utilize the available public datasets. Furthermore, datasets like CLNR \cite{clan_2014}, LCLP \cite{james_LCLP_2014}, and NextGen \cite{shaw_NextGen_2019}, which collect only aggregate load but have a much larger number of users, are rarely used in research, unlike those that have appliance-level power consumption, such as UK-DALE and REDD. High-frequency energy consumption datasets like PLAID \cite{gao_plaid_2014}, WHITED \cite{kahl_whited_2016}, COOLL \cite{thomas_cooll_2016}, and BLOND \cite{kriech_blond_2017} are also gaining attention.

\subsection{Data Pre-processing}
Once a suitable dataset for NILM has been selected, the researcher must preprocess the data for further analysis. Data collected from smart meters or sensors often suffer from missing values and noise, necessitating preprocessing to enhance data quality. The method of processing raw data directly affects the accuracy and robustness of algorithm identification.

\subsubsection{Data Harmonization}
The issue of missing data is unavoidable in wireless sensor networks due to factors such as network communication outages, sensor maintenance, or failure \cite{zhang_ssim_2019}. Missing data can limit or mislead the accuracy and reliability of data analysis results. Data imputation, therefore, is a crucial preprocessing step aimed at estimating identified missing values to avoid under-utilization of data that could lead to biased results. Rather than replacing missing observations with null values \cite{yassine_effective_2020}, many studies have chosen to abandon discontinuous data, opting to use only continuous data for a certain period for further analysis \cite{figueiredo_electrical_2014, zhang_sequence-point_2018, shin_subtask_2019, incecco_transfer_2020, chen_scale_2020, liu_unsupervised_2022, lin_privacy_2022, lin_deep_2022, liu_samnet_2022, liu_single_2022}. Some researchers have imputed missing values with the mean values of relevant data \cite{alcala_event_2017, lin_energy_2022, virt_torch_2022}. The authors in \cite{allik_interpolation_2017} concluded that linear and spline interpolation are the most appropriate approaches for imputing missing electricity consumption data. The methods for recovering missing time series sensor data have been studied in \cite{zhang_ssim_2019}. \textbf{However}, most published works do not mention their preprocessing methods, and even fewer papers study these methods specifically for NILM. Yet, the study of data preprocessing methods for electricity consumption data and their effect on NILM remains largely unexplored.

In addition to missing data, outliers and noise caused by data corruption or unknown appliances should also be carefully investigated. Generally, a high-pass filter \cite{chalmers_detecting_2022} or a median filter \cite{tang_simple_2014} can be applied to mitigate background outliers or noises. Furthermore, according to Equation (\ref{eq:nilm}), the electricity consumption of any appliance $y_{i,t}$ should not be larger than the aggregate load $x_t$, implying $y_{i,t}\le x_t, \forall i, \forall t$. Points where the aggregate load is less than the electricity consumption of an appliance should also be considered outliers, and these can be replaced with a zero value. \textbf{However}, outliers often contain valuable information about the process under investigation or the data gathering and recording process. Before considering the elimination of these points from the data, it is important to understand why they appeared and assess whether similar values are likely to continue to appear.

\subsubsection{Data Augmentation}
With the advancement of smart meter infrastructure, an abundance of building energy consumption data has become available for load monitoring. However, the collection and storage of aggregate load and single-appliance energy consumption data are costly and become impractical with the increased number of sensors \cite{kong_practical_2020}. To artificially enhance the size and quality of datasets with limited data, data augmentation has emerged as an essential approach for generating synthetic data by applying transformations to existing data. Data augmentation has proven effective in improving the performance and outcomes of machine learning models in various fields, such as computer vision and speech processing \cite{georgios_nilm_2022}. Unlike in the image domain, building power consumption datasets for NILM are relatively small. For example, as shown in Table \ref{tab:datasets}, the UK-DALE dataset, widely used in NILM studies, was collected from only five houses over a relatively long period. In contrast, popular image datasets like ImageNet contain 1.2 million training patterns \cite{deng_imagenet_2009}. To fully harness the potential of machine learning methods for NILM, there is a pressing need to develop data augmentation technologies for load consumption data.

The key to data augmentation for NILM lies in acquiring quality training datasets that can represent a variety of real-world situations. Several time series-specific data augmentation methods already exist, such as adding noise to the original data, magnitude warping, and time warping \cite{iwana_time_2021}. By randomly and systematically combining active profiles of target appliances with the aggregate load to create synthetic aggregate data for energy disaggregation, studies like \cite{kong_practical_2020} and \cite{kelly_neural_2015} have shown that data augmentation with synthetic aggregate data effectively improves NILM performance. It is concluded that synthetic aggregate data acts as a regularizer for model training and enhances the network's ability to generalize to unseen houses \cite{kelly_neural_2015}.

\begin{figure}[ht]
\vspace*{-6pt}
\setlength{\abovecaptionskip}{-.1cm} 
\setlength{\belowcaptionskip}{-2cm} 
    \centering
    \includegraphics[width=\linewidth,scale=1.0]{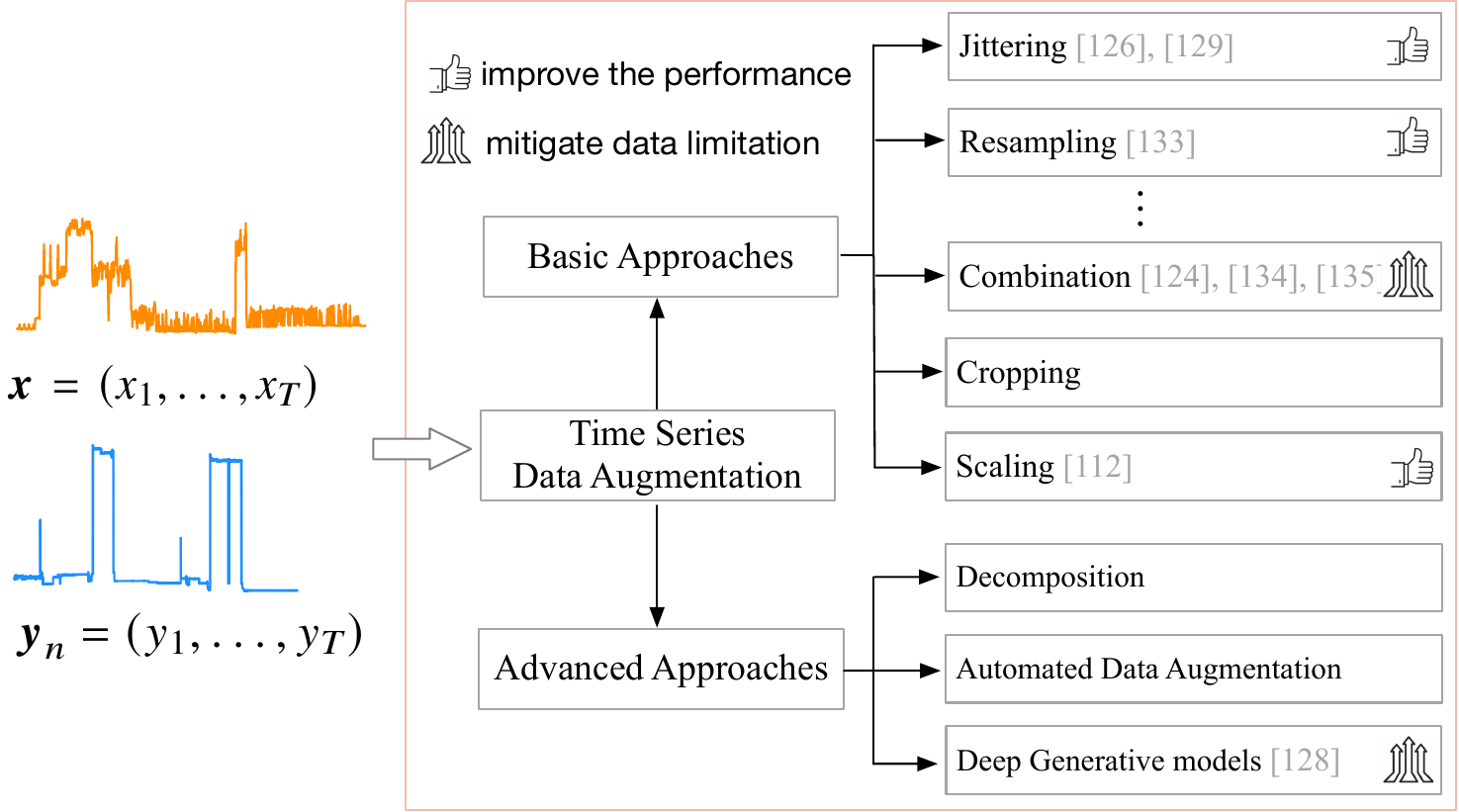}
    \caption{Overview of data augmentation methods for NILM.}
    \label{fig:data_augmentation}
\vspace*{-6pt}
\end{figure}

In recent years, a growing number of data augmentation methods have been proposed to enhance the performance of machine learning in NILM. Fig. \ref{fig:data_augmentation} illustrates several impactful publications on NILM that employ various data augmentation methods. Regarding the purpose of data augmentation in NILM, these methods are primarily designed to \textbf{i) improve the performance of NILM algorithms}, and \textbf{ii) help mitigate data limitations}. From a methodological perspective, data augmentation methods can be categorized into i) basic approaches, such as jittering, and ii) advanced approaches, including decomposition methods, automated data augmentation, and deep generative methods \cite{qing_time_2021}.

To address the variance in on-state power consumption of similar appliances, such as differing peak powers of two fridges from different users or manufacturers, an on-state augmentation method has been proposed in \cite{chen_scale_2020} to disaggregate appliance power more accurately. Given an appliance, the maximum offset values $e_{-}\in \mathbb{R}_{-}$ and $e^+\in \mathbb{R}^+$ are regarded as pri-knowledge. With on-state augmentation method, the aggregate and labeled power consumption $x_t, y_{i,t}$ are replaced by $\bar{x}_t = x_t + e$ and $\bar{y}_{i, t} = y_{i,t} + e$, respectively, where the offset $e \sim U(e_{-}, e^{+})$. This data augmentation, essentially jittering, adds noise (offset) to the input and labeled output. Various experiments in \cite{houidi_comparative_2021} have also demonstrated that data augmentation by adding white Gaussian noise $\eta$ to the aggregate load $\bar{x}_t = x_t + \eta$ can significantly improve the results. 

NILM can be viewed as a classification problem to identify the appliances' ON/OFF states \cite{basu_nonintrusive_2015,singhal_simultaneous_2019, singh_non_2020}. To address the class imbalance problem, where the OFF-state class usually has more samples than the ON-state class, various data augmentation algorithms have been adopted to process imbalanced datasets for NILM and prevent overfitting \cite{fang_specific_2019, chen_scale_2020, liu_samnet_2022}. Beyond improving disaggregation accuracy, data augmentation is also used to enhance the generalizability of deep learning-based NILM by generating synthetic aggregate and sub-meter profiles \cite{hasan_generalizability_2021, harell_tracegan_2021}. \textbf{While} it has been shown that data augmentation effectively improves NILM performance \cite{kelly_neural_2015, fang_specific_2019, kong_practical_2020, chen_scale_2020, liu_samnet_2022}, these conclusions have been limited to specific approaches, and there has been no systematic study on whether the results of data augmentation remain effective across different algorithms and electrical appliances. Furthermore, Table \ref{tab:datasets} and Fig. \ref{fig:pub_dataset} show that there are already a large number of publicly available datasets not yet utilized in NILM research. Researchers should seriously consider whether it is necessary to mitigate data limitations in datasets with very few users and short durations, such as UK-DALE, through data augmentation.

\section{Feature Engineering for NILM}                                                                        
\begin{figure}[ht]
\vspace*{-6pt}
\setlength{\abovecaptionskip}{-.1cm} 
\setlength{\belowcaptionskip}{-2cm} 
    \centering
    \includegraphics[width=\linewidth,scale=1.0]{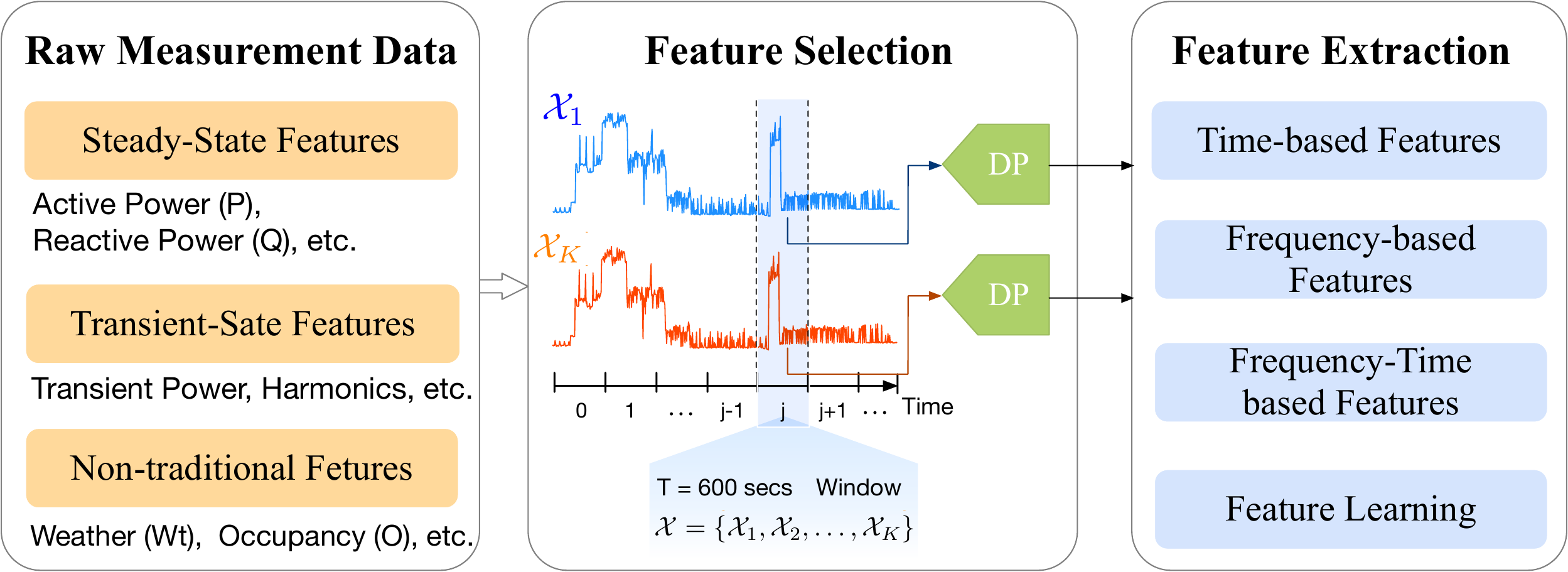}
    \caption{Block diagram of feature engineering for NILM. "DP" means data preprocessing.}
    \label{fig:feature_engineering}
\vspace*{-6pt}
\end{figure}

To efficiently disaggregate appliances from the aggregate load, selecting a set of high-quality features from raw measurement data is crucial. The operational characteristics of an appliance, known as its 'appliance signature' \cite{zeifman_nonintrusive_2011,chen_temporal_2022}, are defined by a specific set of measurable features that reveal information about the appliance's consumption pattern, nature, and operation. It should be noted that 'feature' and 'signature' are two distinct concepts. Features represent an appliance's electrical properties, such as active power, power factor, and current signals, whereas a signature is the unique energy consumption pattern indicated by one or more features. Electrical loads of different appliances exhibit distinct signatures. Appliance signatures can be learned through features observed during transient and steady-state operations of the appliances. As shown in Fig. \ref{fig:feature_engineering}, the measured features can generally be classified into three categories: steady-state features, transient features, and non-traditional features. Different types of features are interconnected and relate to basic electrical quantities like reactive power, current, and voltage in various forms \cite{bao_feature_2022}. Therefore, feature selection, which determines which feature or combination of features to use for NILM, is essential.

As shown in Fig. \ref{fig:feature_engineering}, appliance features are extracted within a time window. These features, whether singular or multiple, are used to distinguish appliances. The feature extraction process involves applying techniques to an initial set of measured data, with the goal of creating a feature vector. This vector is designed to characterize the features of the expected variable, thereby facilitating subsequent learning and generalization steps \cite{georgios_nilm_2022}. Feature extraction can be conducted in various domains, such as the time-based domain, frequency-based domain, or time-frequency domain.

\subsection{Features for NILM}

\begin{figure}[ht]
\vspace*{-6pt}
\setlength{\abovecaptionskip}{-.1cm} 
\setlength{\belowcaptionskip}{-2cm} 
    \centering
    \includegraphics[width=0.95\linewidth,scale=1.0]{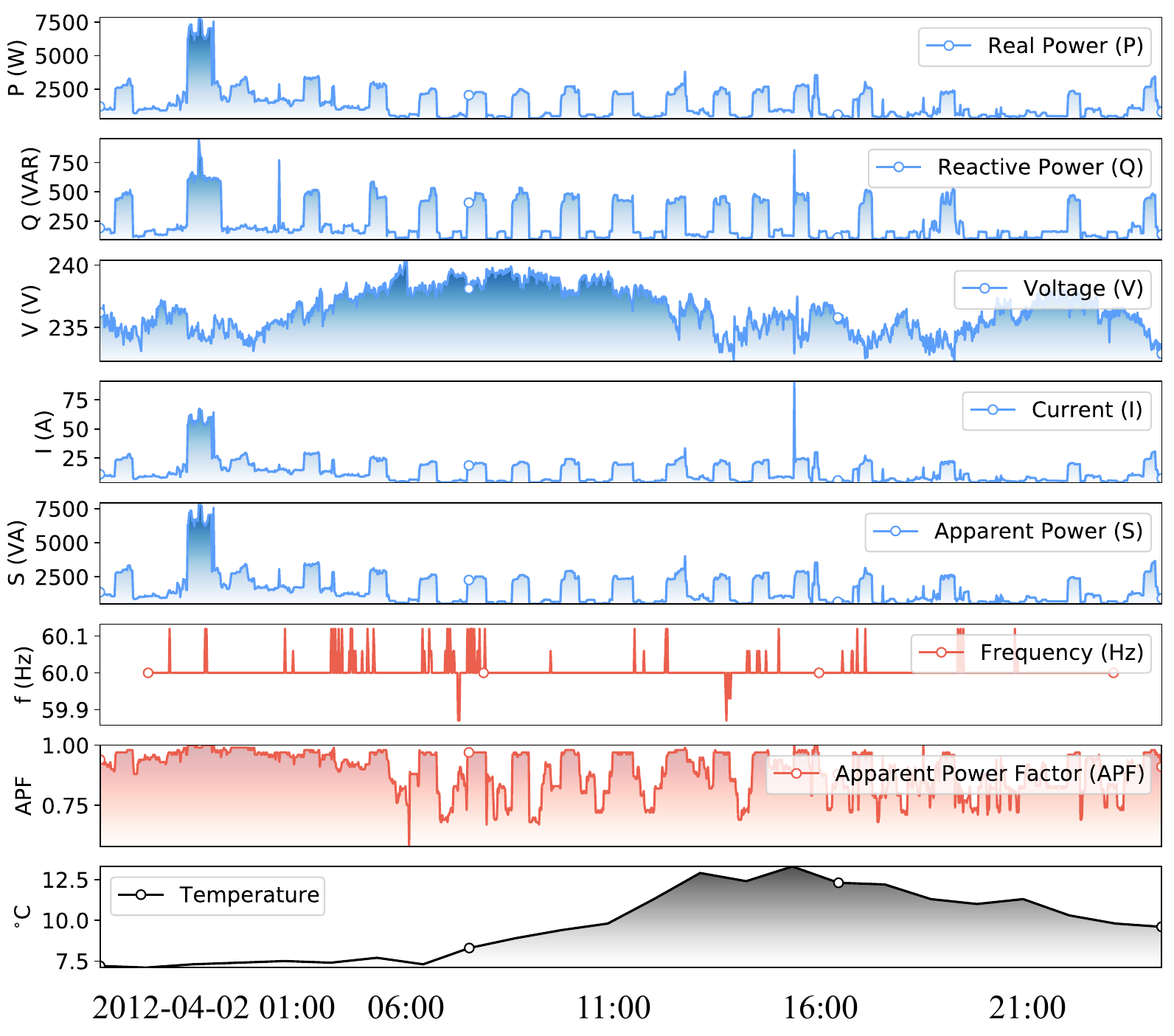}
    \caption{One day of steady-state features for aggregate load and non-traditional temperature features in AMDPs dataset.}
    \label{fig:steady_feature}
\vspace*{-6pt}
\end{figure}

\begin{figure}[ht]
\vspace*{-6pt}
\setlength{\abovecaptionskip}{-.1cm} 
\setlength{\belowcaptionskip}{-2cm} 
    \centering
    \includegraphics[width=0.92\linewidth,scale=1.0]{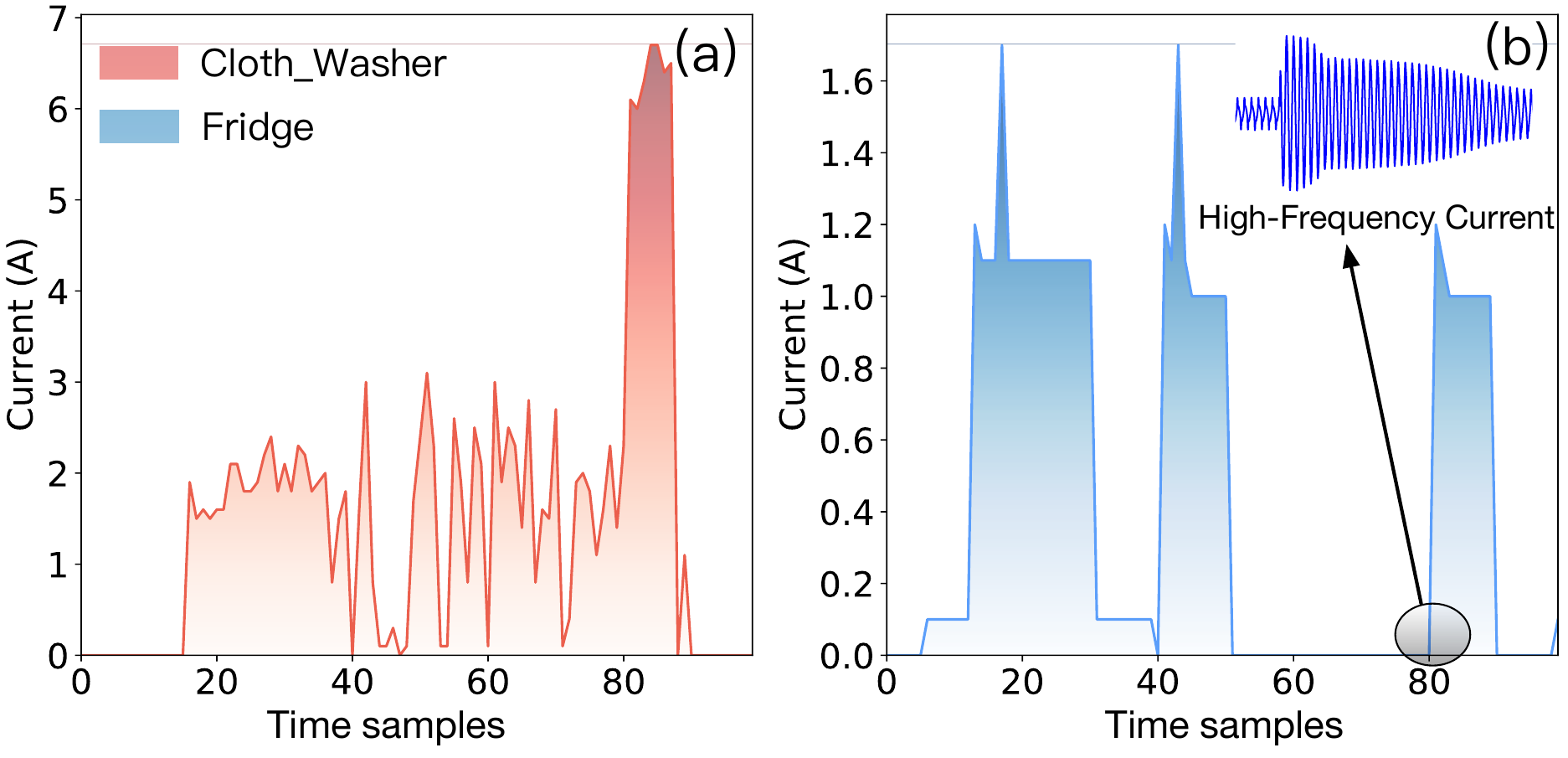}
    \caption{Steady-state current feature sampled at 1 Hz in AMDPs dataset and transient-state feature sampled at 12.8 kHz in SustDataED2 dataset \cite{pereira_sustdataed2_2022}, (a) closth washer and (b) fridge.}
    \label{fig:steady_app_feature}
\vspace*{-6pt}
\end{figure}

\textbf{1) Steady-State Features} are extracted when appliances operate in a steady state, focusing on signal amplitude and its gradual variations from high to low and vice versa, as illustrated in Fig. \ref{fig:steady_feature}. These amplitude changes are not abrupt, thereby not requiring fast sampling. Consequently, steady-state features are preferred for appliances with high power ratings. Primarily, they include active power, reactive power, apparent power, and a set of statistical low-frequency features computed from raw steady-state features (e.g., mean, median, variance, peak, duration values, or energy in a duration period) \cite{karioline_anon_2021}. From Fig. \ref{fig:steady_app_feature}, it is evident that different appliances have distinct steady-state load profiles, which form the basis for their signatures.

The general steady-state features are shown in Table \ref{tab:features}. In recent years, increasing research in NILM has focused on the use of low-frequency data with steady-state features for energy disaggregation. Active power and its temporal variations are the most commonly used steady-state features in NILM studies \cite{zhang_sequence-point_2018, shin_subtask_2019, kong_practical_2020, chen_scale_2020, incecco_transfer_2020, yu_towardsmart_2021, liu_unsupervised_2022, liu_samnet_2022, liu_single_2022}, as they indicate the amount of power being billed. In addition to active power, some studies also incorporate reactive power or apparent power into their proposed NILM algorithms \cite{karioline_anon_2021}. Introducing active-reactive power pairs into the Additive Factorial Hidden Markov Models framework, \cite{roberto_nonintrusive_2017} demonstrates that the algorithm can output disaggregated profiles in both active and reactive power components with a significant improvement. More steady-state features, such as apparent power, current, voltage, power factor, and harmonics, are also considered characteristic loads for energy disaggregation \cite{kumar_nonintrusive_2021}. Ref. \cite{aggelos_load_2017} defined representative current values for the 1st, 3rd, and 5th harmonic orders to be utilized in load signature formulation, concluding that the phase angle of higher harmonic currents should also be considered. 
 
\begin{table*}[htbp]
\vspace*{-10pt}
\setlength{\abovecaptionskip}{-.1cm} 
\setlength{\belowcaptionskip}{-2cm} 
\footnotesize{
    \centering
    \caption{}
    \label{tab:features}
    \begin{adjustbox}{width=\linewidth, center}
    \begin{tabular}{ l| l |l}
    \toprule
    Steady-State Features ($\sim$ 1 Hz) & Transient-State Features ($\ge$ kHz) & Non-Traditional Features\\
    \hline
     $\cdot$ Active Power \cite{chang_feature_2016, enriquez_towards_2017, tabatabaei_toward_2017,roberto_nonintrusive_2017, tang_occupancy_2017, zhang_sequence-point_2018,wittmann_mixedinteger_2018, hamed_linking_2018, diego_interactive_2018, kaselimi_multichannel_2019, ji_factorial_2019,shin_subtask_2019, machlev_monilm_2019,kong_practical_2020, chen_scale_2020, incecco_transfer_2020,ledva_separating_2020, kaselimi_context_2020, karioline_anon_2021,cimen_microgrid_2021, samadi_energy_2021, yu_towardsmart_2021,liu_unsupervised_2022, liu_samnet_2022, liu_single_2022,cimen_online_2022,song_profiles_2022} & $\cdot$ Transient V-I Trajectories\cite{du_electric_2016, kaselimi_multichannel_2019, liu_nonintrusive_2019,zhou_noveltransfer_2021, wang_nonintrusive_2021, ghosh_polluting_2021,faustine_adaptive_2021,bartman_identification_2021, kang_adaptive_2022, zhang_industrial_2022, mills_power_2022, chen_temporal_2022,nolasco_deedfml_2022} & $\cdot$ Weather \cite{enriquez_towards_2017,ledva_separating_2020,song_profiles_2022}\\
    $\cdot$ Reactive Power \cite{chang_feature_2016, tabatabaei_toward_2017,roberto_nonintrusive_2017, wittmann_mixedinteger_2018,kaselimi_multichannel_2019,ji_factorial_2019,machlev_monilm_2019,ledva_separating_2020, karioline_anon_2021,yu_towardsmart_2021} &$\cdot$ Transient Power \cite{ houidi_multivariate_2020,christos_realtime_2021, wang_nonintrusive_2021, ghosh_polluting_2021, mills_power_2022, chen_temporal_2022,cimen_online_2022}  & $\cdot$ Temperature \cite{diego_interactive_2018, samadi_energy_2021} \\
    $\cdot$ Apparent Power \cite{valenti_exploiting_2018,kaselimi_multichannel_2019,kumar_nonintrusive_2021}  & $\cdot$ Transient Harmonics \cite{liu_admittance_2018,aggelos_phase_2019, houidi_multivariate_2020, ghosh_polluting_2021, kang_adaptive_2022, zhang_industrial_2022, chen_temporal_2022} & $\cdot$  Occupancy \cite{tang_occupancy_2017,hamed_linking_2018, samadi_energy_2021} \\
    $\cdot$ V/I Trajectories \cite{chang_feature_2016,aggelos_phase_2019, yang_eventdriven_2020,ledva_separating_2020, kumar_nonintrusive_2021,yu_towardsmart_2021} & $\cdot$ Transient V/I Noise \cite{lee_noise_2018}  & $\cdot$ Cost \cite{cimen_microgrid_2021, cimen_online_2022} \\
    $\cdot$ Power Factor \cite{kumar_nonintrusive_2021}    &$\cdot$ Start-Up/THD Current \cite{faustine_adaptive_2021,bartman_identification_2021} & $\cdot$ Time of Day \cite{ledva_separating_2020, samadi_energy_2021}\\
    $\cdot$ Mean/Variances/Peak/Duration Values \cite{chang_feature_2016, tang_occupancy_2017, karioline_anon_2021, roberto_nonintrusive_2017}  & $\cdot$ Crest/Form Factor \cite{bartman_identification_2021}  & $\cdot$ Day of week \cite{diego_interactive_2018, karioline_anon_2021,samadi_energy_2021}\\
    $\cdot$ Phase Angle \cite{aggelos_load_2017, aggelos_phase_2019} & $\cdot$ Slope/Rise/Fall Time \cite{christos_realtime_2021}& $\cdot$ Appliance usage frequency \cite{cimen_microgrid_2021} \\
    \toprule
    \end{tabular}
\end{adjustbox}}
\end{table*}
\textbf{2) Transient-State Features} are extracted during the transition period between steady states to capture abrupt changes for appliance identification. These features require electrical signals captured at a high sampling rate (in the kHz range). As illustrated in Fig. \ref{fig:steady_app_feature} (b), features extracted during the transient state reveal unique information about the appliance that is not identifiable through steady-state features. Compared to steady-state features, transient-state features pose data storage challenges and computational issues for processing and model training. However, since the transient state discloses unique information for each appliance, these features are crucial, especially when steady-state features face difficulties in identification. They also help distinguish between two or more appliances when they operate simultaneously.

The transient voltage, current, and harmonics are the most widely used transient-state features for NILM. The measured voltage and current data can provide in-depth information for the V-I trajectory load signature, significantly improving load identification accuracy by representing features in image form. Several studies have extended the transient voltage and current features \cite{chang_feature_2016, liu_nonintrusive_2019, mills_power_2022}. V-I mapping methods have been developed to construct binary V-I trajectories \cite{du_electric_2016, kang_adaptive_2022}, gray-scale V-I trajectories \cite{zhou_noveltransfer_2021}, and color-encoded V-I trajectories \cite{liu_nonintrusive_2019, chen_endcloud_2021, wang_nonintrusive_2021, chen_temporal_2022}. These methods have transformed V-I trajectories into image representations and added motion and momentum information to the original V-I trajectory images through color encoding \cite{wang_nonintrusive_2021}. The load current, representing the physical features of electric loads, significantly influences the shape of a V-I trajectory \cite{liu_nonintrusive_2019}. However, active current, having a larger portion of the power factor, can cause appliances in the same category to appear to have similar V-I trajectories. Switching events and synchronized voltage and current waveforms at high frequency have also been explored for multiple load disaggregation in \cite{nolasco_deedfml_2022}.

The harmonic components of current are often used as supplementary information alongside voltage and current for signature identification. In studies such as \cite{liu_admittance_2018}, adaptive linear neurons have been employed to calculate the magnitude and phase of each frequency component based on fundamental and harmonic frequencies. Typically, harmonic components of different orders, such as third, fifth, and seventh, are utilized for load signature detection \cite{aggelos_phase_2019, bartman_identification_2021}. It is crucial to investigate the non-intrusive identification of harmonic loads from a safety perspective to prevent the injection of undesirable harmonics into the system \cite{ghosh_polluting_2021, zhang_industrial_2022}.

In addition to harmonic components, transient power, including active power \cite{ghosh_polluting_2021, wang_nonintrusive_2021, chen_temporal_2022, mills_power_2022}, reactive power \cite{ghosh_polluting_2021, wang_nonintrusive_2021, mills_power_2022}, and apparent power \cite{mills_power_2022}, can provide comprehensive information for load monitoring. Phase noise, a unique characteristic of electrical appliances, has also been adopted in \cite{lee_noise_2018} to extract the electrical energy consumption of individual appliances from the aggregate load. Furthermore, various statistical characteristics, such as start-up current \cite{faustine_adaptive_2021}, current total harmonic distortion (THD) \cite{bartman_identification_2021}, crest factor \cite{bartman_identification_2021}, form factor \cite{bartman_identification_2021}, and rise/fall time \cite{christos_realtime_2021}, learned from transient voltage or current, can significantly enhance the accuracy of NILM when combined with different algorithms.

\textbf{3) Non-Traditional Features} are measurement parameters other than electrical features that can provide additional information for NILM. Weather data, including indoor temperature, outdoor temperature, visibility, humidity, wind speed, sunrise, and sunset time, can serve as input features for NILM algorithms \cite{enriquez_towards_2017, samadi_energy_2021, song_profiles_2022}. Additionally, occupancy information, such as whether the house is occupied by users or not, can assist in energy disaggregation \cite{tang_occupancy_2017, samadi_energy_2021}. By subdividing the occupancy information, different algorithms can be applied to unoccupied and occupied periods to enhance NILM accuracy. Wi-Fi technology has also been employed in \cite{hamed_linking_2018} to count the number of occupants and recognize their identities. Furthermore, non-traditional features encompass start time, time of the day, day of the week, and energy cost \cite{diego_interactive_2018, samadi_energy_2021, karioline_anon_2021, cimen_microgrid_2021, cimen_online_2022}, all of which contribute to improving disaggregation accuracy.

\subsection{Feature Selection}
\begin{figure}[ht]
\vspace*{-6pt}
\setlength{\abovecaptionskip}{-.1cm} 
\setlength{\belowcaptionskip}{-2cm} 
    \centering
    \includegraphics[width=0.8\linewidth,scale=1.0]{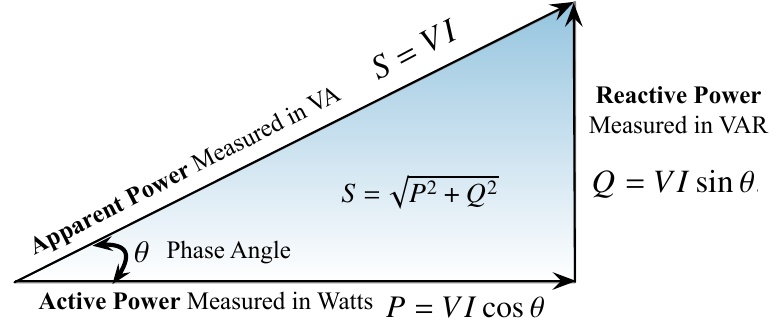}
    \caption{Power triangle to show the relationship of different features.}
    \label{fig:relationship_feature}
\vspace*{-6pt}
\end{figure}

The feature selection process has been rarely considered in existing studies and reviews of NILM. As evident from the previous discussion and Table \ref{tab:features}, numerous features can be chosen for NILM. As illustrated in Fig. \ref{fig:steady_feature}, these various types of features are interconnected and relate to fundamental electrical quantities in diverse ways. The relationship between active power, reactive power, and apparent power in the Alternating Current (AC) component can be depicted using a triangle (refer to Fig. \ref{fig:relationship_feature}). Simultaneously, the Direct Current (DC) component power can be derived from the current and voltage waveforms, as shown in \cite{ravishankar_low_2014}.
\begin{align}
     P_i =\frac{1}{T}\sum_{t=0}^{T} V_{i,t}\cdot I_{i,t}\\
     Q_i = \frac{1}{T}\sum_{t=0}^{T} V_{i,t+\frac{T}{4}}\cdot I_{i,t}\\
     S_n = V_{i, RMS}\cdot I_{i, RMS}
\end{align}
where $V_{i, RMS}$ and $I_{i, RMS}$ are the root mean square of instantaneous value of voltage and current, respectively,
 \begin{align}
     V_{i, RMS} = \sqrt{\frac{\sum_{t=0}^T V_{i,t}^2}{T}}, ~~I_{i, RMS} = \sqrt{\frac{\sum_{t=0}^T I_{i,t}^2}{T}}
 \end{align}
 
Fig. \ref{fig:steady_feature} illustrates that even though the voltage values range from 230 V to 240 V, they remain generally stable. Consequently, active, reactive, and apparent power are all roughly linearly proportional to the current value. From a feature selection perspective, there is significant redundancy among these features. This means that if active power is chosen as the characteristic for NILM, adding additional features like reactive power, voltage, and current will provide relatively little additional information for appliance identification while substantially increasing computational costs \cite{bao_feature_2022}. Let $E=(\mathcal{X}_1, \mathcal{X}_2, \dots, \mathcal{X}_K)$ represent the total steady-state, transient-state, and non-traditional features in the raw measurement data, where $K$ is the total number of features. Feature selection is an essential step to improve learning accuracy and reduce the computational cost of NILM by eliminating irrelevant, redundant, or noisy features from the raw measurement data \cite{miao_selection_2016}. The selected subset of features should adhere to Occam’s Razor principle and optimize performance according to an objective function. 

\subsubsection{\textbf{Single Feature vs Multiple Features}}
In NILM research, active power is the most commonly used feature for steady-state features, whether measured directly or computed from measured current and voltage \cite{chang_feature_2016, enriquez_towards_2017, tabatabaei_toward_2017, roberto_nonintrusive_2017, tang_occupancy_2017, zhang_sequence-point_2018, wittmann_mixedinteger_2018, hamed_linking_2018, diego_interactive_2018, kaselimi_multichannel_2019, ji_factorial_2019, shin_subtask_2019, machlev_monilm_2019, kong_practical_2020, chen_scale_2020, incecco_transfer_2020, ledva_separating_2020, kaselimi_context_2020, karioline_anon_2021, cimen_microgrid_2021, samadi_energy_2021, yu_towardsmart_2021, liu_unsupervised_2022, liu_samnet_2022, liu_single_2022, cimen_online_2022, song_profiles_2022}. However, the choice of using active power as the sole feature for energy disaggregation is, in itself, a feature selection process \cite{zhang_sequence-point_2018, hamed_linking_2018, shin_subtask_2019, kong_practical_2020, chen_scale_2020, incecco_transfer_2020, kaselimi_context_2020, cimen_microgrid_2021, liu_unsupervised_2022, liu_samnet_2022, liu_single_2022, cimen_online_2022}. Some studies have explored the potential benefits of incorporating reactive power as an additional feature when available. It has been found that including reactive power can enhance the NILM process \cite{tabatabaei_toward_2017, wittmann_mixedinteger_2018}. For instance, the integration of reactive power with multi-label K-nearest neighbors (KNN) for NILM resulted in a 36\% improvement over using active power alone \cite{tabatabaei_toward_2017}. Additionally, research by Ref. \cite{ledva_separating_2020} revealed that reactive power measurements can provide more valuable information than active power alone, leading to improved disaggregation accuracy. Moreover, higher-frequency measurements and the consideration of complex voltage at various points within the feeder and at the substation have been shown to further enhance accuracy. A study in Ref. \cite{karioline_anon_2021} investigated the relationship between active and reactive power, highlighting that this relationship is characteristic of specific devices. However, it's worth noting that for purely resistive loads, such as those discussed in Ref. \cite{valenti_exploiting_2018}, the contribution of reactive power may be minimal or even detrimental compared to using active power alone. This suggests that for purely resistive loads, reactive power may introduce noise rather than valuable information.

The incorporation of high-order dynamics has been shown to significantly enhance NILM accuracy. For instance, accuracy improved from 60\% when using only active power to 80\% when utilizing real power, reactive power, and harmonic features. When considering the inclusion of additional steady-state features, especially for appliances with intense harmonic content, it becomes important to also consider the phase angle of higher harmonic currents in the load signature formulation \cite{aggelos_load_2017, le_toward_2021}. \textbf{However, }a systematic investigation into the impact of single or multiple features on different load types remains lacking. There has been no comprehensive comparison to determine whether adding these steady-state attributes as features for specific load types provides valuable information for NILM and should be implemented.

The most commonly used feature for NILM in the transient-state category is the transient V-I trajectory \cite{du_electric_2016, kaselimi_multichannel_2019, liu_nonintrusive_2019, zhou_noveltransfer_2021, wang_nonintrusive_2021, ghosh_polluting_2021, kang_adaptive_2022, zhang_industrial_2022, mills_power_2022, chen_temporal_2022, nolasco_deedfml_2022}. In a notable approach, Ref. \cite{wang_nonintrusive_2021} enhanced the V-I trajectory by incorporating active power information in the red channel, reactive power information in the green channel, and both active and reactive power in the blue channel. This modification not only made the V-I trajectory more informative but also enabled successful utilization by classifiers. Another frequently adopted transient-state feature is transient harmonics \cite{liu_admittance_2018, aggelos_phase_2019, houidi_multivariate_2020, ghosh_polluting_2021, kang_adaptive_2022, zhang_industrial_2022, chen_temporal_2022}. For instance, in \cite{kang_adaptive_2022}, both harmonic current features and V-I trajectory features were extracted for NILM to identify appliances based on their similarity. In contrast, \cite{zhang_industrial_2022} proposed a non-intrusive method for identifying the harmonic sources of industrial users, treating load disaggregation as a single-channel blind problem. It's worth noting that there is a lack of investigation into whether combining multiple transient-steady features can improve the accuracy of energy disaggregation.

Non-traditional features have been combined with steady-state features, primarily active power, to enhance disaggregation accuracy \cite{tang_occupancy_2017, hamed_linking_2018, diego_interactive_2018, ledva_separating_2020, karioline_anon_2021, samadi_energy_2021, song_profiles_2022}. For instance, in a three-phase distribution feeder model, the active and reactive demand of loads were modeled as functions of voltage and outdoor temperature \cite{ledva_separating_2020}. Time of day has also been considered as a feature input for disaggregation algorithms. However, the authors did not provide detailed insights into whether additional non-traditional features like outdoor temperature and time of day would indeed provide valuable information for NILM.

In the context of energy use intensity disaggregation in institutional buildings, various non-traditional features have been employed, including ambient temperature, workday, time of day, daylight length, sunlight intensity, number of occupants, and humidity \cite{samadi_energy_2021}. Additionally, Ref. \cite{tang_occupancy_2017} explored whether the occupancy status of a house or room by users could assist in energy disaggregation. Some studies have incorporated numerous factors such as temperature, day of the year, hour, month, week of the year, and day of the week, in conjunction with active power, to achieve accurate decomposition \cite{diego_interactive_2018}. 

\textbf{However}, there is a gap in existing research when it comes to assessing the effectiveness and necessity of ablation studies to evaluate these additional non-traditional features. This raises important questions for researchers, such as determining which features are essential, which are superfluous, whether more features lead to better results, and how to select the necessary features while eliminating the superfluous ones. Therefore, it is imperative to undertake feature selection as a preprocessing step before advancing non-intrusive load identification algorithms.

\subsubsection{Algorithms for Feature Selection}
\begin{figure}[ht]
\vspace*{-6pt}
\setlength{\abovecaptionskip}{-.1cm} 
\setlength{\belowcaptionskip}{-2cm} 
    \centering
    \includegraphics[width=\linewidth,scale=1.0]{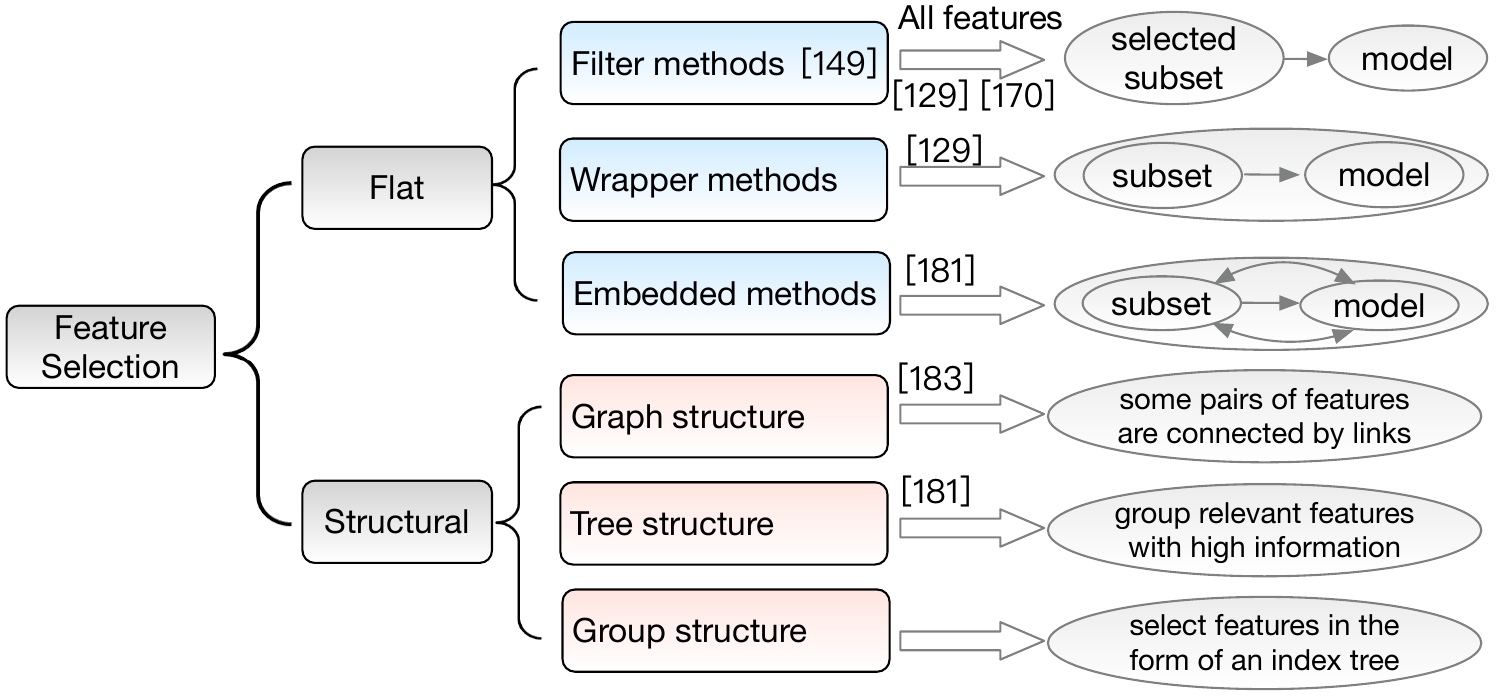}
    \caption{Feature-selection classification taxonomy.}
    \label{fig:feature_slection_algorithms}
\vspace*{-6pt}
\end{figure}

Starting with all features, the feature selection algorithm can
reduce the complexity and computational burden of these models by iteratively eliminating the least important ones until a smaller subset is obtained. It's important to note that the goal isn't to limit the number of features to a specific number, but rather to select a subset of features that performs better or equal to using all of them in terms of discriminative performance \cite{sadeghi_comprehensive_selection_2017}. To put it another way, the iterative feature selection process ends after additional feature reduction fails to improve or degrade appliance identification accuracy. From the features perspective, feature selection methods can be categorized for flat features and structured features \cite{alnuaimi_streaming_2020}. Flat features indicate that the features are independent of each other. As shown in Fig. \ref{fig:feature_slection_algorithms}, feature selection algorithms for flat features can be classified into three methods: filter, wrapper, and embedded methods.

Filter methods select the most discriminative features on the characteristics of the data itself. In general, a feature relevance score is calculated initially to rank all features based on a set of criteria. The features with the highest scores are then chosen \cite{houidi_comparative_2021}. A filter-based backward feature selection method, which utilizes the sum of the trained weights of the first-layer neurons as a relevance score for input features, was proposed for feature selection \cite{houidi_comparative_2021}. By employing both forward and backward processes for feature selection, Ref. \cite{houidi_multivariate_2020} adopted a filter method for feature selection. The forward process starts with an empty set of features, while the backward algorithm begins with the whole set of features. Among the 34 features considered, only a subset consisting of four features corresponding to $P$, $P_H$, $Q_3$, and $Q_9$ were identified as the most relevant.

In wrapper methods, a specific model rather than data is adopted to evaluate the effectiveness of the selected feature subsets. Since creating a model for each potential feature subset is typically computationally expensive, the model is generally 'wrapped' in a heuristic search technique. Ref. \cite{houidi_comparative_2021} introduced an iterative wrapper-based forward feature selection method aimed at maximizing classifier accuracy by adding features one by one. 


Embedded models mainly incorporate feature selection in the process of model construction to reduce the computation time taken up for reclassifying different subsets which is done in wrapper methods. In \cite{sadeghi_comprehensive_selection_2017}, a recursive feature elimination method, an embedded method, was introduced to select feature subsets by ranking all features based on their importance in the model. Since NILM features are not independent and cannot be treated as flat features, a random forest algorithm was employed for feature selection. This systematic reduction of features led to a considerable gain in classification accuracy when compared to models that included all features. Notably, the highest classification accuracy was achieved when a combination of 20 features was used as an appliance signature.

While structural features imply that there are some regulatory relationships between different features \cite{alnuaimi_streaming_2020}, the structural feature selection process can be conceptualized as involving a graph, a group, or a tree to represent the relationships between these features. Ref. \cite{liu_hybrid_selection_2019} adopted a graph structure-based algorithm for feature selection, aiming to minimize redundancy and maximize relevance. They ranked a total of 177 features, and the macro-F1 score peaked at around 45 features. Interestingly, the results from \cite{sadeghi_comprehensive_selection_2017} demonstrated that systematically reducing features significantly increased classification accuracy compared to models using all features. Conversely, in \cite{liu_hybrid_selection_2019}, the addition of redundant features after reaching the peak macro-F1 measure at around 45 features neither improved nor decreased the model's performance.

From the preceding discussion, it is evident that the existing research on feature selection for NILM is quite limited. Most NILM studies have manually selected a very restricted set of features, typically limited to features like active power and reactive power for load disaggregation \cite{chang_feature_2016, enriquez_towards_2017, tabatabaei_toward_2017,roberto_nonintrusive_2017, tang_occupancy_2017, zhang_sequence-point_2018,wittmann_mixedinteger_2018, hamed_linking_2018, diego_interactive_2018, kaselimi_multichannel_2019, ji_factorial_2019,shin_subtask_2019, machlev_monilm_2019,kong_practical_2020, chen_scale_2020, incecco_transfer_2020,ledva_separating_2020, kaselimi_context_2020, karioline_anon_2021,cimen_microgrid_2021, samadi_energy_2021, yu_towardsmart_2021,liu_unsupervised_2022, liu_samnet_2022, liu_single_2022,cimen_online_2022,song_profiles_2022}. The few research efforts that have explored feature selection have primarily focused on filter methods, which assume that different features are independent of each other. \textbf{However,} as discussed earlier with Fig. \ref{fig:steady_feature} and Fig. \ref{fig:relationship_feature}, this assumption is not aligned with common sense. There is a need for more extensive research to investigate the effectiveness of feature selection algorithms based on structural features. \textbf{Additionally}, existing work has been limited to conducting feature selection within either steady-state features \cite{houidi_multivariate_2020, bao_feature_2022} or transient-state features \cite{su_feature_selection_2011,sadeghi_comprehensive_selection_2017,liu_hybrid_selection_2019,houidi_comparative_2021}. However, whether scenarios or appliances can exclusively utilize steady-state features with feature selection for NILM, aiming to reduce data collection, storage, and computational burden, remains unexplored. \textbf{Furthermore}, the potential introduction of transient-steady features as compensating features for identifying load patterns, particularly in cases where two appliances exhibit very similar steady-state load profiles, warrants investigation. \textbf{Lastly}, there is a notable absence of research on non-traditional feature selection. The question of whether the inclusion of non-traditional features in NILM algorithms \cite{enriquez_towards_2017,ledva_separating_2020,song_profiles_2022, diego_interactive_2018, samadi_energy_2021, tang_occupancy_2017,hamed_linking_2018,ledva_separating_2020, karioline_anon_2021} provides valuable information for energy disaggregation demands further study.

\subsection{Feature Extraction}

Feature extraction is a fundamental process in both machine learning and signal processing, involving the careful extraction or transformation of relevant information from raw data to construct a meaningful set of features. These features represent specific characteristics or attributes within the data that are considered informative for a particular task, such as classification or regression. The primary objective behind feature extraction is to extract vital information that holds relevance to the specific problem under consideration. In the context of NILM, feature extraction methods encompass a wide range of techniques, including statistical measures \cite{chang_feature_2016, tabanelli_trimming_2022}, as well as domain-specific approaches tailored to the inherent characteristics of the data \cite{liu_hybrid_selection_2019, de_load_2019, ghosh_improved_2021, tabanelli_trimming_2022, muzaffer_efficient_2022, wu_structured_2022, nolasco_deedfml_2022, yu2023multi, yin2023non}. It's important to note that feature extraction can be applied within both the time and time-frequency domains.

\subsubsection{Time-Domain Feature Extraction}
Feature extraction for NILM predominantly takes place in the time domain. One notable approach is the Hellige distance algorithm proposed in \cite{chang_feature_2016}, which effectively reduces the number of steady-state power features representing load aging signals while maintaining performance integrity. Time series features have proven their superiority for different appliances when compared to widely used real and reactive power features and binary voltage-current trajectory features \cite{liu_hybrid_selection_2019}. In recent years, machine learning-based NILM has witnessed a direct extraction of features from selected time-series data using low-level neural networks \cite{ghosh_improved_2021, tabanelli_trimming_2022, cimen_microgrid_2021, wu_structured_2022, nolasco_deedfml_2022, yu2023multi, yin2023non}. Feature extraction in the time domain offers a straightforward and interpretable analysis of temporal patterns, making it computationally efficient and well-suited for analyzing steady-state signals and transient events.

\subsubsection{Frequency Feature Extraction}
To distinguish between different appliances accurately, it becomes necessary to extract features in the frequency domain, as certain devices exhibit distinct frequency patterns in their electrical signals. For instance, the frequency content of a refrigerator's operation may differ from that of a washing machine. Moreover, frequency-domain analysis enables precise identification and separation of transient events or noise from steady-state signals, thereby improving overall appliance identification.

A noteworthy example can be found in \cite{held2018frequency}, where the original current waveform, serving as a NILM signature, undergoes a frequency-invariant transformation. This transformation, based on periodic signals, aligns uncorrelated sample data with a fixed multiple of the grid frequency. The resulting signal representation fully encapsulates information about the current signal and the phase shift between voltage and current signals for NILM classification. Additionally, events in the frequency domain can be detected using Cepstrum filtering, which has shown promising results for NILM \cite{himeur2020robust}. However, these studies have yet to conduct a comprehensive performance and computational resource comparison for NILM between the time and frequency domains. Such a comparison would help determine the value and necessity of frequency-domain feature extraction. Furthermore, it's important to note that frequency-domain feature extraction typically requires high-resolution input data compared to time-domain feature extraction.
\begin{figure*}[ht]
\vspace*{-6pt}
\setlength{\abovecaptionskip}{-.1cm} 
\setlength{\belowcaptionskip}{-2cm} 
    \centering
    \includegraphics[width=\linewidth,scale=1.0]{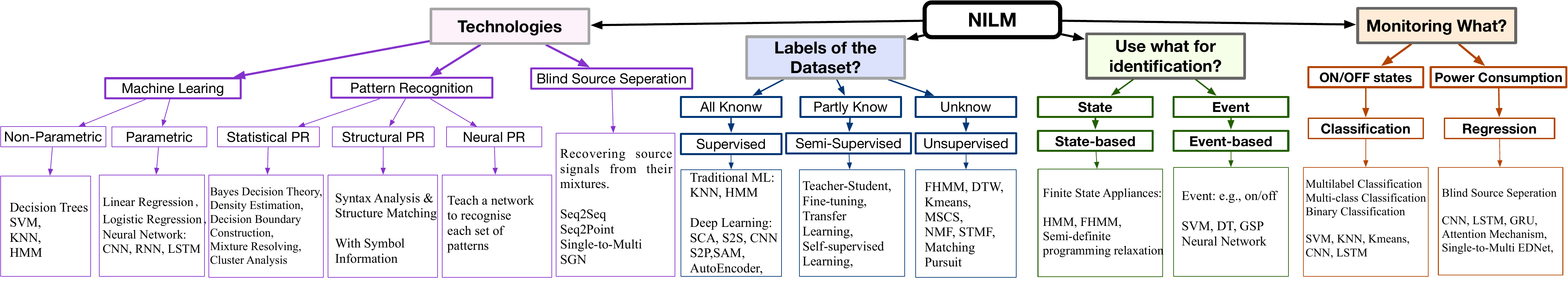}
    \caption{Comparison between supervised, semi-supervised, and unsupervised learning algorithms.}
    \label{fig:algorithms}
\vspace*{-6pt}
\end{figure*}
\subsubsection{Time-Frequency Feature Extraction}
While both time- and frequency-domain features offer unique advantages, it's crucial to take into account the specific application and signal characteristics. In scenarios where appliances exhibit time-varying frequency content, it may become necessary to utilize a combination of time and frequency-domain features for a more comprehensive analysis. A common approach involves using the fast Fourier transform (FFT) filterbank before applying deep neural networks for load disaggregation to extract information from flexible loads \cite{song2021time,chen_temporal_2022}. In some cases, such as when implementing low-latency NILM on resource-constrained microcontroller-based meters, Ref. \cite{tabanelli_trimming_2022} conducted a mean decrease accuracy analysis. This analysis aimed to reduce the feature space while minimizing information loss and identifying the most relevant time- and frequency-domain features for disaggregating load profiles.

The choice between time and time-frequency domain feature extraction depends on the nature of the data and the specific characteristics one aims to capture. Time-domain features are suitable for steady signals, while time-frequency domain features are more appropriate for signals with dynamic frequency content.

\section{Approaches and Metrics for NILM}
NILM aims to develop an effective and efficient solution for extracting appliance-level data, including the state $s_{i,t}$ and load consumption $ y_{i,t}$ for the $i$th appliance at time slot $t$ from an aggregate data $x$ using appropriate algorithms. Load identification, one of the most crucial steps in NILM, relies on features $\widetilde{\boldsymbol{X}}=(\widetilde{x}_1, \dots, \widetilde{x}_T)$ extracted from the measured aggregate load $\boldsymbol{x}=(x_1, \dots, x_T)$. Various algorithms have been developed for load disaggregation. In this work, we conduct a comprehensive analysis of NILM approaches from four perspectives: 1) different problem types of NILM, 2) classifying NILM approaches based on the presence or absence of labels in the dataset, 3) the state/event features adopted for NILM, and 4) the direct monitoring outputs of NILM.

\subsection{NILM Approaches with Different Problem Types}
\begin{figure}[ht]
\vspace{-6pt}
\setlength{\abovecaptionskip}{-.1cm} 
\setlength{\belowcaptionskip}{-2cm} 
    \centering
    \includegraphics[width=\linewidth,scale=1.0]{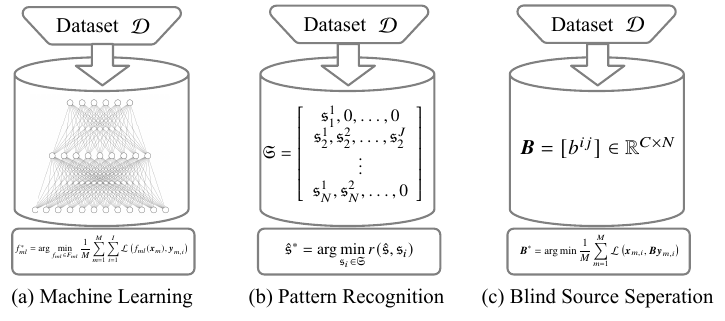}
    \caption{Specific technologies for NILM. (a) Machine Learning, (b) Pattern Recognition, and (c) Blind Source Separation.}
    \label{fig:techonogies}
\vspace{-6pt}
\end{figure}

From the perspective of problem type for load disaggregation, NILM approaches can be categorized into the following three types: 1) \textbf{machine learning-based NILM} for a function optimization problem, 2) \textbf{pattern recognition-based NILM} for a pattern recognition problem, and 3) \textbf{blind source separation-based NILM} for a source separation problem. As depicted in Fig. \ref{fig:techonogies}, all three architectures involve training a model for load disaggregation from different perspectives based on the input dataset $\mathcal{D}=\{\mathcal{X}, \mathcal{Y}\}$. 

\subsubsection{Function Optimization Problem}
The function optimization NILM approaches are centered around simultaneously matching multiple loads to find the set of energized appliances that best fits the compound measurement \cite{zeifman_nonintrusive_2011, wittmann_mixedinteger_2018}. The majority of research efforts in this field utilize machine learning technologies. As illustrated in Fig. \ref{fig:techonogies} (a) for machine learning-based NILM, the goal is to learn a function $f_{ml}$ that approximates, with a reasonable margin of error, the relationship between the input $\boldsymbol{x} \in \mathcal{X}$ and the corresponding load consumption $\boldsymbol{y}_i \in \mathcal{Y}$ (one model for one electric appliance). The accuracy of NILM is evaluated using a loss function $\mathcal{L}(\cdot)$ that assesses the deviation between the estimation output and the ground truth. Without loss of generality, the well-trained model $f_{ml, i}^*$  is the one that minimizes the error.

\begin{equation}
    f_{ml, i}^* = \arg \min_{ f_{ml, i} \in F_{ml}} \frac{1}{M}\sum_{m=1}^{M}\mathcal{L}\left(f_{ml, i}(\boldsymbol{x}_m), \boldsymbol{y}_{m,i}\right)
\end{equation}
where $M$ is the sample number in the dataset $\mathcal{D}=\{\mathcal{X}, \mathcal{Y}\}$, $i$ indicates the $i$th electric appliance, and $F$ is the family of functions for $f_{ml,i}$. Different electric appliances require different models $f_{ml, i}^*(\cdot)$. Generally, the structure is the same for different electric appliances but with different model parameters \cite{zhang_sequence-point_2018, shin_subtask_2019, liu_voltage_2019, chen_scale_2020, incecco_transfer_2020, kaselimi_context_2020, garcia_fully_2021, liu_unsupervised_2022, lin_deep_2022, liu_samnet_2022, chen_temporal_2022, santos_energy_2022}. If $f_{ml}(\cdot)$ is designed for all appliances (one model for multiple electric appliances), the best estimation model $f_{ml}^*$ is as follows,
\begin{equation}
    f_{ml}^* = \arg \min_{ f_{ml} \in F_{ml}} \frac{1}{M}\sum_{m=1}^{M}\sum_{i=1}^{I}\mathcal{L}\left(f_{ml}(\boldsymbol{x}_m), \boldsymbol{y}_{m,i}\right)
\end{equation}
where $I$ is the target electric appliances number concerned in NILM. Usually, NILM is used to monitor multiple electric appliances with the same model only when it is treated as a multi-label classification problem \cite{tabatabaei_toward_2017, singhal_simultaneous_2019, li_residential_2019, singh_non_2020, nolasco_deedfml_2022, singh_multi_2022}. In \cite{liu_single_2022}, the authors designed a single-to-multi model to directly obtain the load consumption of multiple electric appliances rather than their on/off state.

\begin{figure}[ht]
\vspace{-6pt}
\setlength{\abovecaptionskip}{-.1cm} 
\setlength{\belowcaptionskip}{-2cm} 
    \centering
    \includegraphics[width=\linewidth,scale=1.0]{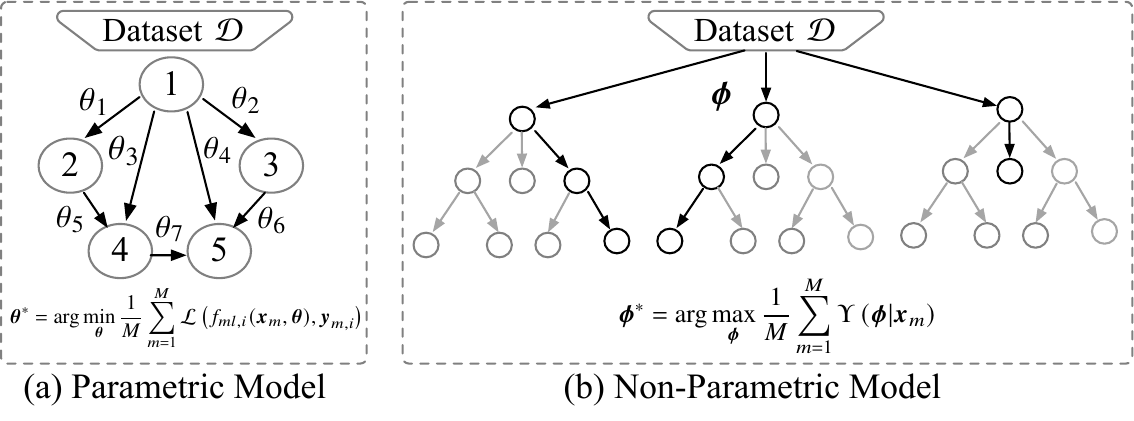}
    \caption{Parametric machine learning vs Non-parametric machine learning.}
    \label{fig:nonparmetric}
\vspace{-6pt}
\end{figure}

To identify the function $f_{ml, i}^*$ or f$_{ml}^*$, the process may be either \textbf{parametric} or \textbf{non-parametric} in general (see Fig. \ref{fig:nonparmetric}). Specifically, a \textbf{parametric machine learning algorithm} assumes that the dataset can be adequately modeled by a probability distribution that has a fixed set of parameters (see Fig. \ref{fig:nonparmetric} (a)).
\begin{equation}
    \boldsymbol{\theta}^* = \arg \min_{\boldsymbol{\theta}} \frac{1}{M}\sum_{m=1}^{M}\mathcal{L}\left(f_{ml, i}(\boldsymbol{x}_m, \boldsymbol{\theta}), \boldsymbol{y}_{m,i}\right)
\end{equation}
where $\boldsymbol{\theta}$ represents the coefficients of the model, and $\boldsymbol{\theta}^*$ represents the optimized parameters of the parametric model. Typical parametric models include linear models such as linear regression, logistic regression, and linear support vector machines. Parametric machine learning models also encompass neural networks with fully connected layers, convolutional layers, and more. In a recent study, a suite of multiple linear regression models estimated by the genetic algorithm and a least-square solver was proposed in \cite{narges_disaggregation_2022} to disaggregate heating and electricity in a non-intrusive manner. Over the years, there has been an increasing use of neural networks, including Convolutional Neural Networks (CNNs) \cite{zhang_sequence-point_2018, shin_subtask_2019, chen_scale_2020, incecco_transfer_2020, kaselimi_context_2020, garcia_fully_2021, liu_unsupervised_2022, liu_samnet_2022, chen_temporal_2022}, Long Short-Term Memory (LSTM) \cite{khodayar_energy_2020, le_toward_2021, hwang_nonintrusive_2022, kaselimi_context_2020}, Gated Recurrent Unit (GRU) \cite{cimen_microgrid_2021}, attention mechanisms \cite{chen_scale_2020, liu_samnet_2022, liu_single_2022, nie_ensemble_2022}, auto-encoders \cite{roberto_denoising_2018, khodayar_energy_2020, garcia_fully_2021}, and more.

Building non-parametric models makes no explicit assumptions about the probability distribution or functional form, as is the case with parametric models. Instead, non-parametric machine learning can be viewed as a process of function optimization that aims to closely match the dataset (see Fig. \ref{fig:nonparmetric} (b)).
\begin{equation}
    \boldsymbol{\phi}^* = \arg \max_{\boldsymbol{\phi}}\frac{1}{M}\sum_{m=1}^{M}\Upsilon\left(\boldsymbol{\phi}|\boldsymbol{x}_m\right)
\end{equation}
where $\boldsymbol{\phi}$ is the hyperparameters of the non-parametric model, $\boldsymbol{\phi}^*$ is the optimized hyperparameters, and $\Upsilon\left(\phi_m|\boldsymbol{x}_m\right)$ denotes the maximum likelihood of the model given gene expression dataset $\mathcal{X}$ with $M$ samples.
Tree-based methods, such as decision trees, are non-parametric models that partition the estimation space into sub-regions and then yield a function based on statistical indicators of the segmented training data \cite{gambella_optimization_2021}. Decision tree mainly adopted as a NILM approach for appliance classification  \cite{liu_hybrid_selection_2019, marco_recognition_2019}. Random forest is the non-parametric technique that creates an ensemble of different decision trees and outputs the mean of each decision tree’s individual prediction. Authors in \cite{zachary_scalable_2021} concluded that random forest regression outperforms other supervised learning regression methods for NILM by taking the mean of each tree’s output to combat the overfitting of individual decision trees. Random forest also has been adopted to reduce the feature space for NILM in \cite{tabanelli_trimming_2022}. Besides, boosting algorithms, such as adaptive boost that uses decision stumps as weak classifiers \cite{hassan_empirical_2014}, boosting framework for continuous online learning and detection \cite{ma_toward_2018}, and extreme gradient boost \cite{ daniel_three_2021}, are also non-parametric models. 

The advantages and disadvantages of parametric and non-parametric machine learning models have been summarized in Table \ref{tab:nonparametric}. Parametric machine learning models are simpler and easier to understand, faster to learn from the dataset, require a smaller amount of data for model training, and are well-suited for simpler problems compared to non-parametric models. However, parametric models are generally limited to addressing simple problems, and the accuracy of their fundamental mapping functions can be questionable. Both parametric models (e.g., linear regression and multilayer perceptron) and non-parametric models (e.g., decision tree and random forest) have been adopted for appliance classification \cite{hassan_empirical_2014, basu_nonintrusive_2015, imran_power_2018, he_non-intrusive_2018, liu_hybrid_selection_2019, saeedi_adaptive_2021, muzaffer_efficient_2022}. Although a study in \cite{imran_power_2018} demonstrated that the k-Nearest Neighbors (KNN) model is most efficient for energy disaggregation compared to decision trees, discriminant analysis, and Support Vector Machine (SVM), it is important to note that these results were based on a specific dataset. To date, there is no definitive conclusion about which of these models is more successful, nor is one type of machine learning (e.g., parametric model) consistently superior to another type of machine learning (e.g., non-parametric model).

\begin{table}
\vspace{-10pt}
\setlength{\abovecaptionskip}{-.1cm} 
\setlength{\belowcaptionskip}{-2cm} 
\centering
\caption{Advantages and disadvantages of parametric and non-parametric machine learning technologies.}
\label{tab:nonparametric}
\begin{adjustbox}{width=\linewidth, center}
\begin{tabular}{c| l| l} 
\hline\hline
 Models & Advantages& Disadvantages   \\ 
\hline
 Parametric& \begin{tabular}[c]{@{}l@{}} $\cdot$ Modest: easy to understand\\
            $\cdot$ Speed: rapid to learn\\
            $\cdot$ Smaller amount of data
            \end{tabular}& \begin{tabular}[c]{@{}l@{}} 
            $\cdot$ Limit to simpler problems\\
            $\cdot$ Deprived fit
            \end{tabular}\\
\hline
 Non-parametric& \begin{tabular}[c]{@{}l@{}} 
            $\cdot$ Large number of function forms\\ 
            $\cdot$ Higher performance models\\
            \end{tabular} &  \begin{tabular}[c]{@{}l@{}} 
            $\cdot$ More data for training\\
            $\cdot$ Slower to train\\
            $\cdot$ Prone to over-fit
            \end{tabular}\\
\hline\hline
\end{tabular}
\end{adjustbox}
\vspace{-15pt}
\end{table}

\subsubsection{Pattern Recognition Problem} 
 Pattern recognition involves the automated identification of patterns and regularities in data using various technologies, such as data science and machine learning. As depicted in Fig. \ref{fig:techonogies} (b), pattern-recognition-based NILM for appliance identification relies on a previously known dictionary matrix $\mathfrak{S}$, which contains a database of appliance load signatures $\boldsymbol{\mathfrak{s}}$ \cite{linda_using_1999, Rahimpour_nonintrusive_2017, cominola_hybrid_2017, welikala_incorporating_2019, kong_practical_2020}. In general, pattern recognition is based on event detection, where events are identified and then separated using established classifiers such as KNN \cite{christos_realtime_2021}, decision trees \cite{anand_emf_2015}, and neural networks \cite{chang_power_2014, chang_nonintrusive_2016, fang_nonintrusive_2020, yang_eventdriven_2020, ciancetta_new_2021, christos_realtime_2021, faustine_adaptive_2021, lorin_uncertainty_2022}. Given that pattern recognition relies on one-to-one matching, significant emphasis is placed on feature extraction strategies \cite{wittmann_mixedinteger_2018}. Another approach involves using a database of appliance signatures $\mathfrak{S}$ that includes different appliances, with pattern recognition NILM aiming to find the closest match between the unknown observed signature $\hat{\mathfrak{s}}$ obtained from a previously learned model and the known signatures within $\mathfrak{S}$.
\begin{equation}
    \hat{\mathfrak{s}}^* = \arg \min_{\mathfrak{s}_i \in \mathfrak{S}} r(\hat{\mathfrak{s}}, \mathfrak{s}_i)
\end{equation}
In this context, $r(\cdot)$ represents the matching algorithm, and $\hat{\mathfrak{s}}^*$ corresponds to the identified load signature for a specific appliance. In cases where the $i$th appliance has multiple functions, such as a washing machine, it may possess several load profiles denoted as $\{\mathfrak{s}_i^1, \mathfrak{s}_i^2, \dots\}$. Some approaches, like the one discussed in \cite{kong_practical_2020}, assume the existence of a signature database or dictionary that encompasses a wide range of appliances from various manufacturers. However, treating the signature database as prior information implies a reliance on a preliminary stage of supervised learning, which entails the intrusive collection of critical details such as the power consumption patterns of each appliance and the manufacturers of different appliances.

Constructing load signatures remains a significant challenge in non-intrusive appliance load monitoring. Numerous researchers are actively working on developing unique appliance signature databases based on various techniques, including graph signal reconstruction \cite{zheng_nonintrusive_2022}, mixed-integer nonlinear programming \cite{marco_mixed_2022}, generative adversarial networks (GANs) \cite{harell_tracegan_2021}, admittance-based algorithms \cite{liu_admittance_2018}, Karhunen-Loève Expansion \cite{dinesh_solar_2017}, and K-means clustering \cite{cominola_hybrid_2017}. With the availability of established load signature databases, pattern recognition-based NILM primarily relies on techniques such as graph signal processing (GSP) \cite{zhao_nonintrusive_2020, majumdar_trainingless_2022}, dynamic time warping (DTW) \cite{basu_time_2015, iwayemi_saraa_2017, liu_dynamic_2017, yang_eventdriven_2020, xiang_ev_2021, luan_nonintrusive_2022}, soft dynamic time warping (sDTW) \cite{schirmer_energy_2020}, Hungarian matching \cite{liu_energy_2019}, global alignment kernel (GAK) \cite{schirmer_energy_2020}, all common subsequences (ACS) \cite{schirmer_energy_2020}, and minimum variance matching (MVM) \cite{jasinski_modelling_2020} for pattern matching.

\begin{figure}[ht]
\vspace*{-10pt}
\setlength{\abovecaptionskip}{-.1cm} 
\setlength{\belowcaptionskip}{-2cm} 
    \centering
    \includegraphics[width=\linewidth,scale=1.0]{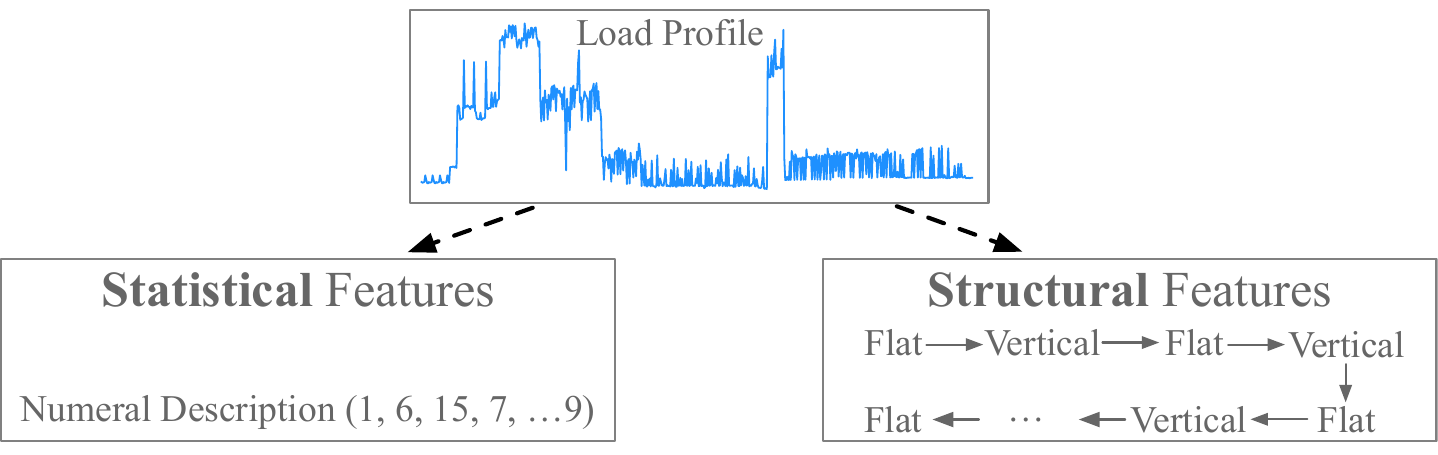}
    \caption{Comparison between statistical and structural pattern recognition algorithms.}
    \label{fig:statistical_structural}
\vspace*{-10pt}
\end{figure}

In today's digital age, patterns are ubiquitous, manifesting themselves statistically through algorithms or in physical forms. Pattern recognition is commonly categorized into two main types: 1) statistical pattern recognition and 2) structural matching. Statistical pattern recognition relies on quantitative features of the data and employs mathematical formulas and models, often coupled with statistical decision theory, to recognize patterns (see Fig. \ref{fig:statistical_structural}). There is a wide array of statistical techniques used in feature extraction, ranging from basic descriptive statistics to intricate transformations. In some cases, pattern matching techniques, as employed in \cite{shaw_nonintrusive_2008}, are used to shape patterns using quantitative vectors that best fit the observed load data. Subsequently, an increasing number of algorithms based on statistical pattern recognition have emerged in the field of NILM, aiming to disaggregate energy from aggregated load data \cite{liang_load_2010, henao_approach_2017, wittmann_mixedinteger_2018, ma_toward_2018, kwak_load_2018, de_load_2019, du_multi_2019}.

Conversely, structural pattern recognition discriminates patterns based on the morphological interrelationships inherent in raw data. Generally, structural pattern recognition in NILM utilizes graph signal processing technologies \cite{he_non-intrusive_2018, li_residential_2019, zhao_nonintrusive_2020, faustine_adaptive_2021, khodayar_spatiotemporal_2021, zheng_nonintrusive_2022, ghaffar_spectral_2022, wu_structured_2022}. Notably, recent advancements in deep learning have led to an increasing number of approaches that employ graph neural networks (GNNs) for energy disaggregation \cite{li_residential_2019, faustine_adaptive_2021, khodayar_spatiotemporal_2021, wu_structured_2022}.

\subsubsection{Blind Source Separation Problem}
\begin{figure}[ht]
\vspace*{-6pt}
\setlength{\abovecaptionskip}{-.1cm} 
\setlength{\belowcaptionskip}{-2cm} 
    \centering
    \includegraphics[width=\linewidth,scale=1.0]{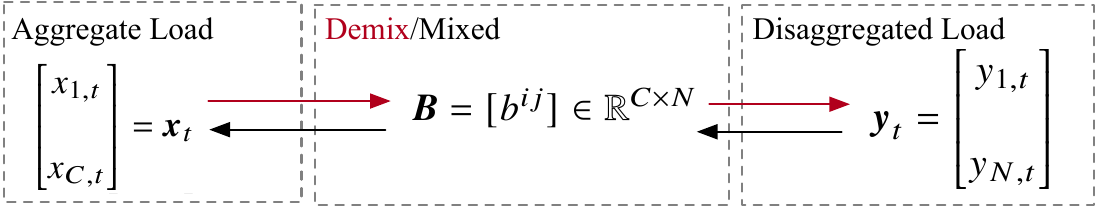}
    \caption{Schematic diagram of blind source separation problem.}
    \label{fig:blind_source_separation}
\vspace*{-6pt}
\end{figure}

NILM can also be viewed as a blind source separation problem, where the goal is to separate a set of source signals from a set of mixed signals, often with limited information about the source signals or the mixing process (see Fig. \ref{fig:techonogies} (c)). This problem assumes that the measured signal is a mixture of unknown source signals, each corresponding to a different appliance. The objective is to recover these source signals from the mixed signal. As depicted in Fig. \ref{fig:blind_source_separation}, given a set of individual source signals, represented as load profiles for various appliances, denoted as \(\boldsymbol{y}_t = (y_{1,t}, \dots, y_{N,t})\) at time \(t\), a set of 'mixed' signals, \(\boldsymbol{x}_t = (x_{1,t}, \dots, x_{C,t})^T\), can be obtained using a matrix \(B=[b_{ij}]\) belonging to \(\mathbb{R}^{C\times N}\).

Since the number of appliances in the aggregate load typically exceeds the number of features (e.g., current, voltage, active power, apparent power, etc.), \(N\) is usually much larger than \(C\). This implies that NILM is an underdetermined problem, and non-linear methods must be employed to effectively recover the unmixed signals. To disaggregate the load signal \(\boldsymbol{y}_i\) for the \(i\)th appliance from the total aggregate load \(\boldsymbol{x}\), the above equation is effectively 'inverted' as follows,
\begin{equation}
    \boldsymbol{B}^* = \arg \max\frac{1}{M}\sum_{m=1}^{M}\mathcal{L}\left(\boldsymbol{x}_m,\boldsymbol{B}\boldsymbol{y}_{m,i}\right)
\end{equation}
A determination of an 'unmixing' matrix \(\boldsymbol{B}^*\) is required to calculate and 'recover' an approximation of the original signal from the aggregate load.

Generally, only the active power/current feature is used for NILM, which requires single-channel blind source separation techniques to disaggregate the source signals. Spectral information \cite{dinesh_residnetial_2016}, deep sparse coding \cite{singh_deep_2018}, generative models \cite{simon_SHED_2018}, Sum-to-k constrained non-negative matrix factorization-based models \cite{Rahimpour_nonintrusive_2017}, dictionary learning \cite{analysis_singh_2019}, contextually supervised models \cite{wang_joint_2022}, wavelet transforms \cite{liu_simultaneous_2022}, the complex local mean decomposition algorithm \cite{zhang_industrial_2022}, and neural network-based models \cite{liu_single_2022} have been developed by regarding NILM as a single-channel blind source separation problem. Although a single channel has a significant advantage in terms of size and scale, a sensor array with multiple features can provide more benefits \cite{xu_artificial_2015}.

\subsection{NILM Approaches with/without Labels}
NILM approaches can be categorized into three groups based on whether labeled data is required for load disaggregation: \textbf{1) supervised approaches},\textbf{ 2) semi-supervised approaches}, and \textbf{3) unsupervised approaches}. Supervised learning aims to develop a function that can closely map input data to target outputs, given a sample of labeled data. Semi-supervised learning, on the other hand, utilizes knowledge gained from a sparse set of labeled data points to label unlabeled data points. In contrast, unsupervised learning does not rely on any labeled outputs and aims to infer the natural structure within a set of data points.
\begin{figure}[ht]
    \centering
    \includegraphics[width=\linewidth,scale=1.0]{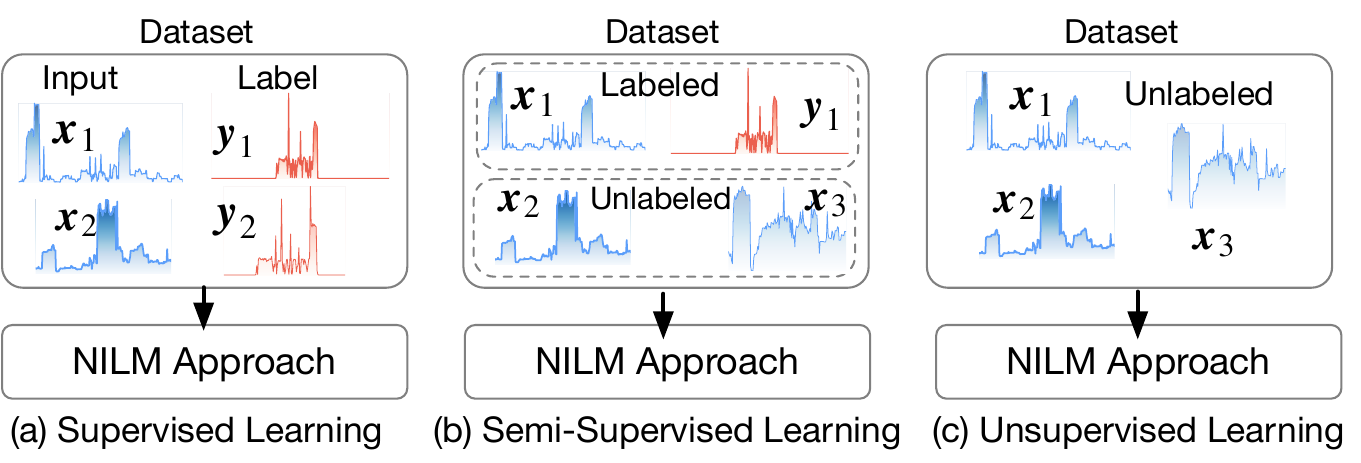}
    \caption{Comparison training processes between supervised, semi-supervised, and unsupervised learning algorithms.}
    \label{fig:algorithms}
\end{figure}
\subsubsection{Supervised Approaches}
As shown in Fig. \ref{fig:algorithms}(a), in supervised learning, the goal is to develop a function that can map input data \(\boldsymbol{x}\) to its corresponding labeled data \(\boldsymbol{y}\). Supervised learning requires labeled data during the training process to guide the learning. In this context, the correct disaggregated load \(\boldsymbol{y}\) for the aggregate load \(\boldsymbol{x}\) is known in advance. Various supervised learning approaches, such as linear regression\cite{saeedi_adaptive_2021}, genetic algorithms\cite{he_efficient_2019,machlev_monilm_2019,narges_disaggregation_2022}, SVM \cite{rahman_power_2018,muzaffer_efficient_2022,wu_structured_2022,chalmers_detecting_2022}, random decision forest\cite{liu_hybrid_selection_2019,saeedi_adaptive_2021,chalmers_detecting_2022,vavouris_low_2022}, hidden Markov models (HMM)\cite{cominola_hybrid_2017,liu_multivariate_2018}, decision trees\cite{imran_power_2018,himeur_effective_2020,saeedi_adaptive_2021}, naive Bayes classifier\cite{mocanu_energy_2016}, kNN\cite{imran_power_2018,christos_realtime_2021, muzaffer_efficient_2022}, and deep learning neural networks \cite{zhang_sequence-point_2018,shin_subtask_2019,chen_scale_2020,du_multi_2019,liu_samnet_2022,santos_energy_2022,castangia_clustering_2022}, have been applied in the NILM literature. As outlined in Table \ref{tab:supervised}, supervised NILM is mainly used for regression and classification tasks. While supervised NILM is generally easy to understand and can achieve high accuracy, it cannot provide information about unknown devices in the training dataset. Therefore, supervised NILM often requires data preprocessing to select suitable examples for training and relies on substantial amounts of labeled data and computational time to improve accuracy.

\subsubsection{Semi-supervised Approaches}
As depicted in Fig. \ref{fig:algorithms} (b), semi-supervised models are trained using a combination of labeled and unlabeled data. Typically, this combination consists of a limited amount of labeled data and a substantial amount of unlabeled data. Semi-supervised learning is particularly valuable in situations where labeling a large dataset is expensive or time-consuming, but some labeled data is available for model training. In this approach, the labeled data helps establish relationships between input features and target labels, while the unlabeled data is used to discover additional patterns in the data. Various methods can be employed in semi-supervised learning, including self-training \cite{kaselimi_context_2020} and co-training \cite{gillis_nonintrusive_2017}.

\begin{figure}[ht]
\vspace*{-6pt}
\setlength{\abovecaptionskip}{-.1cm} 
\setlength{\belowcaptionskip}{-2cm} 
    \centering
    \includegraphics[width=\linewidth,scale=1.0]{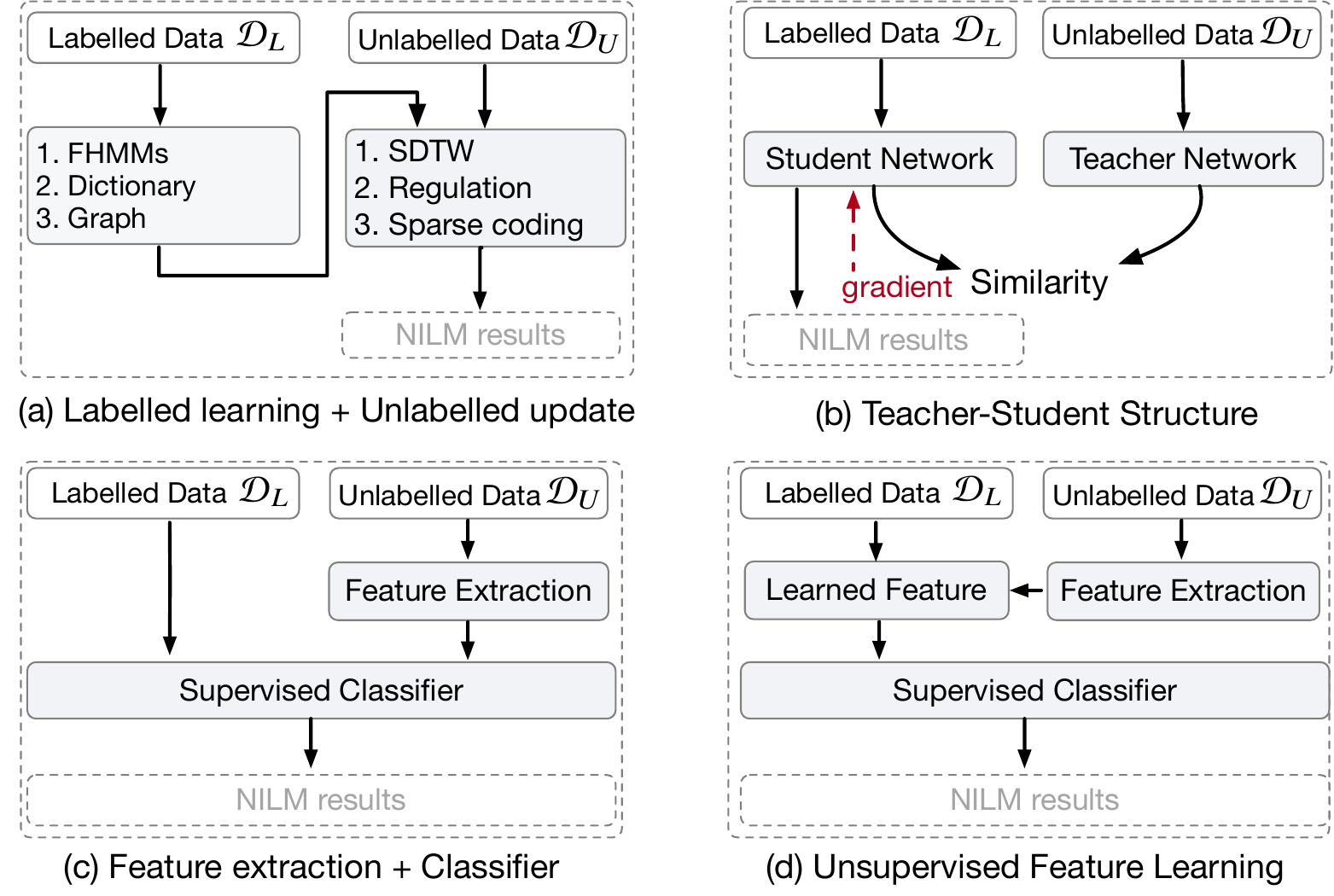}
    \caption{Comparison of different semi-supervised methods for NILM.}
    \label{fig:semi}
\vspace*{-6pt}
\end{figure}

In general, semi-supervised NILM involves utilizing a small amount of labeled observations to enhance the ON/OFF state detection of a much larger unlabeled dataset (refer to Fig. \ref{fig:semi} (a)). This concept has been explored using HMM in \cite{cominola_hybrid_2017}, where Factorial Hidden Markov Models (FHMMs) and Subsequence Dynamic Time Warping (SDTW) were employed. In another approach presented in \cite{liu_towardsmart_2021}, labeled data is used to initialize a dictionary for the feature library. Subsequently, the unlabeled data is leveraged to update the dictionary and identify appliances based on sparse coding. Graph-based technologies have also been utilized for semi-supervised learning. In \cite{li_residential_2019}, three graph-based semi-supervised learning algorithms were extended to multi-label classification. These methods make use of a limited amount of ground truth data and a large pool of unlabeled observations to accurately disaggregate household-level energy demand into the power draws of individual appliances. More recently, a semi-supervised deep learning-based algorithm was developed in \cite{castangia_clustering_2022} to cluster different operation modes of household appliances based on the analysis of their power signatures. An autoencoder neural network was adopted to create a better data representation of the power signatures (labeled data), followed by the use of a K-means algorithm fitted to the new data representation for clustering.

As illustrated in Fig. \ref{fig:semi} (b), Ding et al. developed a temporal CNN in \cite{yang_semisupervised_2020} to automatically extract high-level load signatures for individual appliances and efficiently utilize these signatures to enhance the feature representation capability of the NILM model. Another approach for semi-supervised NILM involves extracting a feature matrix from the unlabeled data to aid in disaggregating appliance-level loads from the labeled data. Feature extraction or feature learning can take a model-free form (refer to Fig. \ref{fig:semi} (c)) or be model-based (refer to Fig. \ref{fig:semi} (d)). In \cite{gillis_nonintrusive_2017}, a co-training semi-supervised approach was proposed, which uses a set of wavelet features to represent energy signals and employs a decision tree for eager learning and nearest-neighbor for lazy learning in the classification process. Notably, there are limited feature learning algorithms for NILM from unlabeled data in existing research. Temporal and spectral feature learning using two-stream CNNs was proposed in \cite{chen_temporal_2022}. This approach explores feature vectors from labeled data in temporal and spectral load signatures. Subsequently, a CNN concatenates these two vectors into one and processes them using fully-connected (FC) or global average pooling (GAP) layers for different appliance types. The technologies employed in semi-supervised NILM are summarized in Table \ref{tab:supervised}. In current research, semi-supervised NILM primarily focuses on the classification task to identify the ON/OFF states of appliances. By combining labeled and unlabeled datasets, semi-supervised algorithms offer high efficiency and significantly reduce the dependence on labeled data volume. \textbf{However}, they have the disadvantages of complex iterative processes and potentially unstable iteration results.

\begin{table}
\vspace{-10pt}
\setlength{\abovecaptionskip}{-.1cm} 
\setlength{\belowcaptionskip}{-2cm} 
\centering
\caption{Advantages and disadvantages of parametric and non-parametric machine learning technologies.}
\label{tab:supervised}
\begin{adjustbox}{width=\linewidth, center}
\begin{tabular}{c | l| l| l } 
\hline\hline
 {Models} & Tasks & Technologies & Pros. v.s. Cons.  \\ 
\hline
 \multirow{13}{*}{Supervised} & \multirow{5}{*}{Regression} & $\cdot$ Linear Regression & \multirow{13}{*}{
 \begin{tabular}[c]{@{}l@{}}
            \textbf{Advantages}:
            \\
            $\cdot$ Powerful learning\\
            $\cdot$ Easy to understand\\
            \\
            \textbf{Disadvantages}:\\
            $\cdot$ Cannot give unknown\\
            information from the\\
            training dataset\\
            $\cdot$ Select good examples \\
            for training\\
            $\cdot$ Large computation time\\
            \end{tabular}}
            \\
            & & $\cdot$ Ridge Regression &  \\
            & & $\cdot$ Lasso regression &  \\
            & & $\cdot$ Neural Network Regression &  \\
            & & $\cdot$ Support Vector Regression &  \\
\cline{2-3}
            & \multirow{8}{*}{Classification} & $\cdot$ Genetic Algorithms &  \\
            & & $\cdot$ Support Vector Machine &  \\
            & & $\cdot$ Random Decision Forest & \\
            & & $\cdot$ Hidden Markov Models &  \\
            & & $\cdot$ Decision Tree & \\
            & & $\cdot$ Navie Bayes Classifier &  \\
            & & $\cdot$ K-Nearest Neighbor &  \\
            & & $\cdot$ Neural Networks Classifier &\\
\hline
 \multirow{7}{*}{Semi-Supervised} & \multirow{7}{*}{Classification} & $\cdot$ HMMs & \multirow{7}{*}{
 \begin{tabular}[c]{@{}l@{}}
            \textbf{Advantages}:
            \\
            $\cdot$ High efficiency\\
            $\cdot$ Small set of labeled data \\
            \\
            \textbf{Disadvantages}:\\
            $\cdot$ Complex iterative process\\
            $\cdot$ Results of are unstable\\
            \end{tabular}}
            \\
            & & $\cdot$ FHMMs & \\
            & & $\cdot$ Dictionary & \\
            & & $\cdot$ Graph Technology & \\
            & & $\cdot$ Autoencoder + K-means & \\
            & & $\cdot$ Teacher-Student Network & \\
            & & $\cdot$ Feature learning + KNN & \\
\hline
 \multirow{6}{*}{Unsupervised} & \multirow{3}{*}{Clustering} & $\cdot$ K-means & \multirow{7}{*}{
 \begin{tabular}[c]{@{}l@{}}
            \textbf{Advantages}:
            \\
            $\cdot$ No labeled data\\
            \\
            \textbf{Disadvantages}:\\
            $\cdot$ Often lesser accuracy\\
            \end{tabular}}
            \\
            & & $\cdot$ Spatial clusteringn & \\
            & & $\cdot$ Subtractive clustering & \\
    \cline{2-3}
    & \multirow{3}{*}{Decoding} & $\cdot$ HMMs &  \\
    & & $\cdot$ FHMMs & \\
    & & $\cdot$ Event + DTW &\\
\hline\hline
\end{tabular}
\end{adjustbox}
\vspace{-15pt}
\end{table}

\subsubsection{Unsupervised Approaches}
As depicted in Fig. \ref{fig:algorithms} (c), an unsupervised model for NILM operates without any labeled training data and is tasked with autonomously uncovering patterns or relationships within the data. In situations where it is either impossible or impractical for a human to propose trends in the data, unsupervised learning can provide initial insights that can then be used to test individual hypotheses. Unsupervised learning models are employed for three primary tasks: 1) clustering, and 2) decoding.

Clustering is a data mining technique used to group unlabeled data based on their similarities or differences. For example, K-means is a popular unsupervised clustering algorithm. In the study by Wang et al. \cite{wang_classification_2015}, K-means was employed to classify the different operational modes of a refrigerator based on its maximum power usage and total energy demand. It successfully grouped the modes, such as normal mode with the least power consumption and defrost/post-defrost modes with higher energy demand, according to their peak power and energy usage. In another approach by Deb et al. \cite{deb_automated_2019}, a density-based spatial clustering of applications with noise (DBSCAN) was proposed to disaggregate heat load from the aggregate load by considering their concentration in the time axis. In contrast to K-means, subtractive clustering is a technique that doesn't require knowledge of the number of classes in advance. Nilson et al. \cite{henao_approach_2017} demonstrated that the unsupervised subtractive clustering algorithm is robust when dealing with measurement or grid power noises in ON/OFF appliance power profiles.

NILM involves the decoding of individual appliance loads from the aggregate load. Decoding-based NILM often relies on HMMs. In the context of NILM, a household's power consumption can be viewed as a superposition of emissions from various Markov chains, each representing a home appliance \cite{kong_improving_2016}. HMMs can enhance the model's adaptability to different households. Ji et al. \cite{ji_factorial_2019} introduced an approach called Additive Factorial Approximate Maximum a Posteriori (AFAMAP) based on Iterative Fuzzy C-Means (IFCM) to decompose aggregated power consumption using independent load models built with HMMs. To mitigate the reliance of traditional HMM-based methods on prior knowledge of appliance working states, Zhao et al. \cite{wu_factorial_2021} proposed an Adaptive Density Peak Clustering (ADPC) algorithm that can automatically determine appliance states based on power consumption patterns. FHMMs have subsequently been employed to reduce the dependency on prior information and enhance usability in real-world scenarios. Unsupervised algorithms have the advantage of not requiring labeled data, but they may necessitate human intervention or prior knowledge (such as a signature database) to interpret patterns and relate them to domain knowledge. Typically, unsupervised algorithms exhibit lower accuracy compared to supervised ones. Majumdar \cite{majumdar_trainingless_2022} demonstrated that their proposed approach outperforms existing unsupervised non-intrusive techniques and performs slightly worse than intrusive supervised approaches. \textbf{However}, there is limited research that directly compares different supervised and unsupervised algorithms for NILM.

To better understand the trends in NILM technology, we conducted an analysis and comparison of how NILM has evolved based on different algorithms over the past decade. We initiated our analysis by searching for publications related to "Non-intrusive load monitoring" or "energy disaggregation" in the Scopus database, which provided annual publication statistics for NILM. To calculate the trends for supervised NILM, we employed keywords such as 'neural network,' 'LSTM,' 'Autoencoder,' and 'learning.' Building on the insights from our discussion on semi-supervised NILM, we used keywords like HMMs, K-means, semi-supervised, event, and clustering in conjunction with 'neural network' or 'learning' to filter semi-supervised algorithm-based work. For unsupervised NILM, we focused on HMMs, K-means, event, and clustering while excluding 'neural network' and 'learning' keywords.

\begin{figure}[ht]
\vspace*{-10pt}
\setlength{\abovecaptionskip}{-.1cm} 
\setlength{\belowcaptionskip}{-2cm} 
    \centering
    \includegraphics[width=0.95\linewidth,scale=1.0]{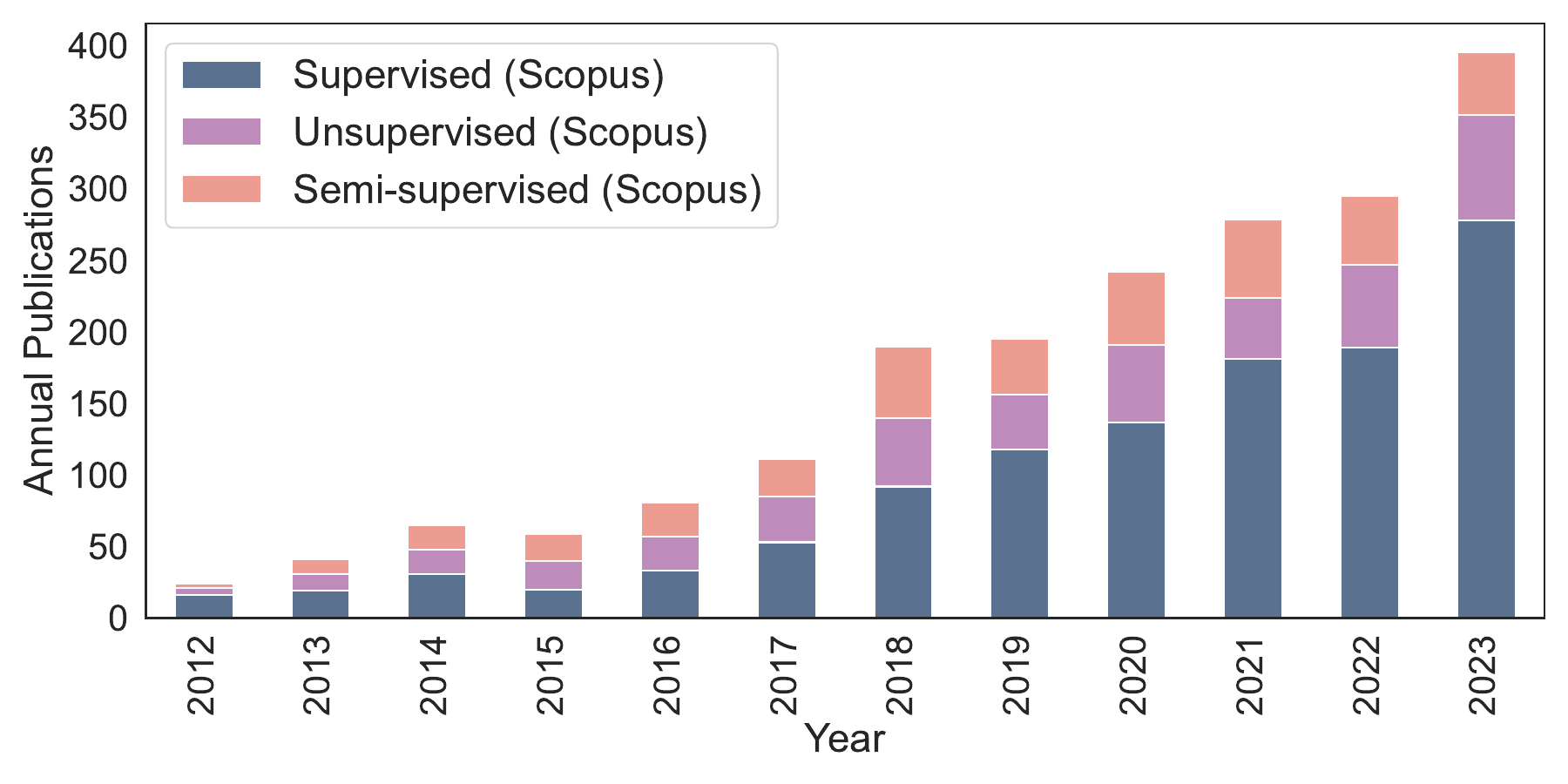}
    \caption{Annual publications of different algorithms for NILM based on Scopus database from 2012 to 2023.}
    \label{fig:annual_algorithm}
\vspace*{-10pt}
\end{figure}

Fig. \ref{fig:annual_algorithm} displays a comparison of the annual publication numbers for different algorithms. Starting from 2012, the number of studies focusing on unsupervised and semi-supervised technologies has remained relatively equal. However, since 2017, there has been a notable surge in research attention towards supervised NILM. In fact, the publication count for supervised algorithms nearly equals the combined count for all other algorithms in the NILM field in 2021 and 2022. This underscores the robust learning capabilities of machine learning technologies in the context of energy disaggregation. It is worth noting that, as of now, there is a lack of research that directly compares the performance of unsupervised, semi-supervised, and supervised algorithms for NILM using the same dataset. Additionally, identifying why certain algorithms perform better for specific electrical appliances than others remains a challenging problem.

\subsection{NILM Approaches with What for Monitoring}
\begin{figure}[ht]
\vspace*{-6pt}
\setlength{\abovecaptionskip}{-.1cm} 
\setlength{\belowcaptionskip}{-2cm} 
    \centering
    \includegraphics[width=0.95\linewidth,scale=1.0]{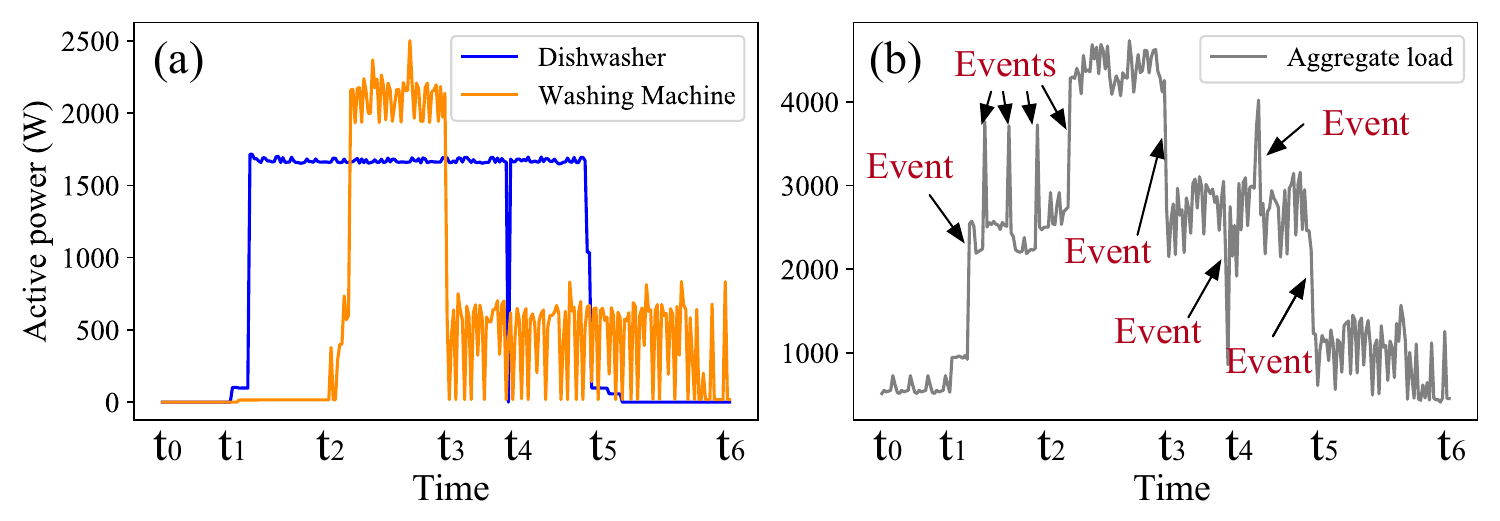}
    \caption{States and events represented with active power for different electrical appliances.}
    \label{fig:state_event}
\vspace*{-6pt}
\end{figure}

NILM approaches can be categorized into state-based and event-based methods, depending on whether they focus on appliance states or events for load monitoring. In this context, '\textbf{state}' is generally characterized by an appliance's power consumption features, including factors like active power, voltage, and current waveforms. The state of an appliance is determined by analyzing its power consumption patterns over time. As illustrated in Fig. \ref{fig:state_event} (a), normally operating electrical appliances typically exhibit specific power consumption levels. For instance, the active power waveform of a dishwasher from $t_1$ to $t_5$ can be considered its state when it's functioning normally. In contrast, appliances like washing machines often have multiple operational modes, each corresponding to distinct states, such as filling, spinning, and draining. The load profiles from $t_2 \rightarrow t_3$ and $t_3 \rightarrow t_6$ may represent different states. In a state-based NILM system, the identification of an electrical appliance's state is achieved through the analysis of these power consumption patterns. 

On the other hand, '\textbf{event}' typically refers to changes in an appliance's power consumption pattern that can be detected within the aggregate power signal. As depicted in Fig. \ref{fig:state_event} (b), events can occur when appliances are switched on or off or when an appliance changes its operational mode.

\subsubsection{State-based NILM}

State-based NILM usually revolves around identifying and monitoring the operational states—such as 'on', 'off', and 'standby'—of each appliance. The fundamental premise of state-based NILM is that each electrical appliance possesses unique characteristics in its power consumption profile when it switches between different operational states \cite{hart_nonintrusive_1992}. This approach usually treats the operation of an appliance as a finite state machine and perform disaggregation based on the state transition model learned during training \cite{he_generic_2019}. By analyzing these unique signatures, the system can infer which appliance is being used, and in which state, without the need for individual appliance meters. Once the state of each appliance has been identified, a NILM system can estimate the energy consumed by each appliance and provide a breakdown of the household's energy consumption. 

In the realm of state-based NILM, techniques such as Combinatorial Optimization (CO) and HMM are commonly utilized to explore the diverse combinations of state sequences exhibited by different appliances \cite{kong_hierarchical_2018, dash2020appliance}. These methods are particularly effective in navigating the complex array of potential state sequences that appliances may demonstrate. To enhance the practicality of these methods, penalties are often incorporated, employing technologies like binary quadratic programming formulation with appliance-specific constraints. This approach is designed to more accurately differentiate the loads of similar appliances, especially in scenarios involving unknown loads \cite{dash2020appliance, marco_mixed_2022}. Furthermore, considering that industrial loads typically operate more consistently in a single state for extended periods and have greater capacity for managing large fluctuations in grid power compared to residential loads, a novel approach involving a double median filtering process has been proposed for initial disaggregation states in load disaggregation. This technique is detailed in \cite{wang2023non}, showcasing its applicability in more complex and dynamic industrial environments.

\textbf{However}, state-based NILM still faces several challenges that impact its effectiveness and accuracy. A primary challenge is the difficulty in distinguishing between appliances with similar energy consumption patterns, particularly when they operate simultaneously, which can lead to misidentification and inaccurate energy usage attribution. This complexity escalates with the increasing number of appliances in a household or facility. Additionally, the presence of unknown or new appliances not accounted for in the system's training data further complicates the disaggregation process. 

\subsubsection{Event-based NILM}
\begin{figure}[ht]
\vspace*{-6pt}
\setlength{\abovecaptionskip}{-.1cm} 
\setlength{\belowcaptionskip}{-2cm} 
    \centering
    \includegraphics[width=0.8\linewidth,scale=1.0]{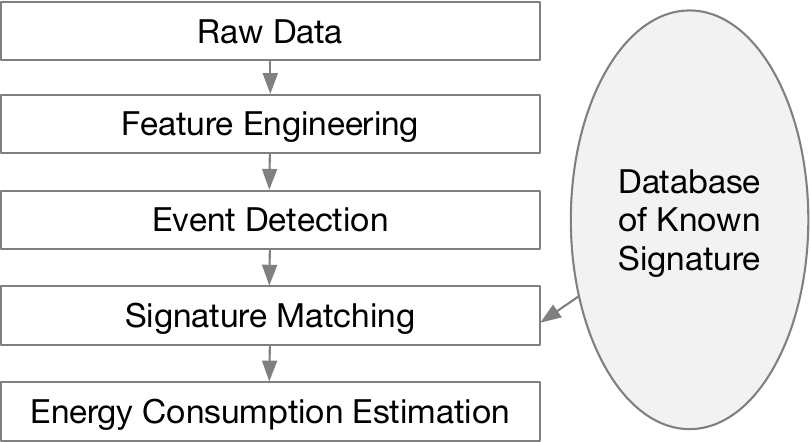}
    \caption{Typical load disaggregation process of event-based NILM.}
    \label{fig:event}
\vspace*{-6pt}
\end{figure}

In the domain of event-based Non-Intrusive Load Monitoring (NILM), the fundamental operational principle is that each activation or deactivation of an appliance generates a unique signature in the overall power profile of a building. These signatures are identifiable as abrupt changes in key electrical parameters such as voltage, current, or power. The ability to detect and analyze these changes enables the identification of specific appliances being used \cite{lu_hybrid_2020}. Fig. \ref{fig:event} illustrates the typical process of load disaggregation in event-based NILM systems. In this process, event detection and signature matching are the most critical steps for the effective functioning of event-based NILM. To reduce dependence on historical data and specific information about individual houses, a novel hybrid approach was introduced in \cite{lu_hybrid_2020}. This method combines a primary algorithm, rigorously designed to identify true events with high accuracy, and two supplemental algorithms specifically tailored to weed out certain types of false alarms. Such a configuration substantially improves the system's capability to discern actual events with greater precision, while effectively minimizing the occurrence of false detections.

Moreover, to address the inherent challenges in setting hyperparameters and detecting multiple simultaneous events—a common limitation in traditional event-based methods—there is a growing adoption of machine learning-based techniques. Notable among these are the wavelet packet tree \cite{himeur2020robust}, the transformer approach \cite{jiao_context_aware_2023}, and CNN \cite{yang_eventdriven_2020}. These advanced machine learning methodologies are employed to more precisely determine the start and stop times of appliances, as well as changes in their operational states, thereby improving the accuracy and reliability of event-based NILM systems.

After detecting an event in the event-based NILM process, it is crucial to perform signature or event matching. This step involves comparing the detected event against a pre-existing signature database to ascertain which appliance triggered the event. To facilitate effective event matching, a two-stage optimization algorithm, as proposed in \cite{liu2022balanced}, has been designed. The first stage of this algorithm employs a depth-first search approach, enhanced with pruning techniques, to generate a set of potential candidate load event sequences. The second stage involves selecting the optimal load event sequences that most accurately match the aggregated load power observed.

Event-based NILM faces significant challenges, primarily due to its reliance on a pre-existing signature or event database for all appliances for energy disaggregation. This approach necessitates prior knowledge of each appliance's unique energy consumption pattern, making the accurate identification and separation of individual appliance events a complex task. One of the main difficulties is dealing with overlapping events, where simultaneous or closely occurring appliance activations can lead to misclassification. The variability in appliance signatures, influenced by factors like age, model, and usage, further complicates the matching process. Additionally, the system's sensitivity to minor power fluctuations can result in false positives, identifying insignificant power changes as appliance events.

\subsection{NILM Approaches with Monitoring What}

\begin{figure}[ht]
\vspace*{-6pt}
\setlength{\abovecaptionskip}{-.1cm} 
\setlength{\belowcaptionskip}{-2cm} 
    \centering
    \includegraphics[width=\linewidth,scale=1.0]{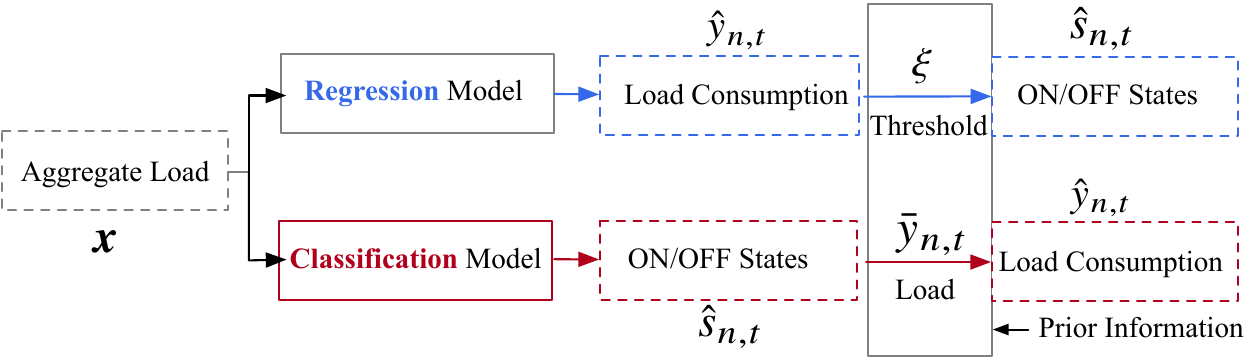}
    \caption{Schematic diagram of regression and classification NILM.}
    \label{fig:regcls}
\vspace*{-6pt}
\end{figure}

According to the model output is continuous or non-continuous values, NILM approaches can be grouped into two main types: regression and classification.
When NILM is approached as a regression problem, the load consumption values $y_{i,t}$(e.g., 1.5 kW) can be directly estimated by the regression model (see Fig. \ref{fig:regcls}). To deduce the ON/OFF states from the estimated load consumption $\hat{y}_{i,t}$, a threshold value $\xi$ is required. For regression supervised learning, linear regression\cite{narges_disaggregation_2022}, ridge regression \cite{sato_energy_2020}, lasso regression, support vector regression \cite{leiria_methodology_2023}, and more and more neural network regression\cite{sato_energy_2020, zhang_sequence-point_2018,shin_subtask_2019, liu_single_2022} have been applied for NILM. 
In contrast, classification NILM tries to estimate the ON/OFF states $s_{i,t}$ of $n$th appliance at time $t$. With the estimate ON/OFF states $\hat{s}_{i,t}$, the prior information of load consumption $\bar{y}_{i,t}$ when the appliance is at work can be used to obtain the estimated load consumption $\hat{y}_{i,t}$. Note that $\bar{y}_{i,t}$ is generally a fixed value. For example, the load consumption of the kettle is $3$kW when it is at work. However, this method might cause large errors for multiple-function appliances (i.e., washing machines) as they consume different amounts of power in different working states. 

\subsubsection{Classification-based NILM}
\begin{figure}[ht]
\vspace*{-6pt}
\setlength{\abovecaptionskip}{-.1cm} 
\setlength{\belowcaptionskip}{-2cm} 
    \centering
    \includegraphics[width=\linewidth,scale=1.0]{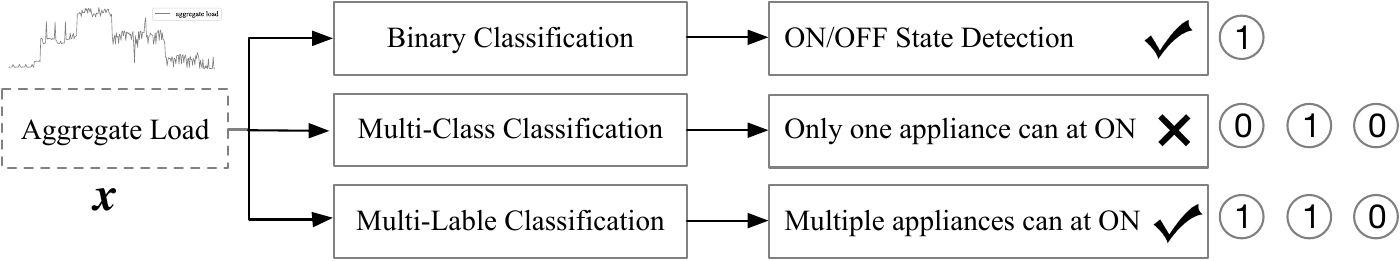}
    \caption{Category of the classification-based NILM.}
    \label{fig:cls}
\vspace*{-6pt}
\end{figure}

Classification-based NILM aims to train models that are adept at distinguishing between different states or activities of appliances, as highlighted \cite{tabanelli_trimming_2022}. These models primarily produce the ON/OFF states of appliances as their output. They achieve this by classifying the aggregated electricity consumption data, determining whether each appliance is in an active (ON) or inactive (OFF) state at any given time \cite{incecco_transfer_2020}. The methodology of classification-based NILM is broadly categorized into three types based on the nature of the output: binary classification, multi-class classification, and multi-label classification (see Fig. \ref{fig:cls}).

In binary classification, the NILM model is specifically designed to differentiate between two states – typically the ON and OFF states of appliances. This model processes the aggregated electricity consumption data to classify each appliance's state as either active (ON) or inactive (OFF) \cite{machlev2018modified, held2018frequency, devlin_nonintrusive_2019, incecco_transfer_2020, le_toward_2021}. This approach is particularly effective for identifying the operational status of specific appliances, rather than detecting multiple appliances operating simultaneously. In contrast, the multi-class classification in NILM.

Multi-class classification in NILM is a sophisticated approach designed to categorize aggregated electricity consumption data into multiple classes, each corresponding to a different appliance or load type. This method extends beyond the basic binary classification of just identifying appliances as being in ON or OFF states. Instead, it involves recognizing and distinguishing between various appliances based on their unique energy consumption patterns \cite{christos_realtime_2021, faustine_adaptive_2021, nie_ensemble_2022}. Unlike multi-label classification, multi-class classification in NILM involves categorizing energy usage data into exclusive classes, where each class represents a single appliance type. In contrast, multi-label classification in NILM assigns multiple labels to the data, indicating the concurrent operation of various appliances.

In multi-label classification NILM, the classification-based model is trained to assign multiple labels or states to the aggregated electricity consumption data \cite{tabatabaei_toward_2017, li_residential_2019, singhal_simultaneous_2019}. Unlike multi-class classification, this approach allows for identifying multiple appliances or load states that may be simultaneously active \cite{singh_non_2020}. The model can recognize and label the presence of multiple appliances or load combinations within the observed data. This enables a more comprehensive analysis of energy usage in complex scenarios where appliances may operate concurrently or exhibit overlapping energy consumption patterns \cite{singh_multi_2022, tanoni_multilabel_2023}.

In classification-based NILM, when estimating the energy consumption of an appliance, it's necessary to have pre-existing knowledge about the typical energy usage of that appliance, which is usually considered as a fixed value. However, this assumption can lead to substantial errors in energy disaggregation, as it may not accurately reflect the actual, variable energy consumption of the appliance in different contexts or over time.

\subsubsection{Regression-based NILM}

Unlike classification methods that categorize data into discrete states like ON or OFF, regression-based NILM focuses on estimating actual energy usage figures, typically in kilowatts or similar units (see Fig. \ref{fig:regcls}). This approach is characterized by its ability to estimate power usage in continuous numerical values (e.g., watts or kilowatts), rather than just classifying appliances into binary states or identifying appliance types. Regression-based NILM employs sophisticated algorithms, such as sequence-to-point CNN \cite{zhang_sequence-point_2018}, encoder-decoder mechanisms \cite{liu_samnet_2022}, dynamic mirror descent \cite{Ledva_realtime_2018}, scale- and context-aware CNN \cite{chen_scale_2020}, bidirectional LSTM \cite{kaselimi_context_2020}, auto-encoder \cite{garcia_fully_2021}, and multi-task learning that combines classification \cite{incecco_transfer_2020, cimen_microgrid_2021, liu_samnet_2022}. These algorithms analyze overall energy consumption patterns, such as the active power measured by a smart meter, and disaggregate this data to estimate the real-time consumption of individual appliances.

A crucial component common to both classification-based and regression-based NILM is the process of feature learning. This involves extracting specific characteristics from the energy usage data, such as peaks, steady states, and fluctuations, and transforming them into a high-dimensional space for model training \cite{chen_temporal_2022, liu2023self}. The strength of regression-based NILM lies in its capability to provide detailed and precise insights into the power consumption of each appliance, thereby enhancing energy management and efficiency analysis. However, this method requires a detailed and accurately labeled dataset for effective training. The complexity of the modeling process escalates with the diversity of appliances and their varying consumption patterns. In environments where many appliances have overlapping or similar energy signatures, this approach becomes particularly challenging, necessitating sophisticated modeling techniques to ensure high accuracy in the results.

\section{Performance Assessment}
There are numerous performance metrics that can be used for NILM performance assessment. These metrics offer a quantitative evaluation of NILM accuracy in identifying and disaggregating individual appliance energy consumption, providing crucial insights into its reliability and effectiveness. By validating performance under varied scenarios, these metrics enable researchers and engineers to compare and optimize different NILM algorithms, contributing to advancements in the field. Accurate performance metrics instill confidence among users, whether homeowners or businesses, ensuring that the NILM system can be trusted for informed energy management decisions. Additionally, these metrics guide ongoing research and development efforts, measuring progress and facilitating a cost-benefit analysis of deploying NILM in real-world environments. Performance metrics serve as a vital tool for the continuous improvement and adaptation of NILM systems in the realm of energy monitoring and management.

\begin{table*}[htbp]
\vspace*{-10pt}
\setlength{\abovecaptionskip}{-.1cm} 
\setlength{\belowcaptionskip}{-2cm} 
\footnotesize{
    \centering
    \caption{Metrics for performance assessment of State/Event estimations}
    \label{tab:metric_class}
    \begin{adjustbox}{width=\linewidth, center}
    \begin{tabular}{ l| l |l}
    \toprule
    \hline
    Metric & Equation & Variables\\
    \hline
    Precision	& $\frac{TP}{TP+FP}$	&$TP$ is the true positive number, $FP$ is the false positive points
    \\ \hline
    Recall	& $\frac{TP}{TP+FN}$	& $FN$ is false negative points
    \\  \hline
    F1	& $\frac{2 \times TP}{2\times TP + FN + FP}$	& -
    \\  \hline
    Accuracy	& $\frac{TP+TN}{TP+FP+TN+FN}$	& $TN$ is the true negative points
    \\ \hline
    Detection Accuracy \cite{liang2009load}	& $\eta_{det}=\frac{N_{dis}}{N_{det}}$	& 
    \begin{tabular}[c]{@{}l@{}}
    $N_{dis}$ is the total number of events that is accurately recognized\\
    $N_{det}$ is the total detected events
    \end{tabular} 
    \\ \hline
    Disaggregation Accuracy \cite{liang2009load}	& $\eta_{dis}=\frac{N_{dis}}{N_{det}-N_{wro}}$	& $N_{wro}$ is the number of wrongfully detected events
    \\
    \hline
    Overall Accuracy \cite{liang2009load}	& $\eta_{all}=\frac{N_{dis}}{N_{det}-N_{wro}+N_{miss}}$	&$N_{miss}$ is the number of missed events
    \\ \hline
    Hamming Loss \cite{batra2014nilmtk}	& $HL = \frac{1}{T}\sum_t \frac{1}{N}\sum_i XOR\left(s_t^i, \hat{s}_t^i\right)$	& $s_t^i$ is the ground truth state from appliance $i$ at time t
    \\ \hline
    Inaccurate portion of TP \cite{makonin2015nonintrusive}	&$Inacc=\sum_{t=1}^T\frac{|\hat{s}_t^i - s_t^i|}{K^i}$	 &$K^i$ is the number of states from appliance $i$
    \\ \hline
    Pprecision  \cite{makonin2015nonintrusive}	&$\frac{TP - inacc}{TP+FP}$ & -
    \\ \hline
    Precall \cite{makonin2015nonintrusive}	& $\frac{TP-inacc}{TP+FN}$	& -
    \\ \hline
    Identification accuracy \cite{dinesh_residnetial_2016}	& $A_{STP} = \frac{n_{ow}}{n_{tow}}$	& \begin{tabular}[c]{@{}l@{}}
    $n_{ow}$ is the number of observation window (OW)\\
    $n_{tow}$ is the percentage turned-on appliance combination
    \end{tabular} 
    \\ \hline
    Informedness	& $Bi=Recall + TNR -1 $	& $TNR = \frac{TN}{TN+FP}$
    \\ \hline
    Markedness	&$Ms = Precision + TNA - 1$	& $TNA = \frac{TN}{TN+FN}$
    \\ \hline
    Matthews Correlation Coefficient \cite{barsim2018feasibility}	& $MCC = \sqrt{Bi\cdot Ms}$	& - 
    \\ \hline
    Average F1 for multiclass \cite{zhang2019new}	& $F_{macro} = \frac{1}{N} \sum_{i=1}^N F(TP_i, TN_i, FP_i, FN_i)$	& -
    \\ \hline
    Micro F1 for multiclass \cite{zhang2019new}	& $F_{micro} = F\left(\sum TP_i, \sum TN_i, \sum FP_i, \sum FN_i,\right)$	& -
    \\ \hline
    \hline
\toprule
    \end{tabular}
\end{adjustbox}}
\end{table*}

\begin{table*}[htbp]
\vspace*{-10pt}
\setlength{\abovecaptionskip}{-.1cm} 
\setlength{\belowcaptionskip}{-2cm} 
\footnotesize{
    \centering
    \caption{Metrics for performance assessment of Power/Energy estimations}
    \label{tab:metric_energy}
    \begin{adjustbox}{width=\linewidth, center}
    \begin{tabular}{ l| l |l}
    \toprule
    \hline
    Metric & Equation & Variables\\ \hline
    Mean absolute error	& MAE$ = \sum_{i=1}^N{\frac{\hat{y}_{i,t} - y_{i,t}}{N}}$ & $N$ is the total appliance number
    \\ \hline
    Symmetric Mean Absolute Percentage Error\cite{delfosse2020deep}	& SMAPE $=\frac{2}{N}\sum_{i=1}^{N}\frac{|\hat{y}_{i,t}-y_{i,t}|}{|\hat{y}_{i,t} + y_{i,t}|}$ & $y_{i,t}$ and is the actual power of $i$th appliance
    \\ \hline
    Total energy correctly assigned \cite{kolter_redd_2011}	& TECA = $1- \frac{\sum_{t=1}^{T}\sum_{i=1}^{N}|\hat{y}_{i,t} - y_{i,t}|}{2\sum_{t=1}^T{y_{i,t}}}$ & $\hat{y}_{i,t}$ is the estimated power of $i$th appliance
    \\ \hline
    Error in total energy assigned \cite{batra2014nilmtk}	&  $\text{ETEA}_i =|\sum_t y_{i,t} - \sum_t \hat{y}_{i,t}|$ & -
    \\ \hline
    Normalised error in assigned power\cite{batra2014nilmtk}	& $\text{NEAP}_i = \frac{\sum_t |y_{i,t} - \hat{y}_{i,t}|}{\sum_t y_{i,t}}$ & - 
    \\ \hline
    Root mean square error \cite{batra2014nilmtk}	& $\text{RMSE} = \sqrt{\frac{1}{T} \sum_t \left(y_{i,t} - \hat{y}_{i,t}\right)^2}$ & $T$ is the total time slots
    \\ \hline
    RMSE \cite{figueiredo_electrical_2014}	& $\sqrt{\frac{\sum_{t=1}^T\sum_{d=1}^m\left(X-\hat{X}\right)^2}{T\times m}}$ & -
    \\ \hline
    Disaggregation Error \cite{figueiredo_electrical_2014}	& $\text{DE} = \sum_{i=1}^{N}\frac{1}{2}\parallel y_i - \hat{y}_i\parallel_{F}^2$ & -
    \\ \hline
    Disaggregation percentage \cite{bonfigli2015unsupervised}	& $D=\frac{\sum_{i=1}^N E_i}{E_{tot}}$ &  $E_{i}$ is the estimated energy of $i$th appliance
    \\ \hline
    Fraction of total energy assigned correctly \cite{uttama_DRED_2015}	& $\text{FTE} = \sum_{i} \min \left(\frac{\sum_i y_{i,t}}{\sum_{i,t} y_{i,t}}, \frac{\sum_i \hat{y}_{i,t}}{\sum_{i,t} \hat{y}_{i,t}}\right)$ & -
    \\ \hline
    Number of appliances identified correctly \cite{uttama_DRED_2015}	&$J_a = \frac{J_a^P \bigcap J_a^a}{J_a^P \bigcup J_a^a }$ & \begin{tabular}[c]{@{}l@{}}
    $J_a^P$ is the predicted set of appliances\\
    $J_a^a$ is the actual set of appliances
    \end{tabular}
    \\ \hline
    Number of states identified correctly \cite{uttama_DRED_2015}	& $J_s = \frac{J_s^P \bigcap J_s^a}{J_s^P \bigcup J_s^a }$ & \begin{tabular}[c]{@{}l@{}}
    $J_s^P$ is the predicted set of appliance states\\
    $J_s^a$ is the actual set of appliance states 
    \end{tabular}
    \\ \hline
    Proportion error per appliance \cite{uttama_DRED_2015}	&$P_e = \left|\sum_t y_{i,t} - \hat{y}_{i,t} \right|$ & -
    \\ \hline
    Normalized error per appliance \cite{uttama_DRED_2015}	& $N_e = \frac{\sum_t|y_{i,t} - \hat{y}_{i,t}|}{\sum_t y_t^i}$ & -
    \\ \hline
    Correct assignment rate \cite{kong_improving_2016}	& $\text{CAR} = 1 - \frac{\sum_{t=1}^T \sum_{i=1}^N|\hat{y}_{i,t} - y_{i,t}|}{2\sum_{t=1}^T y_t}$ & -
    \\ \hline
    Precision of appliance \cite{roberto_nonintrusive_2017}	& $P^i = \frac{\sum_{m=1}^M \min \left(\hat{y}_{m,t}^i, y_{m,t}^i\right)}{\sum_{m=1}^M \hat{y}_{m,t}^i}$	& $M$ is the total number of samples.
    \\ \hline
    Recall of appliance \cite{roberto_nonintrusive_2017}	& $R^i = \frac{\sum_{m=1}^M \min \left(\hat{y}_{m,t}^i, y_{m,t}^i\right)}{\sum_{m=1}^M {y}_{m,t}^i}$ & -
    \\ \hline
    Overall Precision \cite{roberto_nonintrusive_2017}	& $P = \frac{1}{N} \sum_{i=1}^N P^i$ & -
    \\ \hline
    Overall Recall \cite{roberto_nonintrusive_2017}	& $R =  \frac{1}{N} \sum_{i=1}^N R^i$ & -
    \\ \hline
    Normalized Disaggregation Error \cite{roberto_nonintrusive_2017}	& $\text{NDE} = \sqrt{\frac{\sum_{m,i} \left(y_{m,i} - \hat{y}_{m,i}\right)}{\sum_{m, i} \left(\hat{y}_{m,t}^i\right)^2}}$ & -
    \\ \hline
    Energy True Positive \cite{bao2018enhancing}	& $TP^E = \sum_{t=1}^T \min \left(\hat{y}_t, y_t\right)$& -
    \\ \hline
    Energy False Positive \cite{bao2018enhancing}	& $FP^E = \sum_{t=1}^T \max\left(\hat{y}_t - y_t, 0\right)$ & -
    \\ \hline
    Energy False Negative \cite{bao2018enhancing}	& $FN^E = \sum_{t=1}^T \max\left( y_t - \hat{y}_t, 0\right)$ & -
    \\ \hline
    Energy True Negative  \cite{bao2018enhancing}	& $TN^E = \sum_{t=1}^T \min \left(y^{max} - \hat{y}_t, y^{max} - y_t\right)$ & -
    \\ \hline
    Energy-based Precision \cite{bao2018enhancing}	& $P^E = \frac{\sum_{t=1}^T \min \left(\hat{y}_t, y_t\right)}{\sum_{t=1}^T \hat{y}_{i,t}}$ & -
    \\ \hline
    Energy-based Recall \cite{bao2018enhancing}	& $R^E = \frac{\sum_{t=1}^T \min \left(\hat{y}_{i,t}, y_{i,t}\right)}{\sum_{t=1}^T y_{i,t}}$ & -
    \\ \hline
    Energy-based F1 Score \cite{bao2018enhancing} &	$F_1^E = 2\times \frac{P^E \times R^E}{P^E \times R^E}$ & -
    \\ \hline
    Balanced accuracy \cite{bao2018enhancing}	& $\text{BACC}^E = \frac{1}{2} \left(\frac{TP^E}{TP^E + FN^E} + \frac{TN^E}{TN^E + FP^E}\right)$ & -
    \\ \hline
    Average normalized error \cite{he_non-intrusive_2018}	& $\text{ANE} = \frac{|\sum_{i=1}^N y_i - \sum_{i=1}^N \hat{y}_i|}{\sum_{i=1}^N y_i}$
    \\ \hline
    Signal aggregate error \cite{zhang_sequence-point_2018}	 & $SAE = \frac{|\hat{E}_{tot} - E_{tot}|}{E_{tot}}$ & $E_{tot}$ is the total energy 
    \\ \hline
    Disaggregation Error Measurement \cite{zhao2018improving}	& $\text{DEM} = \frac{\sum_{i=1}^N\left|y_i - \hat{y}_{base} - \sum_{m=1}^M \hat{y}_{m,i}\right|}{\sum_{i=1}^N P_i}$ & -
    \\ \hline
    Intersection over Union \cite{cui2019estimation}	&$IoU = \frac{\sum_t^T \min \left(y_t^i, \hat{y}_t^i\right)}{\sum_{t}^T \max \left(y_t^i, \hat{y}_t^i\right)}$ & $\hat{E}_d$ is the estimated daily energy
    \\ \hline
    Energy error per day \cite{incecco_transfer_2020}	&$\text{EpD} = \frac{1}{D}\sum_{d=1}^D|\hat{E}_d - E_d|$ & $E_d$ is the actual daily energy
    \\ \hline
    Estimated energy fraction index \cite{kaselimi_context_2020}	&$\text{EEFI(i)} = \sqrt{\frac{\sum_t \hat{y}_{i,t}}{\sum_t\sum_i\hat{y}_{i,t}}}$ & 
    \\ \hline
    Actual energy fraction index \cite{kaselimi_context_2020}	&$\text{AEFI(i)} = \sqrt{\frac{\sum_t y_{i,t}}{\sum_i\sum_t \hat{y}_{i,t}}}$ & -
    \\ \hline
    Absolute difference of energy \cite{kaselimi_context_2020}	& $DEFI(i) = |EEFI(i) - AEFI(i)|$ & - 
    \\ \hline
    \multirow{2}{*}{Mean quantile score \cite{lin_privacy_2022}}	& $QS(y_{i,m}, \hat{y}_{i,m}, q) = \left\{\begin{matrix}
    (1-q)(\hat{y}_{i,m}^q - {y}_{i,m}^q ), ~~\hat{y}_{i,m}^q \geqslant  {y}_{i,m}^q  \\ 
    q({y}_{i,m}^q - \hat{y}_{i,m}^q),~~\hat{y}_{i,m}^q< {y}_{i,m}^q
    \end{matrix}\right.$  & $q$ is the quantile score \\
    &$MQS = \frac{1}{M\times |Q|\sum_{m=1}^{M}\sum_{q\in Q}QS(y_{i,m}, \hat{y}_{i,m}, q)}$ & $Q$ is the quantile set
    \\ \hline
    Winkler score \cite{lin_privacy_2022}	& $WS = \left\{\begin{matrix}
    \bigtriangleup_{i,m} + 2(y_{i,m} - y_{i,m, max})/(1-\alpha ), ~~ y_{i,m}>y_{i,m, max}\\ 
    \bigtriangleup_{i,m}, ~~ y_{i,m, min}\leqslant y_{i,,}\\ 
    \bigtriangleup_{i,m} + 2(y_{i,m, min} - y_{i,m})/(1-\alpha), ~~ y_{i,m}<y_{i,m, min}
    \end{matrix}\right.$ & -
    \\ \hline
    \hline
\toprule
    \end{tabular}
\end{adjustbox}}
\end{table*}
\subsection{Metrics for Classification NILM}
NILM can directly estimate the binary ON/OFF state or multi-class states of an appliance. The classification metrics can quantify and communicate how well a model is performing in terms of correctly classifying instances. The most widely used metrics for classification NILM are accuracy and $F_1$ score \cite{kong_practical_2020, liu_samnet_2022, kaselimi_context_2020}, 

\begin{align}
     F_1=\frac{2\times \text{Precision} \times \text{Recall}}{\text{Precision} + \text{Recall}}\\
     \text{Accuracy} = \frac{TP+TN}{TP+FP+TN+FN}\\
     \text{Precision} = \frac{TP}{TP+FP}, ~~~\text{Recall} = \frac{TP}{TP+FN} 
\end{align}
where \textit{TP}, \textit{FP}, \textit{FN} are the true positive, false positive and false negative, respectively. 
$F_1$ score evaluates how well a classifier estimates the correct labels for a given set of data. It can handle imbalanced datasets. However, it assumes that precision and recall are equally important. The accuracy is the ratio of correct estimations to all estimated results. It works well only if there are an equal number of samples belonging to each class. Table \ref{tab:metric_class} summarized the metrics used to evaluate the performance of NILM for state and event detection. \textbf{However}, the current study does not indicate which evaluation metric is more effective in assessing the strengths and weaknesses of the model.

\subsection{Metrics for Power Estimation}
In addition to identifying the state or event of an appliance, NILM can also be utilized for estimating the power consumption of individual appliances. If the estimated power is greater than zero or exceeds a small threshold value, the appliance can be identified as operational. Table \ref{tab:metric_energy} summarizes the metrics used for power estimation. Mean Absolute Error (MAE) measures the average absolute difference between the estimated and actual power consumption, providing a straightforward indication of the error magnitude. Root Mean Squared Error (RMSE) is similar to MAE but penalizes larger errors more significantly due to the squared term, making it sensitive to outliers. Total energy correctly assigned expresses accuracy as a percentage of the actual power consumption, offering insight into the relative accuracy of the estimation \cite{kolter_redd_2011}. When assessing the performance of NILM algorithms, it's common to use a combination of these metrics to gain a comprehensive understanding of their strengths and weaknesses in estimating power consumption. The choice of metrics may depend on the specific goals and requirements of the monitoring application.

\section{Selective Key Applications}
\begin{figure}[ht]
\vspace*{-6pt}
\setlength{\abovecaptionskip}{-.1cm} 
\setlength{\belowcaptionskip}{-2cm} 
    \centering
    \includegraphics[width=\linewidth,scale=1.0]{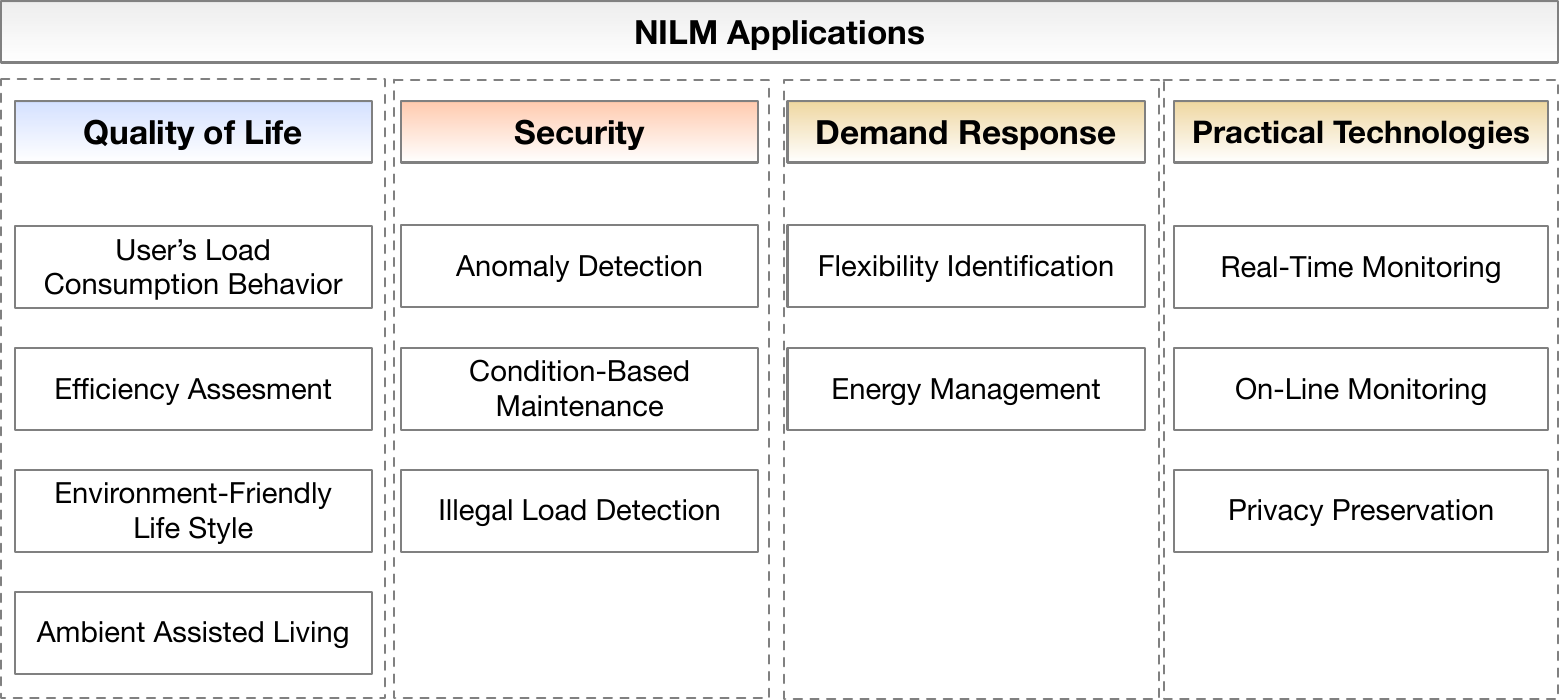}
    \caption{Selective key applications of NILM.}
    \label{fig:applications}
\vspace*{-6pt}
\end{figure}

NILM has various applications due to its ability to disaggregate and analyze electricity consumption without the need for individual appliance-level sensors. It's a valuable tool for enhancing energy efficiency, promoting sustainability, and supporting various applications within both residential and commercial settings (see a summary in Fig. \ref{fig:applications}).

\subsection{Quality of Life}
NILM is crucial for enhancing the quality of life by promoting energy efficiency, reducing costs, supporting sustainability efforts, and empowering individuals and communities to make informed decisions about their energy consumption.
\subsubsection{User Behavior Understanding}
NILM proves invaluable in deciphering user behavior regarding energy consumption. By disaggregating energy data at the appliance level, NILM provides detailed insights into usage patterns, revealing daily routines, appliance interactions, and trends in energy consumption over time \cite{kong_practical_2020,gopinath_energy_2020, jiang2021home, cimen_microgrid_2021, cimen_online_2022,han2023unknown}. This level of granularity allows for a deep understanding of when and how users engage with their devices. Moreover, NILM offers opportunities for personalized feedback, empowering users with actionable insights to enhance their energy consumption habits \cite{ghosh_improved_2021}. The system's ability to analyze and understand user behavior allows for the development of customized recommendations \cite{luo_personalized_2021}, fostering a culture of energy awareness and efficiency \cite{jiang2021home}. Researchers also leveraged NILM data for behavioral studies, contributing to a comprehensive understanding of human interactions with energy-consuming devices and facilitating the development of user-centric solutions for smart homes and buildings \cite{rafsanjani2018linking, ruano2019nilm, brucke2021non,liu2022home}.

\subsubsection{Efficiency Assessment}
Energy efficiency assessment refers to the process of evaluating and analyzing the energy performance of a system with the goal of identifying opportunities for improvement and implementing measures to enhance energy efficiency. The objective is to optimize energy use, reduce waste, and achieve a more sustainable and cost-effective operation. In recent years, methods for energy efficiency assessment based on NILM have been developed \cite{kong2020household, jiang2021home, yao2022user, garcia2020nilm}. These studies have demonstrated that by combining the analysis results of energy efficiency indicators, household users can identify weaknesses in their energy efficiency and improve overall electricity efficiency by adjusting their energy-use patterns. Ref. \cite{rafati2022fault} discussed that even though NILM could be successfully implemented for the energy efficiency evaluation of HVACs, and enhance the performance of these techniques, there are many research opportunities to improve or develop NILM-based methods to deal with real-world challenges. However, existing methods primarily rely on information such as family population, household income, and housing area, which is related to user privacy and can be difficult to obtain for evaluating user energy efficiency. A smarter method should be developed based solely on smart meter data to assess user energy efficiency.

\subsubsection{Environment-Friendly Life Style}
NILM is indispensable for cultivating an environmentally friendly lifestyle by providing individuals with detailed insights into their energy consumption patterns. The ability to customize energy-saving strategies based on individual habits promotes sustainable practices tailored to specific lifestyles. Moreover, NILM supports the integration of renewable energy sources, encourages behavioral changes to reduce energy waste, and contributes to lowering carbon footprints \cite{saeedi_adaptive_2021, lin_privacy_2022}. Beyond individual impact, NILM can inform sustainable building design and promote the adoption of eco-friendly technologies, collectively contributing to global environmental goals \cite{hamed_linking_2018}. In essence, NILM serves as a catalyst for positive environmental change, aligning personal choices with broader efforts toward energy efficiency and sustainability.
\subsubsection{Ambient Assisted Living}
Ambient Assisted Living (AAL) integrated with NILM represents a pioneering approach to enhancing the quality of life for individuals, particularly the elderly, by seamlessly blending technology and personalized care. AAL leverages smart devices, sensors, and assistive technologies to create living environments that actively support occupants in their daily activities while promoting independence. The integration of NILM into AAL systems further augments their capabilities by enabling the non-intrusive monitoring of energy consumption patterns. Various health features, including inactivity, sleep disorders, memory issues, variations in activity patterns, and low-activity routines, can be inferred from smart meters through the use of NILM technologies. A comprehensive overview of NILM in the application fields of HEMS and AAL has been discussed in \cite{ruano2019nilm}. Understanding the load consumption behavior of users can further be applied to monitor the activities of the elderly and detect deviations from their daily routines \cite{alcala2017assessing}. Furthermore, scalability for identifying potential milestones was studied in \cite{alcala2017sustainable} by simplifying parameterization for sustainable homecare monitoring.

\subsection{Security}
NILM can contribute to security systems by detecting occupancy patterns in a building. Unusual patterns or unexpected changes in energy consumption may indicate unauthorized access or occupancy, triggering security alerts.

\subsubsection{Anomaly Detection}
By establishing baseline patterns for normal operation, NILM can raise alerts when unexpected variations occur, such as unusual energy spikes, irregular appliance behaviors, or changes in consumption patterns. In general, the proposed approach initially learns the normal operation of an appliance and subsequently monitors its energy consumption for anomaly detection \cite{rashid2019evaluation}. In Ref. \cite{azizi2021appliance}, three features were proposed: (i) the distance of a given event from the predefined classes; (ii) the distribution of power values after load disaggregation, and (iii) the participation index of each appliance for anomaly detection. However, the authors in \cite{rashid2019can} discussed that the output of NILM is often not accurate enough for identifying anomalies. Therefore, they advocate for the development of anomaly-aware NILM methods, emphasizing the adoption of post-processing techniques to improve the accuracy of NILM.

\subsubsection{Condition-Based Maintenance}
 Unlike traditional time-based maintenance approaches, condition-based maintenance with NILM focuses on real-time monitoring and analysis of electrical signatures and load patterns associated with machinery and appliances \cite{bucci2021state}. NILM allows for the detection of anomalies and deviations from normal operating conditions, enabling the early identification of potential issues or deterioration in equipment performance. This proactive approach to maintenance helps organizations move away from fixed schedules, allowing for more efficient resource allocation and reducing unnecessary downtime. The use of NILM as a viable tool for condition-based maintenance has been demonstrated in \cite{aboulian_nilm_2019, green_nilmdashboard_2020}. However, the authors discuss that the development of accurate fault detection and anomaly identification is crucial to ensure the practicality and cost-effectiveness of condition-based maintenance.
 
\subsubsection{Illegal load detection}
Illegal load detection can be considered a narrower scope of anomaly detection, specifically focusing on identifying loads that violate rules or regulations. Illegal load detection with NILM involves utilizing NILM technology to identify and flag unauthorized or suspicious electrical loads within a system. NILM analyzes electrical signatures and load patterns to establish a baseline of normal usage. Deviations from this baseline, such as unexpected or irregular load patterns, can indicate the presence of illegal loads. These could include unauthorized appliances, tapping into power sources without permission, or any activity that violates the established electrical usage norms. NILM for illegal load detection is primarily applied to combat illicit activities such as the unauthorized charging of electric bicycles \cite{wu2022electric} and unauthorized charging of electric bicycle lithium batteries \cite{wangresearch_2023}. In \cite{wu2022electric}, a non-intrusive online charging load identification method based on spectrum characteristics at the edge, and a 'confirmed charging' state identification method based on tracking the entire charging process in the cloud, are proposed for inspecting illegal charging of electric bicycles. Real-world condition experiments conducted in offices, factories, and laboratories demonstrate that the accuracy of lithium battery detection can reach 93\%, yielding satisfactory results \cite{wangresearch_2023}.

\subsection{Demand Response in Smart Grids}
\subsubsection{Flexibility Identification}
Appliance flexibility is the realizable increase or decrease of energy consumption \cite{lucas2019load}. Energy flexibility is crucial for the efficient operation and optimization of smart grids, as it allows for dynamic adjustments in energy consumption and production to match the varying demands on the grid. Flexibility enables the integration of intermittent renewable energy, such as solar and wind, by allowing users to adapt their energy consumption patterns to align with periods of abundant or scarce energy supply. Household loads, such as heating systems, refrigerators, or freezers, are considered to have high potential for providing flexibility. Generally, NILM-based methods for energy flexibility characterization can be divided into two stages. The first stage involves estimating the consumption patterns of shiftable appliances using NILM. Subsequently, energy flexibility is characterized by determining the earliest and latest start times of these appliances at the second stage \cite{lucas2019load, azizi2021residential}. Additionally, residential electric vehicle charging loads constitute a significant part of flexible loads and are valuable for demand response in smart grids. Reference \cite{zhao2019quantifying} has demonstrated that the charging power levels of users' electric vehicles can be automatically identified and extracted using NILM results.
\subsubsection{Energy Management}
Home and building energy management serve as cornerstones of smart grids, offering a myriad of benefits, including demand response, grid stability, cost savings, and environmental sustainability. As smart grids evolve, the effective integration of energy management at the individual building level becomes increasingly essential for achieving a resilient, sustainable, and intelligent energy infrastructure. With end-users becoming more proactive due to the adoption of distributed energy resources such as solar PV, integrating users' load consumption behavior into energy management systems becomes necessary in smart grids. One common application of NILM in energy management is to minimize users' inconvenience when scheduling day-ahead loads for energy savings \cite{cimen_microgrid_2021, liu2022home, cimen_online_2022}.

\subsection{Technologies Towards Practical Applications}
Recognizing the significance of Non-Intrusive Load Monitoring (NILM) in the context of the smart grid, there is a natural inclination to delve into methods for enhancing its practical application. The goal is to explore avenues that make NILM more feasible and effective in real-world scenarios, aligning with the evolving needs of smart grid implementations.
\subsubsection{Real-Time Monitoring}
Real-time monitoring technologies are integral to the efficacy of NILM, playing a pivotal role in enhancing the immediacy and responsiveness of energy management systems. The instantaneous feedback provided by real-time monitoring allows users to make prompt decisions and adjustments to optimize their energy consumption. The real-time NILM algorithm is utilized to identify the combination of currently active appliances and disaggregate their power levels at the present moment. Total power and voltage measurements of a house/building can be used for instantly identifying the turned ON/OFF appliances, which mainly includes, event detection and model for appliance classification \cite{welikala2019implementation}. Once the appliance has been turned-on, the system can calculate its power in real-time \cite{athanasiadis2021scalable, christos_realtime_2021,yu_towardsmart_2021}. 
The proposed work on real-time energy disaggregation for solar generation \cite{li2020real} and plug-in electric vehicles \cite{ebrahimi2021real} involves developing methods to instantly analyze and distinguish the energy consumption patterns related to these specific domains. 

\subsubsection{On-Line Monitoring}
On-line monitoring emphasizes the continuous and interconnected nature of the monitoring system, suggesting that it is part of a larger online network where data may be shared or accessed in real-time or near real-time. On-line NILM typically involves processing short durations of aggregate input data through a cloud-based implementation of the model within a smart metering infrastructure \cite{mengistu2018cloud, asres2021computational}. Additionally, on-line NILM can be integrated into a microgrid to comprehend users' load consumption patterns for an energy management system \cite{cimen_online_2022}. However, on-line or real-time NILM is only useful for immediate demand response and is not necessary for day-ahead load scheduling in energy management.
\subsubsection{Privacy Preservation}
\begin{figure}[ht]
\vspace*{-12pt}
\setlength{\abovecaptionskip}{-.1cm} 
\setlength{\belowcaptionskip}{-2cm} 
    \centering
    \includegraphics[width=0.7\linewidth,scale=1.0]{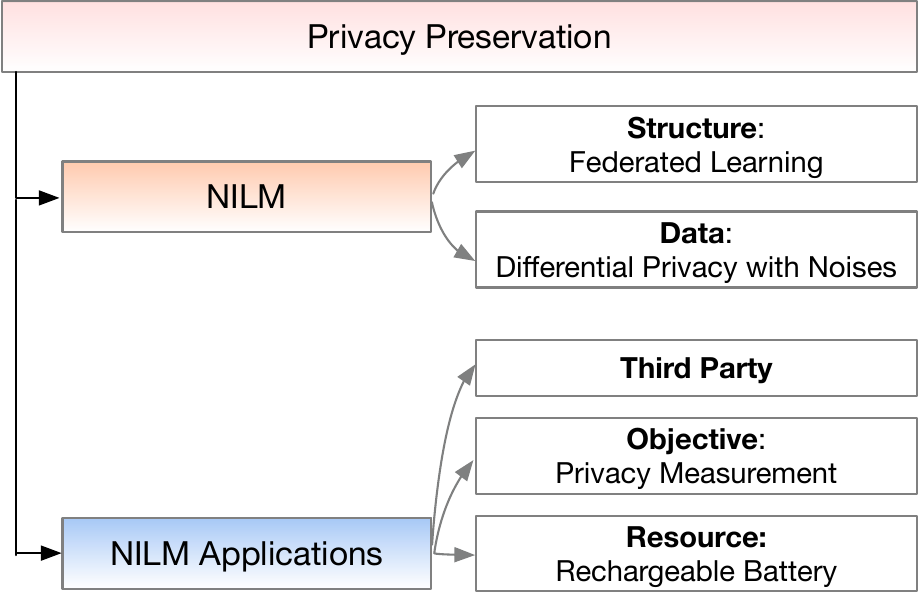}
    \caption{Privacy-preserving methods for NILM and its applications.}
    \label{fig:privacy}
\vspace*{-12pt}
\end{figure}

As NILM involves monitoring and analyzing the electrical signatures of appliances to infer user activities, ensuring the privacy of individuals is paramount when considering the application of NILM in the real world. This becomes particularly crucial in smart home environments, where granular insights into appliance usage patterns may inadvertently reveal sensitive information about occupants' routines and lifestyles. As illustrated in Fig. \ref{fig:privacy}, to safeguard user privacy in NILM, protective measures can be implemented either by adjusting the model structure (e.g., federated learning \cite{zhang2022fednilm, wang2023blockchain}) or by incorporating differential privacy through the injection of noise into the smart meter data \cite{wang2020privacy, zheng2021decentralized}. In these methods, the smart meter captures the user's real power consumption. For NILM applications, particularly in smart grid energy management, there are generally three approaches to protect user privacy. One involves engagement with a trusted third party. However, it proves ineffective when the adversary gains access to the original smart meter measurements through unauthorized sensor installation or by compromising a third party \cite{you2022non}. Another method involves incorporating privacy considerations into the objective function to balance the trade-off between energy savings \cite{you2022non}. A widely used alternative is to introduce rechargeable batteries or other energy storage units as physical noise to obscure the private information within user load curves \cite{li2023research}. However, rechargeable batteries are relatively expensive, and energy storage units must primarily respond to customer demands \cite{zheng2021decentralized}.

\section{Challenges and Prospects of NILM in the Future Smart Grids}
The challenges and prospects of NILM in future smart grids encapsulate a dynamic landscape of opportunities and obstacles. On the one hand, the increasing complexity of modern energy systems poses challenges for NILM algorithms, requiring them to adapt to diverse appliance behaviors, variations in user preferences, and emerging technologies. The integration of renewable energy sources, electric vehicles, and smart appliances further amplifies the intricacy of load disaggregation. Privacy concerns and the need for robust security measures also pose challenges in the context of the growing volume of sensitive energy data, even though some technologies have already been proposed to address these problems. The prospects are equally compelling, with the potential for advanced machine learning techniques, artificial intelligence, and data analytics to refine NILM algorithms. Furthermore, the evolution of communication infrastructures and the proliferation of smart devices offer avenues for enhanced data accessibility and accuracy. 

\subsection{Single-Label and Multi-Label NILM}
One challenge with existing machine learning-based NILM algorithms is the necessity to train a separate model for each target appliance. Despite sharing a common architecture, these models require distinct series of parameters to effectively differentiate and identify different appliances \cite{wu_structured_2022,chen_temporal_2022,cimen_microgrid_2021,incecco_transfer_2020, kaselimi_context_2020}. These methods focus on single-label classification, where the task is to assign a specific label (appliance) to a segment of the energy consumption signal. However, in real-world scenarios, multiple appliances may be active simultaneously, leading to a need for multi-label NILM. Besides, the individualized training approach not only demands significant computational resources and time but also raises scalability concerns as the number of target appliances grows. In recent years, researchers have paid attention to this challenge by developing single-to-multiple disaggregation algorithms based on encoder-decoder architecture \cite{liu_single_2022} and multi-label classification \cite{singh_non_2020, singh_multi_2022, tanoni_multilabel_2023}. The load consumption of multiple appliances can be directly estimated as discussed in \cite{liu_single_2022}. However, for appliances with small power profiles, like fridges, they may be overwhelmed by other appliances, leading to significant disaggregation errors. Treating NILM as a classification problem allows the adoption of multi-label classification methods to identify the ON/OFF state of multiple appliances. However, multi-label classification often involves more complex algorithms compared to their single-label counterparts. To our knowledge, no study has investigated whether multi-label NILM suffers from lower accuracy compared to single-label classification. Additionally, the increased computational complexity can impact training time and resource requirements. Overcoming this challenge requires further studies to achieve higher or similar performance than single-label NILM with much smaller model sizes.

\subsection{Complicated States of Type II Appliances}
\begin{figure}[ht]
\vspace*{-12pt}
\setlength{\abovecaptionskip}{-.1cm} 
\setlength{\belowcaptionskip}{-2cm} 
    \centering
    \includegraphics[width=\linewidth,scale=1.0]{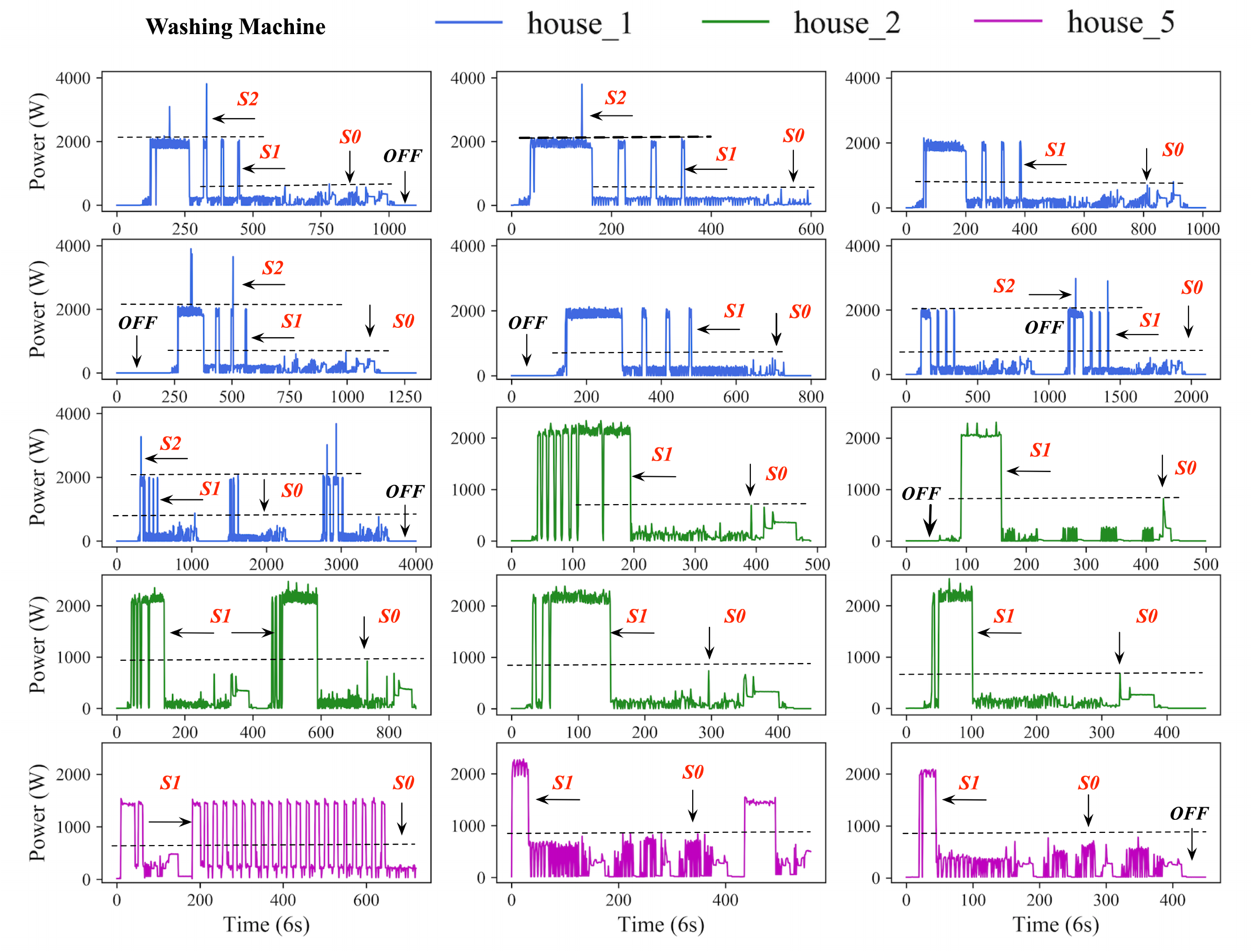}
    \caption{Examples of complicated states of Type II appliance - Washing Machine.}
    \label{fig:typeii}
\vspace*{-12pt}
\end{figure}

Unlike simpler Type I appliances with distinct on/off patterns, Type II appliances may have variable power levels, intermittent usage, or irregular load signatures, making it challenging for traditional NILM algorithms to precisely identify their behavior \cite{kong_practical_2020}. Fig. \ref{fig:typeii} illustrates the load profiles of the washing machine, a typical Type II appliance with multiple functions, from different houses in the UK-DALE dataset. It is evident that the timing and magnitude of power consumption vary significantly among different houses, possibly due to variations in brand or other factors. Type II appliances typically exhibit complex load patterns that make it more difficult to accurately discern their operational states and energy consumption from the overall electrical signal and also affect the generalization performance of the proposed NILM methods. Establishing a universal state database for Type II appliances holds the potential to enhance the accuracy and applicability of NILM in monitoring and managing the energy consumption of these complex devices across various settings.

\subsection{Absence of Universal Baseline for NILM}
\begin{figure}[ht]
\vspace*{-6pt}
\setlength{\abovecaptionskip}{-.1cm} 
\setlength{\belowcaptionskip}{-2cm} 
    \centering
    \includegraphics[width=0.8\linewidth,scale=1.0]{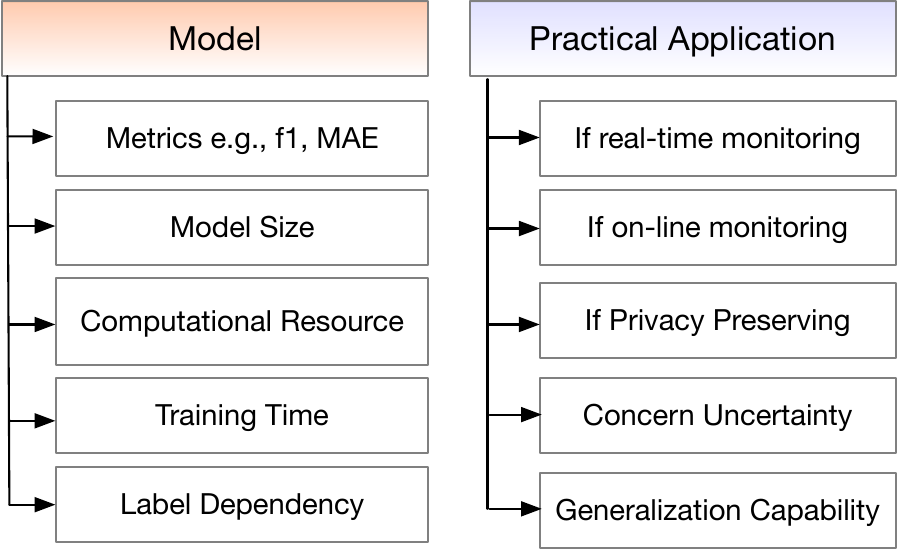}
    \caption{Factors that should be considered when evaluating a NILM algorithm.}
    \label{fig:universal_baseline}
\vspace*{-6pt}
\end{figure}

Despite the development of a large number of machine learning-based algorithms for NILM, there is an absence of a standardized or universally accepted reference point or baseline for evaluating these algorithms. As illustrated in Fig.\ref{fig:universal_baseline}, a comprehensive evaluation of a NILM model should encompass considerations beyond simple metrics like the $f_1$ score or MAE. Factors such as model size, computational resources required for training, training time, and dependency on the number of labeled data should also be taken into account. However, there is currently no established baseline for these metrics and factors to discern the advantages and disadvantages of proposed NILM algorithms. Moreover, the practical capabilities of NILM algorithms, including real-time or online monitoring and privacy preservation considerations, should be thoroughly assessed when introducing a novel algorithm. Additionally, given that individual appliance energy consumption is estimated from aggregated load data, it is crucial to consider the uncertainty of NILM results for practical applications. Lastly, establishing a universal metric and dataset becomes imperative to evaluate the generalization capability of NILM models and assess their performance in real-world scenarios.

\subsection{Generalization Capability Enhancement}

The generalization capability of NILM revolves around the model's ability to extend learned patterns and behaviors to new, unseen environments and appliances. Challenges arise due to inherent variability in household characteristics, electrical infrastructure, and user behaviors. Achieving a high level of generalization requires addressing these variabilities to ensure the model's effectiveness across diverse settings. Some studies have developed algorithms to improve the generalization capability of NILM models, making them more versatile and applicable across a wide range of homes and appliances. By leveraging knowledge gained from one task or domain to enhance the performance of a model on a different, but related task, transfer learning has been adopted to improve the model's generalization capability \cite{liu_nonintrusive_2019, liu_samnet_2022, incecco_transfer_2020}. Instead of training a model from scratch for a specific NILM task, a pre-trained model, often on a larger and more diverse dataset or a different but related NILM task, is used as a starting point. The knowledge gained from pre-training is transferred to the target NILM task, enabling the model to generalize better to new environments, appliances, or households. Another widely adopted technology for enhancing generalization capability is meta-learning, which involves learning to learn for new tasks \cite{luo2023generalizable}. Despite recent improvements, the generalization capability of NILM approaches to different houses as well as the disaggregation of multi-state appliances are still major challenges. 

\begin{figure}[ht]
\vspace*{-10pt}
\setlength{\abovecaptionskip}{-.1cm} 
\setlength{\belowcaptionskip}{-2cm} 
    \centering
    \includegraphics[width=\linewidth,scale=1.0]{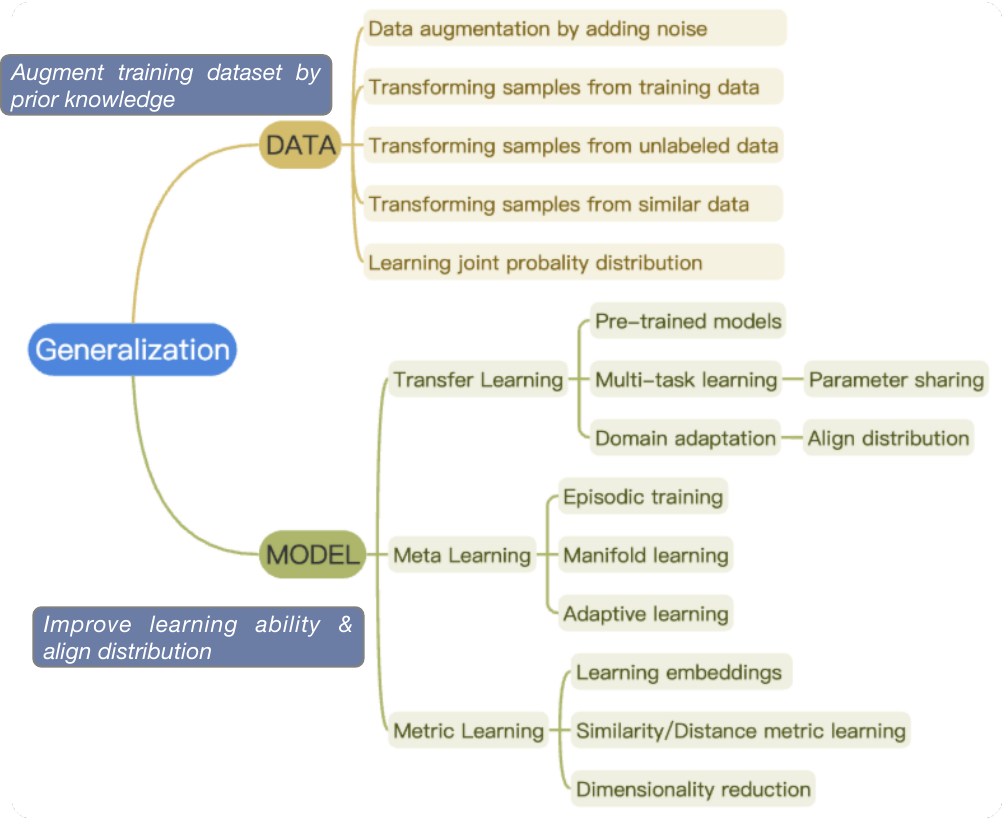}
    \caption{Methods for generalization capability enhancement.}
    \label{fig:generalization}
\vspace*{-8pt}
\end{figure}

Fig. \ref{fig:generalization} summarizes potential methods to enhance the model's generalization capability from data to model aspects. When a model is trained with a small amount of labeled data, it may become overly specialized and struggle to adapt to diverse or unexpected scenarios, compromising its generalization capability. From the data perspective, data augmentation is a technique commonly used in machine learning to artificially increase the diversity of a training dataset by applying various transformations to the existing data to improve model's generalization capability. Applying data augmentation to the training data can enhance the model's ability to handle diverse load patterns, user behaviors, and environmental conditions, contributing to improved generalization when the model is deployed in real-world situations where the data might exhibit variations not fully captured in the original training set \cite{hasan_generalizability_2021}.
From the model and algorithm perspectives, strategies such as transfer learning with pretrained models, multi-task learning with parameter sharing, and domain adaptation to align distributions across different domains, along with meta-learning and metric learning, can be explored to enhance the generalization capability of NILM. Additionally, it's essential for the generalization capability enhancement to consider the detection of new appliances \cite{zhang2019new} or unknown appliances \cite{han2023unknown} for further applications, such as smart home management.

\subsection{Limited Studies for the Applications of NILM}
One challenge arises from the fact that certain applications of NILM, such as anomaly detection and the implementation of NILM for demand response, may not have been thoroughly investigated. Additionally, there are limited studies on the practical applications of NILM in commercial and industrial fields. However, these gaps also present opportunities for researchers to expand the horizons of NILM, explore uncharted use cases, and collaborate with industries for practical implementations. As technological advancements progress, integrating NILM into emerging trends like the Internet of Things (IoT) and smart grids holds promising prospects. Furthermore, customizing NILM applications for specific industries and addressing the disparity between academic research and industry needs can facilitate more effective and targeted implementations, unlocking the untapped potential of NILM in diverse domains.

\section{Conclusion}
This study provides a thorough review of NILM, encompassing a wide range of topics essential to the field. It covers an extensive collection of datasets available for NILM research, delving into the intricacies of feature engineering, which is a key component in NILM analysis. The review also examines various NILM methodologies from multiple perspectives, offering a holistic understanding of the different approaches employed in this area. Additionally, it includes a detailed summary of metrics used for assessing performance, crucial for evaluating NILM systems. Furthermore, the study discusses the applications of NILM, outlining its importance in the context of smart grids. It also addresses the challenges faced in NILM research and anticipates future prospects, thereby highlighting the significant role NILM plays in advancing smart grid technologies. The insights and contributions from various aspects of NILM can be summarized as follows:

\begin{enumerate}[leftmargin=*]
    \item There are already numerous datasets available for NILM research. However, the scope of popular datasets frequently used in NILM research, like UK-DALE, is somewhat restricted in terms of the number of homes covered and the duration of monitoring. This situation suggests that current NILM research is limited to a small group of users and does not fully leverage the extensive range of publicly available datasets. There is significant potential for further research utilizing these underused datasets.
    
    \item In the field of feature engineering for NILM, data augmentation has been shown to improve performance, but this has primarily been observed in specific approaches without a systematic study across various algorithms and appliances. Some research indicates that incorporating reactive power as an additional feature can be beneficial, but there's a lack of thorough investigation into how different features, whether single or multiple, affect various types of loads. Current research on feature selection for NILM is quite limited, often manually selecting a narrow set of features, such as active power, focused on load disaggregation. Furthermore, while studies have explored feature extraction in both time and frequency domains, they have not extensively compared their performance and computational demands to ascertain the most effective approach for NILM.

    \item In this work, NILM approaches are categorized from four perspectives: i) The technologies used for load monitoring, including various machine learning technologies. ii) The classification of NILM approaches based on the presence or absence of labels in the dataset, covering methods like supervised, semi-supervised, and unsupervised NILM. iii) The state/event features utilized in NILM. iv) The direct monitoring target of NILM, such as binary classification for ON/OFF state detection and regression-based NILM for energy consumption estimation.

    \item Regarding NILM approaches, there is limited research that directly compares different supervised and unsupervised algorithms for NILM. Both state- and event-based NILM methods face significant challenges, mainly because they depend on a pre-existing database of signatures or events for all appliances to effectively perform energy disaggregation.

    \item When assessing the performance of NILM algorithms, it's common to use a combination of these metrics to gain a comprehensive understanding of their strengths and weaknesses in estimating power consumption. The choice of metrics may depend on the specific goals and requirements of the monitoring application.

    \item NILM still faces numerous challenges, including the need in current machine learning-based NILM algorithms to train a separate model for each specific appliance. Additionally, managing the complex states of Type II appliances presents significant difficulties. Moreover, the lack of a universal baseline for NILM adds to these challenges, making standardization and comparison across different studies and approaches problematic.
\end{enumerate}

\section*{Acknowledgments}
The authors acknowledge and appreciate the use of Artificial Intelligence -Generated Text (ChatGPT) as a helpful tool for grammar revision.

\bibliographystyle{IEEEtran}

\bibliography{references.bib}

\end{document}